\documentclass[aps,prx,reprint,groupedaddress,superscriptaddress,floatfix]{revtex4-2}

\usepackage[T1]{fontenc}
\usepackage{newtxtext}
\usepackage{newtxmath}
\usepackage{bm}

\usepackage{titlesec}

\usepackage{amsmath}
\usepackage{dsfont}
\usepackage{mathtools}
\usepackage{amsfonts}
\usepackage{bbold}
\usepackage{physics}

\usepackage{color}
\usepackage{xcolor}
\usepackage{soul}
\usepackage[colorlinks,citecolor=blue,linkcolor=blue,urlcolor=blue]{hyperref}

\usepackage{graphicx}
\usepackage[all]{hypcap}

\usepackage{enumitem}

\usepackage{tabularx}
\usepackage{makecell}

\graphicspath{{./img/}}

\usepackage{hyperref}
\usepackage{lineno}
\usepackage{cleveref}
\usepackage{etoolbox}

\makeatletter
\let\addcontentslineOriginal\addcontentsline
\let\addcontentsline\@gobblethree
\makeatother

\begin{document}

\title{Study of double kicked top: a classical and quantum perspective}

\author{Avadhut V. Purohit}
\email{avdhoot.purohit@gmail.com}
\affiliation{Department of Physics, Visvesvaraya National Institute of Technology, Nagpur 440010, India}
 \author{Udaysinh T. Bhosale}
\email{udaysinhbhosale@phy.vnit.ac.in}
\affiliation{Department of Physics, Visvesvaraya National Institute of Technology, Nagpur 440010, India}
\date{\today}

\begin{abstract}
    The double kicked top (DKT) is studied as an extension of the standard quantum kicked top (QKT) model. The model allows us to study the transition from time-reversal symmetric to broken time-reversal symmetric dynamics. A transformation in the kick strength parameter space $(k, k') \to (k_r, k_\theta)$ reveals interesting features. The transformed kicked strength parameter $k_r$ drives a higher growth of chaos and is equivalent to the standard QKT, whereas $k_\theta$ leads to a weaker growth. Fixed points and their stability are analysed both analytically and computationally by using the largest Lyapunov exponent (LLE) and the Kolmogorov-Sinai entropy (KSE). Exact solutions are obtained for 2- to 4-qubit version of the DKT, including eigenvalues, eigenvectors, and entanglement dynamics. Criteria for periodicity of the entanglement dynamics is obtained. Quantum correlations are investigated in both deep quantum and semi-classical regime. Signatures of phase-space structure are numerically shown in the long-time averages of the quantum correlations. The homoclinic point and its associated state are also investigated within the semi-classical regime. Our model can be realised experimentally as an extension of the standard QKT.
\end{abstract}

\maketitle
\section{Introduction}
The seminal model of kicked top \citep{haake1987classical} has played an important role in the understanding of classical and quantum chaos \citep{haake1992lyapunov,Haake1991Quantum,bohigas1984characterization,wang2004entanglement,Lombardi2011,ruebeck_entanglement_classical,Bandyopadhyay2004,bandyopadhyay2002testing,madhok2015signatures,bhattacharya2000continuous,bhosale_signatures_bifurcation,Kumari2018,dogra2019quantum,chaudhary_quantum_signatures,krithika2023nmr,krithika2022quantum,kumari_untangling_entanglement,bhosale_periodicity_correlations,neil2016ergodic,Poggi,Meier2019}. It consists of a top constantly precessing around a magnetic field and undergoing a periodic sequence of impulsive non-linear kicks. For small values of the kick strength $k$, it shows regular behaviour in the phase-space and gradually becomes chaotic for higher values of $k$. Applications of random matrix theory (RMT) \citep{wigner1951statistical,wigner1955characteristic} showed that the QKT follows Poisson statistics in the regular region and the Gaussian orthogonal ensemble statistics when it is fully chaotic \citep{mehta2004random,BhosaleUniversalScaling,Haake1991Quantum}. The Hilbert space dimension of QKT is finite; thus, it does not suffer from the different statistical properties arising due to different truncation schemes employed for systems having infinite-dimensional Hilbert space \citep{haake1987classical,Izrailev1989}.

The RMT reveals that the statistics of eigenvalues of the Floquet operator $U$ for the kicked top models fall under different universality classes depending on the existence or absence of non-conventional time-reversal symmetry \citep{haake1987classical,bohigas1984characterization,berry1986statistics,mehta2004random}. Although a broken time-reversal symmetric extension of the standard QKT was proposed \citep{haake1987classical} to study quadratic level repulsion, not much classical, quantum, or semi-classical analysis has been conducted.

The advantage of the standard QKT model is that it can also be written as a spin Hamiltonian where the nature of interaction is Ising and all-to-all \citep{dogra2019quantum,Ghose2008}. This property facilitates its analysis from the perspective of quantum information theory. The quantum correlations as signatures of chaos have been investigated earlier \citep{ruebeck_entanglement_classical,wang2004entanglement,Madhok2015,neil2016ergodic,krithika2022quantum,krithika2023nmr,chaudhary_quantum_signatures,Meier2019}. Studies have shown that infinite-time averages of the quantum correlations reproduce coarse-grained phase-space structures of the kicked top \citep{madhok_dogra_correlations}. Such correspondence is observed even for two qubits \cite{ruebeck_entanglement_classical}. Remarkably, 2- to 4-qubit systems are exactly solvable and show signatures of chaos \citep{dogra2019quantum}. The QKT has been implemented in various experimental setups \citep{neil2016ergodic,krithika2022quantum,krithika2023nmr,chaudhary_quantum_signatures,Meier2019}.

Among the two major approaches to quantum chaos, one looks for the quantum analogue of the Lyapunov exponent \citep{Li2017,Carlos2019,Vedika2018,Trunin2023}. On the contrary to the general understanding some models (such as inverted harmonic oscillator, Lipkin-Meshkov-Glick and Dicke model) show that the regular dynamics can have exponentially growing out-of-time-correlator \citep{Hashimoto2020,Cameo2020}. The other approach looks for the phase-space trajectory-independent description of quantum chaos \citep{Zarum1998,Madhok2015,Kumari2018,Yan2022,Trail2008,Lombardi2011,Madhok2015comment,Lombdardi2015,Akshay2018}. In this approach, various quantum correlations such as the entanglement entropy, have been explored as signatures of chaos. The phase-space averaged entanglement entropy shows excellent agreement with the classical chaos indicators \citep{Zarum1998,neil2016ergodic,madhok_dogra_correlations,ruebeck_entanglement_classical}. The quasi periodic behaviour of quantum discord has been observed in the regular region. Whereas in the chaotic region neither periodic nor quasi periodic behaviour is observed \citep{Madhok2015}. Quantum-correlations measures are capable of showing signatures of bifurcations \citep{bhosale_signatures_bifurcation}.

Controlling classical and quantum chaos is one of the major reasons to study it \citep{grebogi1990controlling,Tomsovic2023Jan,Tomsovic2023Oct,Beringer2024,Gruebele2007,Eastman2022Tuning}. This includes atmospheric systems, biological systems, and kicked rotor \citep{ferreira2011chaos, schiff1996controlling, Tomsovic2023Jan, Tomsovic2023Oct}. The most widely used technique to control chaos is known as \textit{targeting} \citep{Tomsovic2023Jan,Tomsovic2023Oct,Eastman2022Tuning,Beringer2024}. The procedure involves the application of a weak perturbation to the chaotic system. Continuous quantum measurements have also been used to control chaos \citep{Eastman2022Tuning}.

In this work, we study the extension of QKT by breaking its time-reversal symmetry \citep{haake1987classical}. Here, the QKT further suffers periodic sequences of impulsive non-linear kicks of strength denoted by $k'$, perpendicular to both the precession axis and the first non-linear kick. Our transformation $(k, k')\to (k_r, k_\theta)$ eases the classical dynamics by dividing it into two distinct parts. The dynamics corresponding to $k_r$, for a fixed value of $k_\theta$, is similar to that of the standard kicked top. Whereas the parameter $k_\theta$ twists the phase-space regions around the trivial fixed points. This parameter enables us to investigate the transition from time-reversal symmetry to its lack thereof. Here, we investigate the fixed points and their stability. The results are then compared with the computed LLE \citep{p-spin_poggi,Constantoudis} and the KSE \citep{Zarum1998}. It is shown that when the standard kicked top (i.e., the DKT with $k'=0$) exhibits chaotic behaviour for a certain kick strength $k$, the choice of another kick strength $k'=-k$ minimizes classical chaos. 

Similar to QKT \citep{sharma2024exactly,dogra2019quantum,Ghose2008}, the DKT can also be expressed as a system of qubits. We exactly solve for eigenvalues, eigenstates, and entanglement dynamics of the 2- to 4-qubit system. The study extends exploration of quantum correlations as signatures of chaos to the broken time-reversal symmetry. Conditions under which the DKT shows periodic entanglement dynamics in 2- to 4-qubit systems \citep{owusu2008link,doikou2010introduction,gubin2012quantum,yuzbashyan2013quantum} are obtained here.
If the entanglement dynamics of QKT (DKT with $k'=0$) is not periodic for a particular value of the kick strength $k$. Then, in the corresponding DKT, it can be made periodic by choosing $k'$ such that $k_r = (k + k')/2=j \pi/2$. This will be discussed in the further parts of the paper.

In the semi-classical analysis, we numerically investigate the behaviour of von Neumann entropy, quantum discord, and concurrence as functions of $k_r$ and $k_\theta$. The obtained results are first compared among these quantum correlations, then with those from the deep quantum regime, and finally with the KSE. The high-spin systems show a fine-grained phase-space structures. Nevertheless, quantum effects remain significant for certain values of $j$, $k_r$ and $k_\theta$. Our findings indicate that $k_\theta$ shows features that are absent in the corresponding classical dynamics. In addition, we study homoclinic point to see if quantum effects change the stability of the corresponding state.

The structure of this paper is as follows: Sec. \ref{sec:dkt} defines the Floquet operator and the classical dynamics of our model. The fixed points and their stability are examined in Subsection \ref{subsec:stability}. The symmetries are covered in Subsection \ref{subsec:symm}. To characterise chaos, Subsection \ref{subsec:lle} computes the LLE and the KSE. In Sec. \ref{sec:mbfs} the many-body version of our model having all-to-all and Ising interaction is defined. In Sec \ref{sec:qm} relevant measures of quantum correlations are defined. In Secs. \ref{sec:2qubit}, \ref{sec:3qubit}, and \ref{sec:4qubit} entanglement and its dynamics are studied using linear entropy for 2, 3, and 4 qubits (deep quantum regime), respectively. Sec. \ref{sec:3qubit} also studies quantum correlations, such as von Neumann entropy, concurrence, and quantum discord. The von Neumann entropy is computed in Secs. \ref{sec:2qubit} and \ref{sec:4qubit} to check the consistency with the results of Sec. \ref{sec:3qubit}. Sec. \ref{sec:hs} analyses these quantum measures in the semi-classical regime to compare results obtained with the classical chaos indicators. Results are summarised in Sec. \ref{sec:r}, and Sec. \ref{sec:d} concludes the paper with a discussion.

\section{Double kicked top model}\label{sec:dkt}
\begin{figure}
    \includegraphics[width=\linewidth]{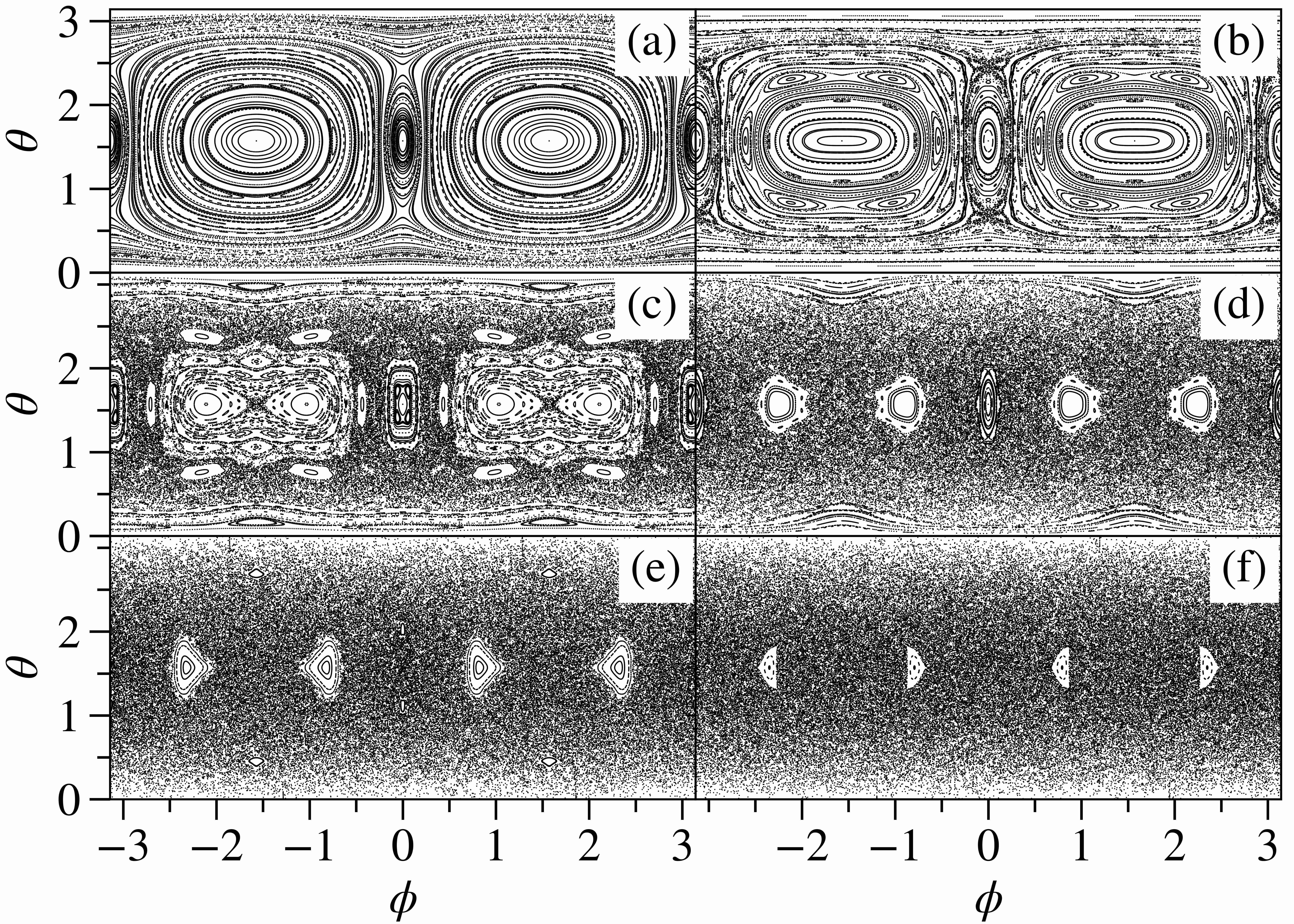}
    \caption{The phase-space portrait with $25\times 25$ initial conditions and 200 kicks is presented for $k_\theta = 0$, (a) $k_r = 0.75$, (b) $k_r = 1.0$, (c) $k_r = 1.25$, (d) $k_r = 1.5$, (e) $k_r = 1.75$ and (f) $k_r = 2.0$.}\label{ppkr}
\end{figure}

The classical dynamics corresponding to the Floquet operator is studied in this section. As discussed in the earlier section, the first non-linear kick, denoted by $k$, produces a torque around the $z$-axis. The second kick, denoted by $k'$, produces a torque around the $x$-axis, immediately following the first. The Floquet operator for this DKT is given by
\begin{equation}\label{U}
    \mathcal{U} = \exp\left(- i \frac{k'}{2j}J_x^2\right)\exp\left(- i \frac{k}{2j}J_z^2\right)\exp\left(- i \frac{\pi}{2} J_y\right).
\end{equation}
This operator leads to the classical map $\mathbf{X}'=\mathbf{F}(\mathbf{X})$, described by the following equations (see supplementary material for detailed derivation \citep{supplementary2025}):
\begin{widetext}
  \begin{align}
    X' =& Z \cos(k X) + Y \sin(k X),\\
    Y' =& \left[Y \cos(k X) - Z \sin(k X)\right] \cos\left[k'\left[Z \cos(k X) + Y \sin(k X)\right]\right] + X \sin\left[k'\left[Z \cos(k X) + Y \sin(k X)\right]\right],\\
    Z' =& - X \cos\left[k'\left[Z \cos(k X) + Y \sin(k X)\right]\right] + \left[Y \cos(k X) - Z \sin(k X)\right] \sin\left[k'\left[Z \cos(k X) + Y \sin(k X)\right]\right],
  \end{align}
\end{widetext}
where $X=\frac{J_x}{j}$, $Y=\frac{J_y}{j}$, and $Z=\frac{J_z}{j}$. The transformation given by
\begin{equation}\label{kkpTransform}
    k_{r} = \frac{k + k'}{2}, \quad k_{\theta} = \frac{k - k'}{2},
\end{equation}
resolves the dynamics by identifying two distinct kinds of motions. Here, the transformed kick strength parameter $k_r$ is responsible for bifurcations, as shown in Fig.~\ref{ppkr}. This motion as a function of $k_r$ can be considered as a \textit{radial} in the sense of outwards bifurcations of the stable orbits. The other parameter, $k_\theta$, \textit{rotates} the phase-space structures around the fixed points $(0, \pm 1, 0)$ and $(\pm 1, 0, 0)$, as illustrated in Fig.~\ref{ppktheta}.

The special case $k_\theta = k_r$ can be translated to $(k' = 0, k \neq 0)$ gives us the standard kicked top \citep{haake1987classical}:
\begin{equation}\label{Tsymm1}
    \mathcal{U} = e^{-i \frac{k}{2j}J_z^2} e^{-ipJ_y}.
\end{equation}
Conversely, the case of $k_\theta = -k_r$ implies $(k = 0, k' \neq 0)$, gives us the following kicked top:
\begin{equation}\label{Tsymm2}
    \mathcal{U} = e^{-i \frac{k'}{2j}J_x^2} e^{-ipJ_y}.
\end{equation}
The transformation in Eq. (\ref{kkpTransform}) establishes a connection between time-reversal symmetric kicked top with the broken time-reversal symmetric kicked top. 
\begin{figure}
    \includegraphics[width=\linewidth]{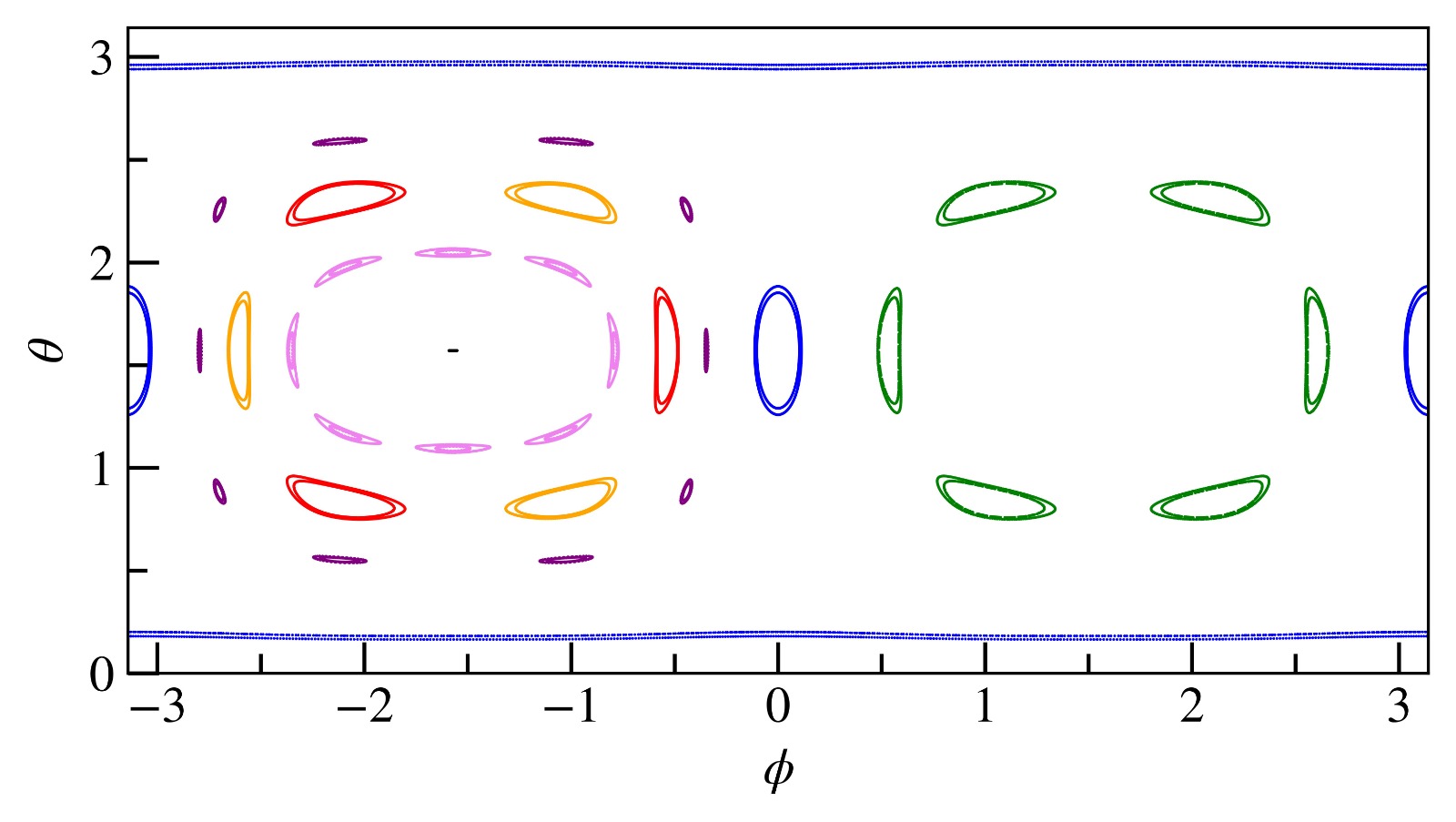}
    \caption{The phase-space portrait show pairs of red and orange coloured period-3 orbits in the negative $\phi$ domain. The period-4 orbits are shown in blue colour. The positive $\phi$ domain show period-6 orbits in green colour. The period-8 orbits and the period-10 orbits are shown in violet and purple colour respectively. Here, the kick strength parameters are $k_\theta = 0$ and $k_r = 1.0$.}\label{ppkr1}
\end{figure}
\begin{figure*}
    \includegraphics[width=1\linewidth]{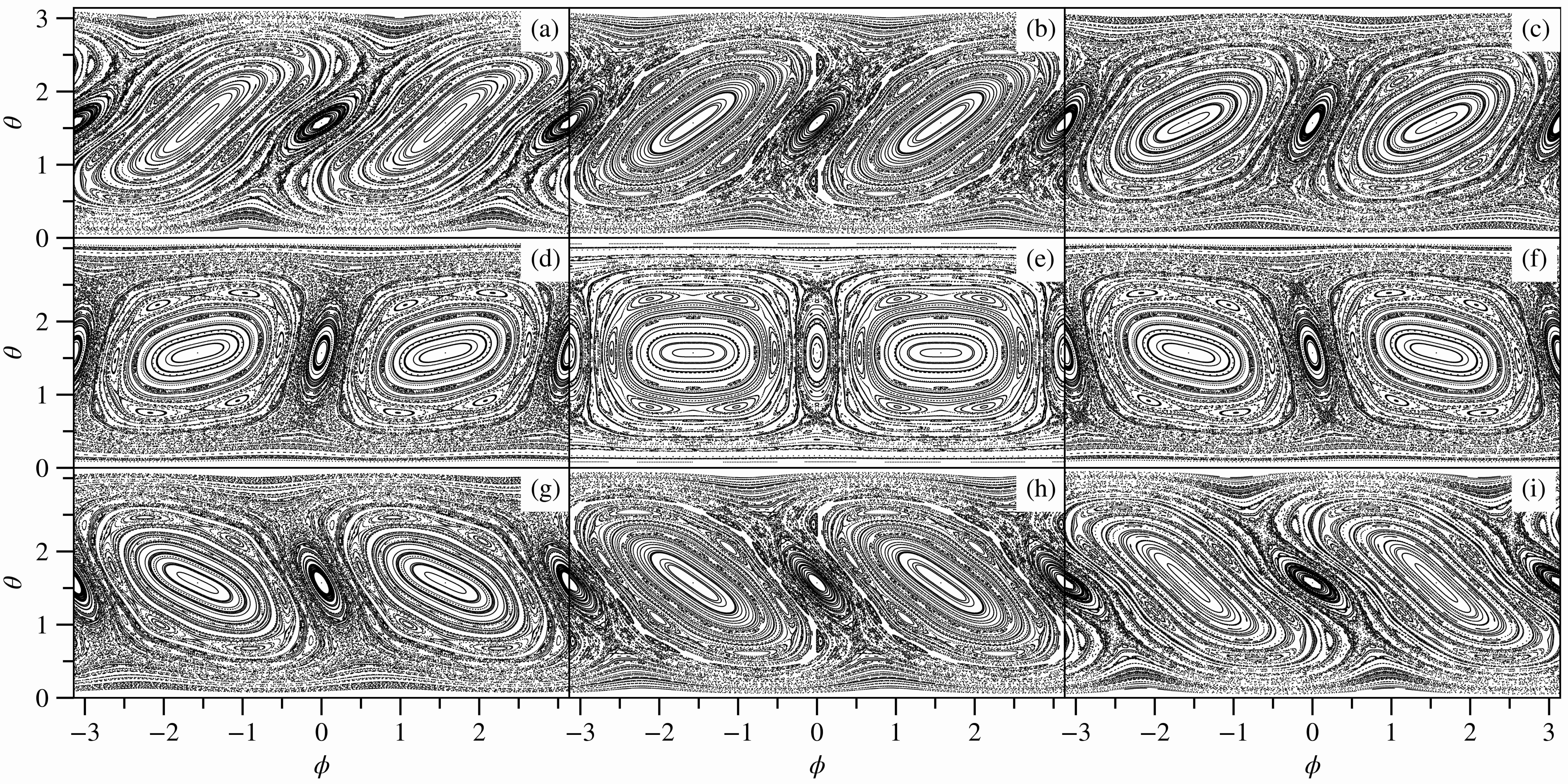}
    \caption{The phase-space portrait with $25\times 25$ initial conditions and 200 kicks is presented for $k_r = 1$, (a) $k_\theta = -1$, (b) $k_\theta = -0.75$, (c) $k_\theta = -0.5$, (d) $k_\theta = -0.25$, (e) $k_\theta = 0$, (f) $k_\theta = 0.25$, (g) $k_\theta = 0.5$, (h) $k_\theta = 0.75$ and (i) $k_\theta = 1$.}\label{ppktheta}
\end{figure*}

The phase-space at $k_r = 1$ shows a complex and rich structure as shown in Fig.~\ref{ppkr}(b). Therefore, we fix $k_r = 1$ while analysing $k_\theta$ (see Fig.~\ref{ppktheta}). It can be observed that points in the phase-space rotate clockwise around the fixed points $(\pm 1, 0, 0)$ and $(0, \pm 1, 0)$ as $k_\theta$ increases from $-1$ to $1$. This is clearly illustrated in Fig.~\ref{ppktheta}. In the subsection \ref{subsec:symm}, we show that Figs.~\ref{ppktheta}(a), \ref{ppktheta}(b), \ref{ppktheta}(c), and \ref{ppktheta}(d) are isomorphic to Figs.~\ref{ppktheta}(i), \ref{ppktheta}(h), \ref{ppktheta}(g), and \ref{ppktheta}(f), respectively. Notably, the chaotic boundaries of the period-4 cycles are narrowest when $k_\theta = 0$, as shown in Fig.~\ref{ppktheta}(e), and these boundaries become broader with $|k_\theta|$.

As we shall see later in this section, the case $k_\theta = 0$ is more stable compared to other cases, i.e., $k_\theta \neq 0$. Hence, we fix $k_\theta = 0$ while analysing the dynamics for $k_r$ (see Fig.~\ref{ppkr}). At small values of $k_r$, the kicked top shows regular behaviour (see Fig.~\ref{ppkr}(a)). For $k_r = 1$, the fixed points at $Y = \pm 1$ are marginally stable. Around these bifurcated fixed points, a pair of red and orange coloured period-3 orbits can be observed in the negative $\phi$ domain (see Fig.~\ref{ppkr1}). Due to $R_y(\pi)$ symmetry, the positive $\phi$ domain has a corresponding green-coloured period-6 orbit. The higher-order orbits including purple-coloured period-8 orbits and violet-coloured period-10 orbits emerge, occupying significant portions of the phase-space (see Fig.~\ref{ppkr1}). Thin chaotic regions (see Fig.~\ref{ppkr}(b)) begin to form at the boundaries of blue-coloured period-4 orbits (see Fig.~\ref{ppkr1}). Points $(\theta_0 = \pi/2, \phi_0 = \pm \pi/2)$ become hyperbolic at $k_r = 1.25$ with chaotic neighbourhood as shown in Fig.~\ref{ppkr}(c). Although the region around bifurcated orbits, period-3 orbits, and period-6 orbits shrink, orbits remain stable. Chaos becomes dominant at $k_r = 1.5$, leaving only the bifurcated orbits and the equatorial period-4 cycles as stable features (see Fig.~\ref{ppkr}(d)). By the time kick strength reaches $k_r = 2$, even the equatorial period-4 cycles lose their stability, leaving only a narrow regular region around the bifurcated fixed points (see Fig.~\ref{ppkr}(f)). For $k_r > 3$, the system transitions into a fully chaotic state.

\subsection{Stability Analysis}\label{subsec:stability}
The trivial fixed points of our map correspond to poles:
\begin{equation}
    X = Z = 0, \, Y = \pm 1.
\end{equation}
These poles remain invariant under the transformation $\mathbf{F}$ for any value of $k$ and $k'$. Non-trivial fixed points, however, emerge for sufficiently large $k$ and/or $k'$ and are determined by solving $\mathbf{F(X)} = \mathbf{X}$. Thus, the fixed point $\mathbf{X} = (X, Y, Z)$ is given by
\begin{align}\label{fixedpoints0}
    Z =& -X \sin\left(\frac{k-k'}{2}X\right)\csc\left(\frac{k+k'}{2}X\right),\nonumber \\
    Y =& X \cos\left(\frac{k-k'}{2}X\right)\csc\left(\frac{k+k'}{2}X\right) \, \text{and} \\
    f(X) =& \frac{\sin^2\left(\dfrac{k+k'}{2}X\right)}{1 + \sin^2\left(\dfrac{k+k'}{2}X\right)} - X^2 = 0.\nonumber
\end{align}
These fixed points occur in pairs, with one member of each pair generated from the other by an $R_y(\pi)$ rotation. Thus, it suffices to consider only the fixed points with $X > 0$. 

Since $f(X)$ involves transcendental equations, one can find the fixed points through graphical analysis. We approximate $f(x)$ for small values of $x = \left(\dfrac{k+k'}{2}\right)X \ll 1$, to get $f(x) \approx x^2 - {\dfrac{4}{{(k + k')}^2}}x^2$. Thus, the first non-trivial solution is given by
\begin{equation}
    {\left(k + k'\right)}_0 = 2.
\end{equation}
Additional pairs of non-trivial fixed points ${\left(k + k'\right)}_m$ can be determined (graphically) by solving $f'(X)=0$ in conjunction with $f(X)=0$ and ensuring $X^2 + Y^2 + Z^2 = 1$.

To analyse the stability of the non-trivial fixed points, we first linearise the map $\mathbf{F}$. This involves evaluating the tangent map at the fixed point $\mathbf{X}_n$, whose stability is to be determined.
\begin{equation}
    \mathbf{M}(\mathbf{X}_n) = \frac{\partial \mathbf{F}\left(\mathbf{X}_{n}\right)}{\partial \mathbf{X}_{n}}.
\end{equation} 
Among three eigenvalues of the above tangent map, one is always equal to unity due to $|\mathbf{X}|^2 = 1$. The modulus of other two eigenvalues is also unity if they satisfy
\begin{equation}
    \bigg| \left(k+k'\right) X \cot\left(\frac{k+k'}{2}X\right)  + \cos\left[(k + k')X\right]  - 1\bigg| < 2.
\end{equation}
Since we know that the first term in the $f(X)$ is maximum at $x = \left(\dfrac{k+k'}{2}\right)X = \dfrac{\pi}{2} + m \pi$ for any integer $m$, solving $f(X) = 0$ gives us the maximum value of $\widetilde{\left(k + k'\right)}_0 = \sqrt{2}\pi$ after which, the emerging fixed points lose their stability.   

This shows that the non-trivial fixed points emerging at $\left(k+k'\right) \geq 2$ remain stable until $\left(k + k'\right) \leq \sqrt{2}\pi$. Among the pair of fixed points appearing at $\left(k+k'\right) > {\left(k+k'\right)}_m$, the fixed points with small $X$ are unstable, while the others are stable for 
\begin{equation}
    {\left(k+k'\right)}_m \leq \left(k+k'\right) \leq \widetilde{\left(k + k'\right)}_m = (2m+1)\sqrt{2}\pi.
\end{equation}
Beyond this range, these fixed points lose stability and bifurcate into period-2 solutions (see bifurcated orbits in the positive $\phi$ domain of Fig.~\ref{ppkr}(c)), obtained as follows:
\begin{align}
    X_1 &= \frac{\left(2j+1\right)\pi}{\left(k+k'\right)} = - Z_1,\;
    Y_1 = \sqrt{1 - 2\frac{{\left(2j+1\right)}^2 \pi^2}{{\left(k+k'\right)}^2}}, \nonumber  \\ 
    X_2 &= \frac{\left(2j+1\right)\pi}{\left(k+k'\right)} = - Z_2, \;
    Y_2 = -\sqrt{1 - 2\frac{{\left(2j+1\right)}^2 \pi^2}{{\left(k+k'\right)}^2}}. \nonumber 
\end{align}
As $\left(k+k'\right)$ increases further, these orbits lose stability, leading to period-4 orbits. For even higher values of $\left(k+k'\right)$, a cascade of period-doubling bifurcations occurs. A blue-coloured period-4 cycles (see Fig.~\ref{ppkr1}) exists on the equator, with $\mathbf{X} = (0, 0, \pm 1)$ and $\mathbf{X} = (\pm 1, 0, 0)$. These period-4 orbits remain stable as long as $(k + k')$ satisfies the following condition:
\begin{equation}
    \bigg| \frac{{\left(k+k'\right)}^2}{2} \sin^2\left(\frac{k+k'}{2}\right) + 4 \left[\sin(k+k') + \cos(k+k')\right] \bigg| < 4.
\end{equation}
Following this analysis, one can find higher $n$-cycles, including the period-3 cycles. However, due to their complexity, the resulting expressions are not presented here. We can recover the standard QKT results \citep{haake1987classical} by setting $k' = 0$ in the above analysis.

\subsection{Symmetries}\label{subsec:symm}
Our kicked top model can be divided into four quadrants in the parameter space of $(k, k')$. The first quadrant corresponds to $(k > 0, k' > 0)$, the second quadrant to $(k > 0, k' < 0)$, the third quadrant to $(k < 0, k' > 0)$, and the fourth quadrant to $(k < 0, k' < 0)$. Due to the symmetry of the map $\mathbf{F}$ under a rotation around the precession axis by an angle $\pi$, the first and fourth quadrants $(k, k') \cong (-k, -k')$ are isomorphic, as are the second and third quadrants $(k, -k') \cong (-k, k')$. However, the first quadrant $(k, k')$ is not isomorphic to the second quadrant $(k, -k')$.

To show this distinction between two quadrants discussed above, we rewrite Eq. (\ref{fixedpoints0}) in terms of $(k_r, k_\theta)$ as follows:
\begin{align}\label{fixedpointsSet1}
    Z =& -X \sin\left(k_\theta X\right)\csc\left(k_r X\right),\nonumber \\
    Y =& X \cos\left(k_\theta X\right)\csc\left(k_r X\right) \; \text{and} \\
    f(X) =& \frac{\sin^2\left(k_r X\right)}{1 + \sin^2\left(k_r X\right)} - X^2 = 0.\nonumber
\end{align}
The second quadrant is related to the first quadrant by the transformation $(k, k') \to (k, -k')$, which is equivalent to the transformation $(k_r, k_\theta) \to (k_\theta, k_r)$. Therefore, in the second quadrant, equations for the fixed points become:
\begin{align}\label{fixedpointsSet2}
    Z =& -X \sin\left(k_r X\right)\csc\left(k_\theta X\right),\nonumber \\
    Y =& X \cos\left(k_r X\right)\csc\left(k_\theta X\right) \; \text{and} \\
    f(X) =& \frac{\sin^2\left(k_\theta X\right)}{1 + \sin^2\left(k_\theta X\right)} - X^2 = 0.\nonumber
\end{align}
It is clear that $(k, k') \ncong (k, -k')$, meaning that we have two distinct classes of solutions, one of which is related to the other by the transformation $(k_r, k_\theta) \to (k_\theta, k_r)$.

\begin{figure}
    \includegraphics[width=1\linewidth]{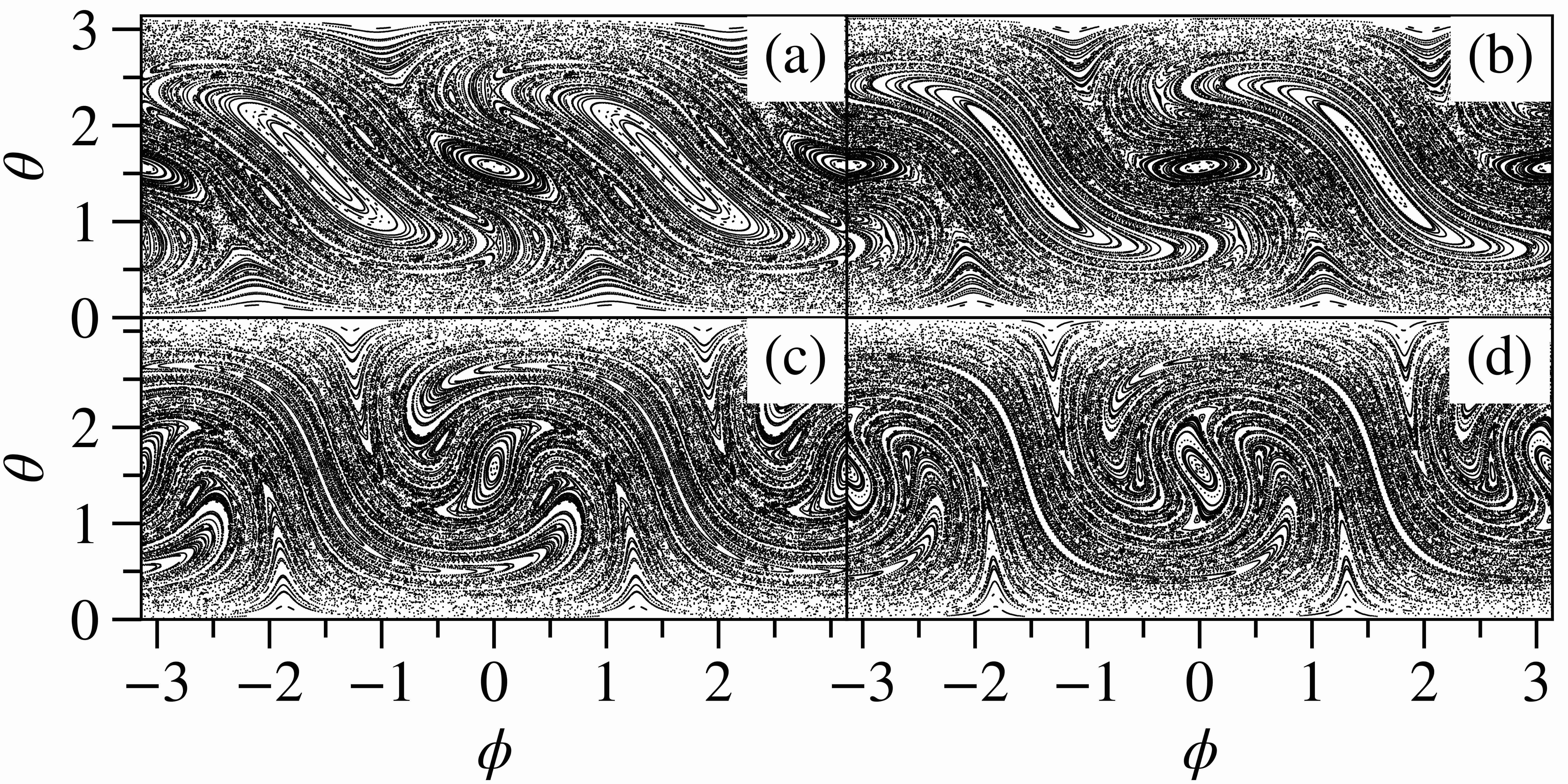}
    \caption{The phase-space portrait with $25\times 25$ initial conditions and 200 kicks is presented for $k_r = 1.0$, (a) $k_\theta = 1.25$, (b) $k_\theta = 1.75$, (c) $k_\theta = 3.0$ and (d) $k_\theta = 3.75$.}\label{ppkthetaTwist}
\end{figure}

In the first quadrant, with $k_\theta \in (-k_r, k_r)$, all points in the phase-space rotate with the same angular velocity around the trivial fixed points. However, in the second quadrant, when $k_\theta > k_r$, torsion occurs around trivial fixed points, as shown in Fig.~\ref{ppkthetaTwist}.

As mentioned earlier, the classical map $\mathbf{F}$ is invariant under rotations around the precession axis by an angle $\pi$:
\begin{equation}
    R_x\begin{pmatrix}
        X \\ Y \\ Z
        \end{pmatrix} = \begin{pmatrix}
            X \\ -Y \\ -Z
        \end{pmatrix},
    R_y\begin{pmatrix}
        X \\ Y \\ Z
    \end{pmatrix} = \begin{pmatrix}
        -X \\ Y \\ -Z
    \end{pmatrix}.
\end{equation}
This yields the following symmetry relations:
\begin{align}
    R_y \cdot \mathbf{F} =& \mathbf{F} \cdot R_y, \\
    \mathbf{F} \cdot R_x =& R_x \cdot \mathbf{F} \cdot R_y \, \text{and} \nonumber  \\
    \mathbf{F}^2 \cdot R_x =& R_x \cdot \mathbf{F}^2.
\end{align}
The $R_y$-image of every $n$-cycle of $\mathbf{F}$ is an $n$-cycle of $\mathbf{F}$. Furthermore, if a $(2n+1)$-th cycle is not symmetric under $R_y$, then every such $(2n+1)$-cycle, along with its $R_y$-image, is mapped by $R_x$ into a $2(2n+1)$-cycle of $\mathbf{F}$.

The Floquet operator with $(k \neq 0, k' = 0)$ is the standard QKT \citep{haake1987classical} and shows a non-conventional \citep{Haake1991Quantum} time-reversal symmetry:
\begin{align}
        T \cdot \left(e^{-i \frac{k}{2j} J_z^2} e^{-ip J_y}\right) \cdot T^{-1} =& e^{ip J_y} e^{i \frac{k}{2j} J_z^2}, \nonumber  \\ 
        =& \left(e^{-i \frac{k}{2j} J_z^2} e^{-ip J_y}\right)^\dagger,
\end{align}
where $T = \left(e^{ip J_y} e^{i \pi J_z}\right) K$ is the anti-unitary, non-conventional time-reversal operator (see supplementary material for the derivation \citep{supplementary2025}), and $K$ is the conjugation operator. We found absence of the non-conventional time-reversibility of our kicked top :
\begin{align}\label{TReversal}
    T \cdot \mathcal{U} \cdot T^{-1} =& \exp(ip J_y) \exp(i \frac{k'}{2j} J_x^2) \exp(i \frac{k}{2j} J_z^2)  \nonumber\\  
    \neq& \mathcal{U}^\dagger, \hspace*{1cm} \text{for } k \neq k' \neq  0.
\end{align}
The time-reversal symmetry can be achieved for $k = 0$ or $k' = 0$. However, the case $k' = k$ also has time-reversal symmetry for the following non-conventional time-reversal operator:
\begin{equation}
    T = \left(e^{ip J_y} e^{i\frac{\pi}{2} J_y} e^{i \pi J_z}\right) K.
\end{equation}

Due to the observation of a linear level repulsion for the case $k' = k$, Haake et al. \citep{haake1987classical} suggested the existence of a time-reversal operator. But its explicit expression was not given. In supplementary material, we derive the above non-conventional time-reversal operator \citep{supplementary2025}.

\subsection{LLE results}\label{subsec:lle}
In this subsection, we examine the chaotic behaviour of the system by calculating the largest Lyapunov exponent (LLE), denoted by $\lambda_+$, which quantifies how close initial points in the phase-space diverge over time. Computing the full spectrum of Lyapunov exponents is computationally intensive. However, the LLE dominates the system's behaviour and provides sufficient insight into its chaotic nature. Therefore, we use the LLE unless a more sensitive measure of chaos is needed. This method relies on the divergence of trajectories in phase-space, where the Lyapunov exponent quantifies the rate at which nearby trajectories separate. Following the work of Ref. \cite*{Constantoudis,Robnik}, we define:
\begin{equation}
  \lambda_+ = \lim_{n \rightarrow \infty} \frac{1}{n} \ln \left[\frac{||\delta \mathbf{X}_n||}{||\delta \mathbf{X}_0||}\right].
\end{equation}
Here, $\delta \mathbf{X}_n$ is the tangent vector evolved at time $n$, which depends on the product of tangent maps. These tangent maps govern the divergence of nearby trajectories, reflecting the system's chaotic nature. The time-evolved tangent vector $\delta \mathbf{X}_n$ is given by
\begin{equation}
  \delta \mathbf{X}_n = \mathcal{T}\left[\mathbf{X}_{n-1}\right]\cdot \delta \mathbf{X}_{n-1} = \prod_{l=0}^{n-1} \mathcal{T}\left[\mathbf{X}_{l}\right]\cdot \delta \mathbf{X}_{0},
\end{equation}
where,
\begin{align}
    \mathcal{T}\left[\mathbf{X}_{n-1}\right] =& \frac{\delta \mathbf{X}_n}{\delta \mathbf{X}_{n-1}}, \\
    \prod_{l=0}^{n-1} \mathcal{T}\left[\mathbf{X}_{l}\right] =& \mathcal{T}\left[\mathbf{X}_{n-1}\right]\cdot \mathcal{T}\left[\mathbf{X}_{n-2}\right] \dots \mathcal{T}\left[\mathbf{X}_{0}\right]. \nonumber
\end{align}
We compute the LLE for a set of kick strengths to compare the phase portraits presented earlier. 

\begin{figure}
    \includegraphics[width=\linewidth]{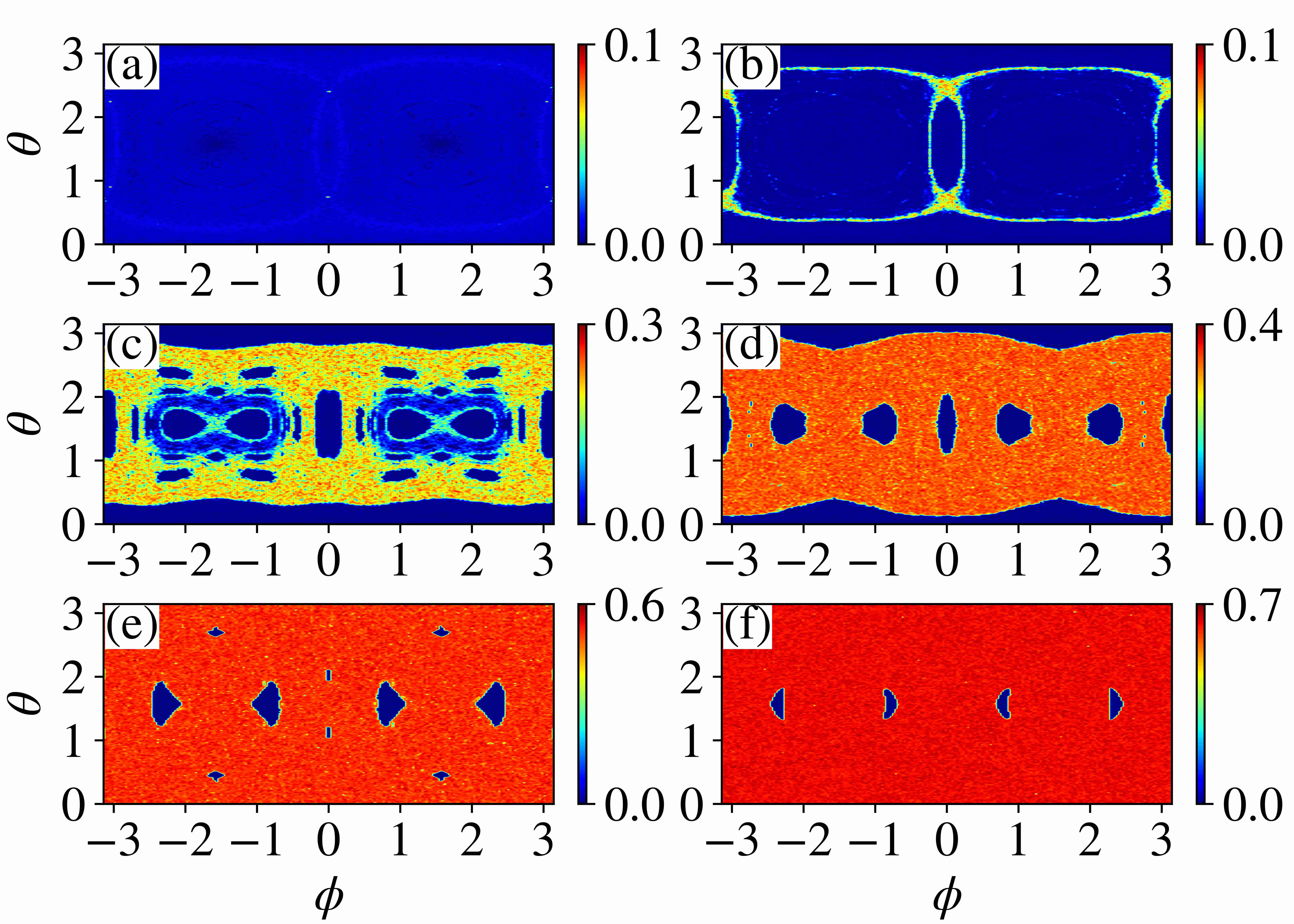}
    \caption{The LLE ($\lambda_+$) in the unit phase-space of $200\times 200$ grid points is presented. Each point in the phase-space is evolved for 1500 kicks keeping $k_\theta = 0$, (a) $k_r = 0.75$, (b) $k_r = 1.0$, (c) $k_r = 1.25$, (d) $k_r = 1.5$, (e) $k_r = 1.75$ and (f) $k_r = 2.0$. The values of $\lambda_+$ are colour coded using the colour map shown by the site.}\label{fig3}
\end{figure}
The $\lambda_+ \approx 0$ indicated by the blue region for $k_r = 0.75$ shows regular motion (see Fig.~\ref{fig3}(a)). The yellow boundaries of the period-4 orbits become chaotic for $k_r = 1.0$ (see Fig.~\ref{fig3}(b)). Chaotic neighbourhood (green colour) arise near hyperbolic trajectories at $k_r = 1.25$. The LLE supports the existence of pairs of 3-cycles and higher-order cycles for this $k_r$. It increases to 0.4, occupying most of the region shown in red colour except for bifurcated blue regions for $k_r = 1.5$ (see Fig.~\ref{fig3}(d)). The equatorial period-4 cycles further bifurcate, and the rest of the region becomes chaotic, as shown in Fig.~\ref{fig3}(e). At $k_r = 2.0$, the dynamics become fully chaotic, except for a pair of bifurcated islands.

\begin{figure}
    \includegraphics[width=\linewidth]{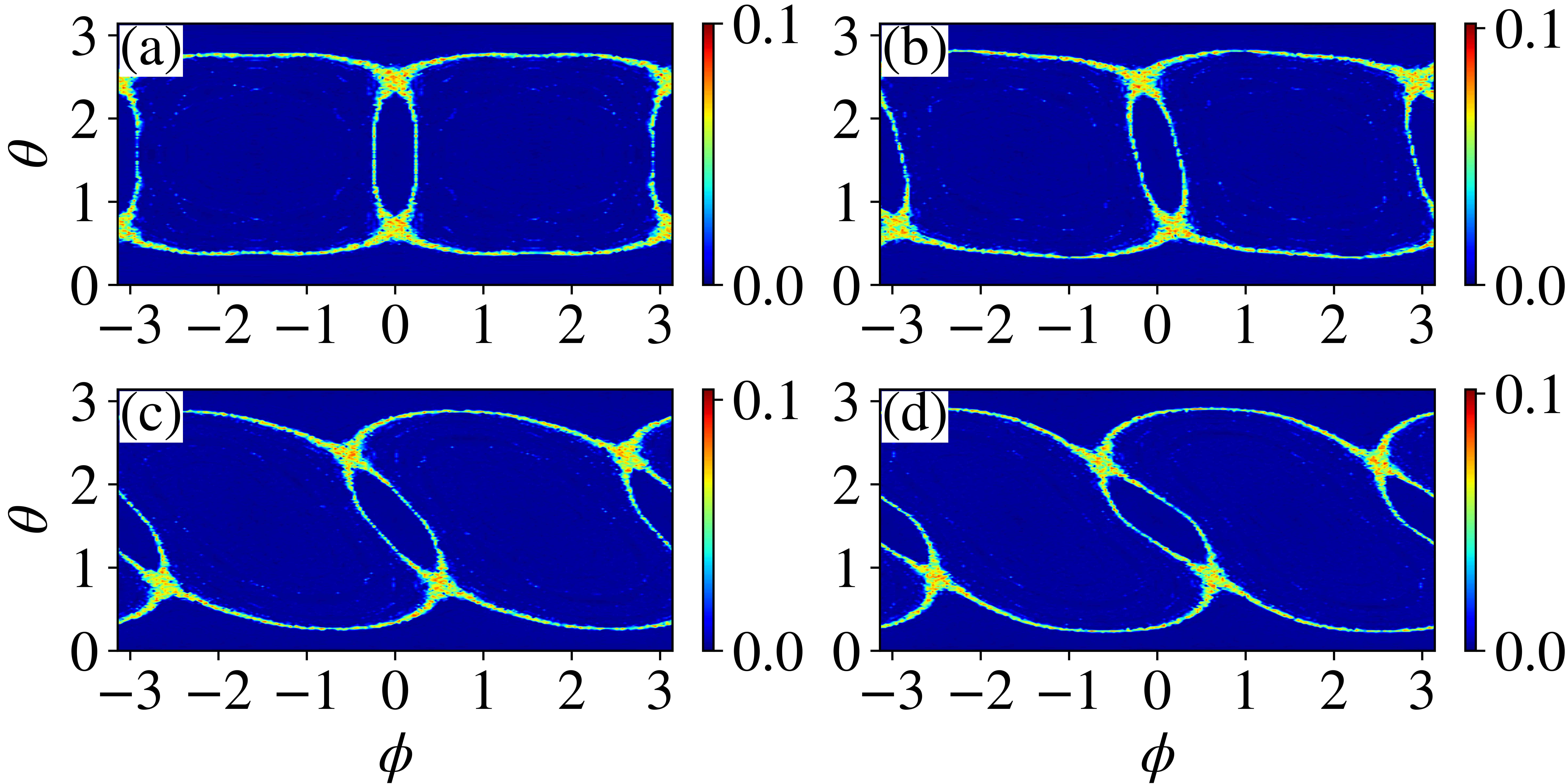}
    \caption{The LLE ($\lambda_+$) in the unit phase-space of $200\times 200$ grid points is presented. Each point in the phase-space is evolved for 1500 kicks keeping $k_r = 1.0$, (a) $k_\theta = 0$, (b) $k_\theta = 0.25$, (c) $k_\theta = 0.75$ and (d) $k_\theta = 1.0$. The values of $\lambda_+$ are colour coded using the colour map shown by the site.}\label{fig4}
\end{figure}
\begin{figure}
    \includegraphics[width=\linewidth]{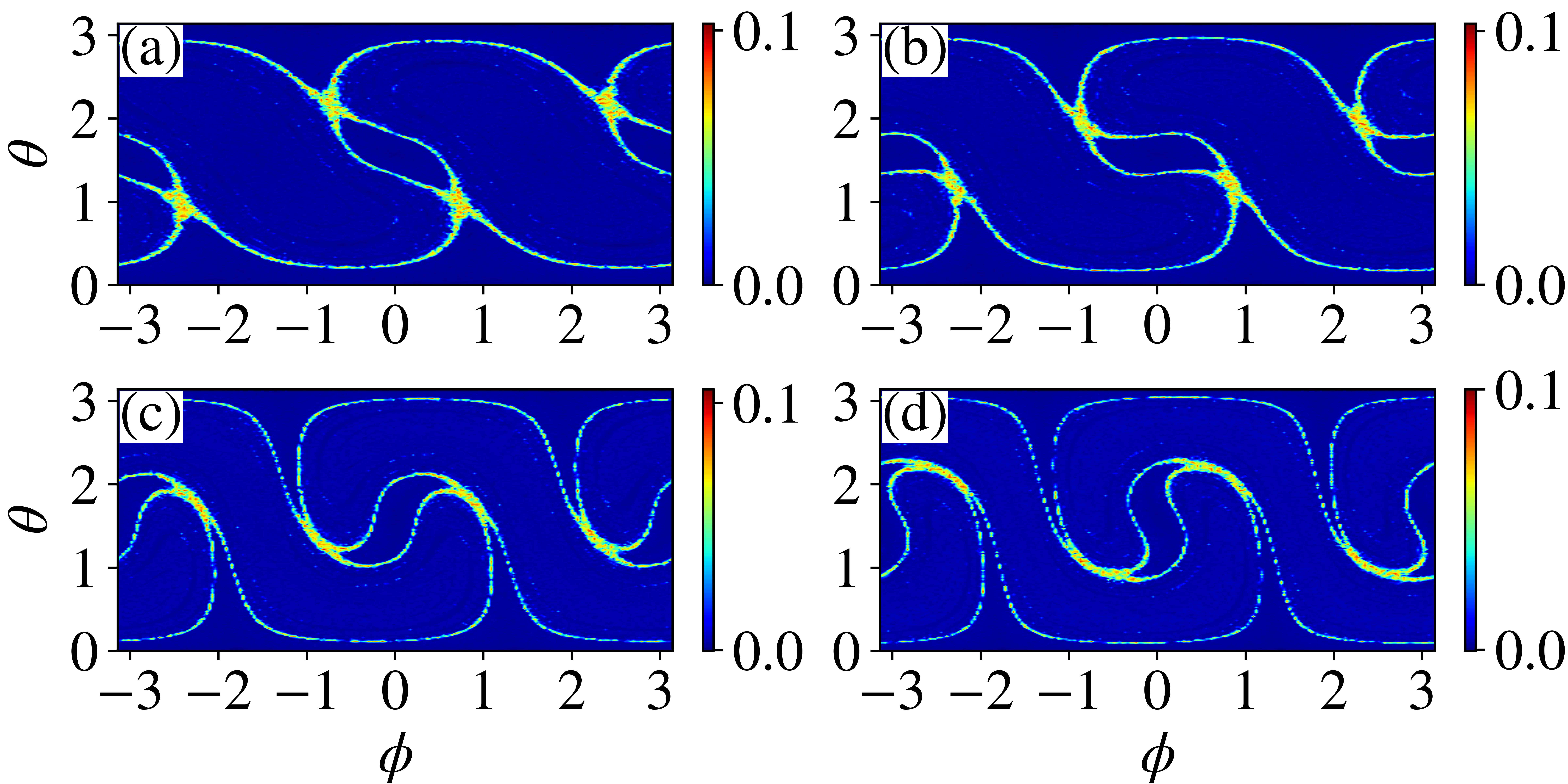}
    \caption{The LLE ($\lambda_+$) in the unit phase-space of $200\times 200$ grid points is presented. Each point in the phase-space is evolved for 1500 kicks keeping $k_r = 1$, (a) $k_\theta = 1.25$, (b) $k_\theta = 1.75$, (c) $k_\theta = 3.0$ and (d) $k_\theta = 3.75$ is presented. The values of $\lambda_+$ are colour coded using the colour map shown by the site.}\label{ktTwist}
\end{figure}
\begin{figure}
    \includegraphics[width=0.52\linewidth]{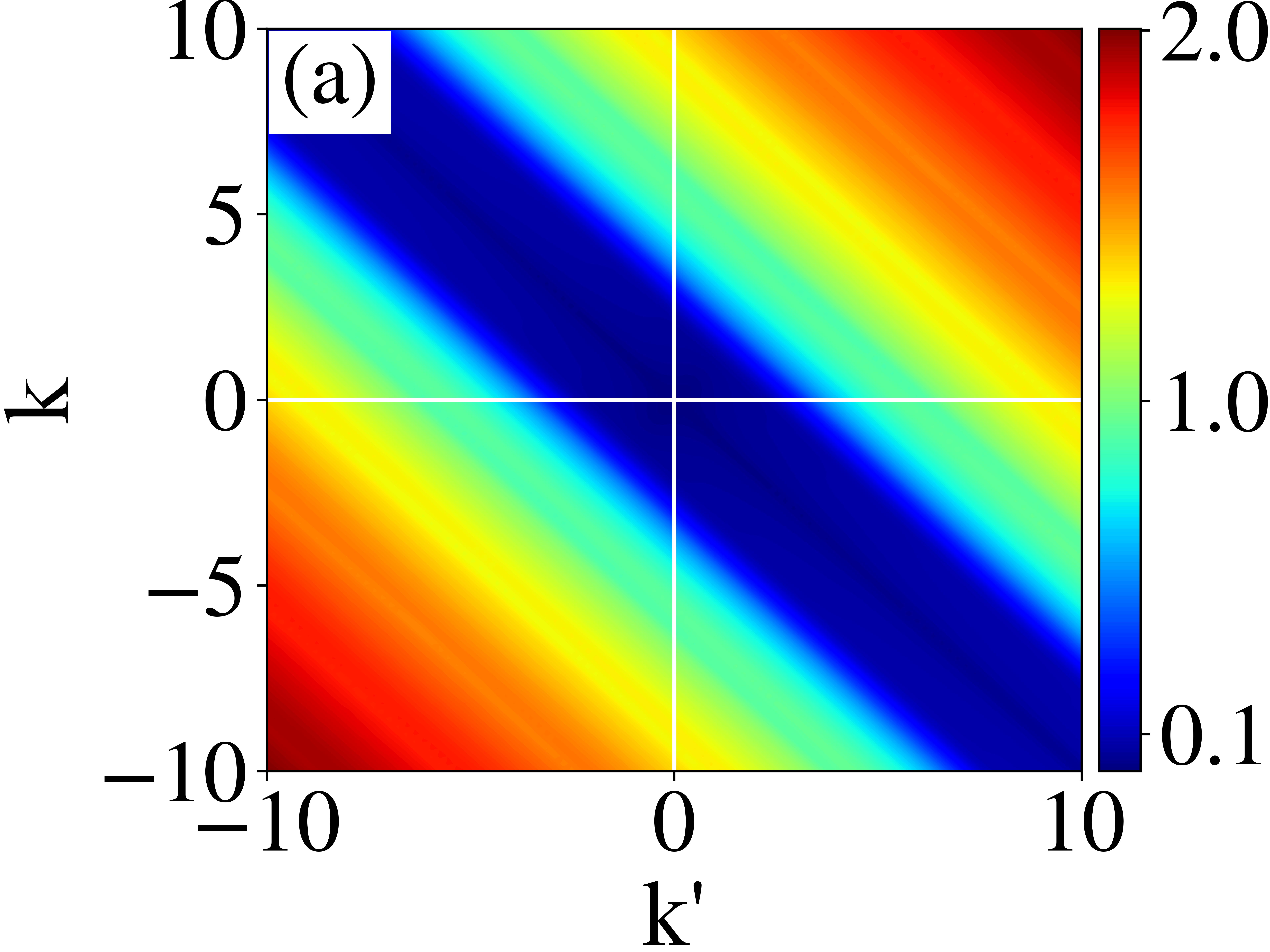}
    \includegraphics[width=0.47\linewidth]{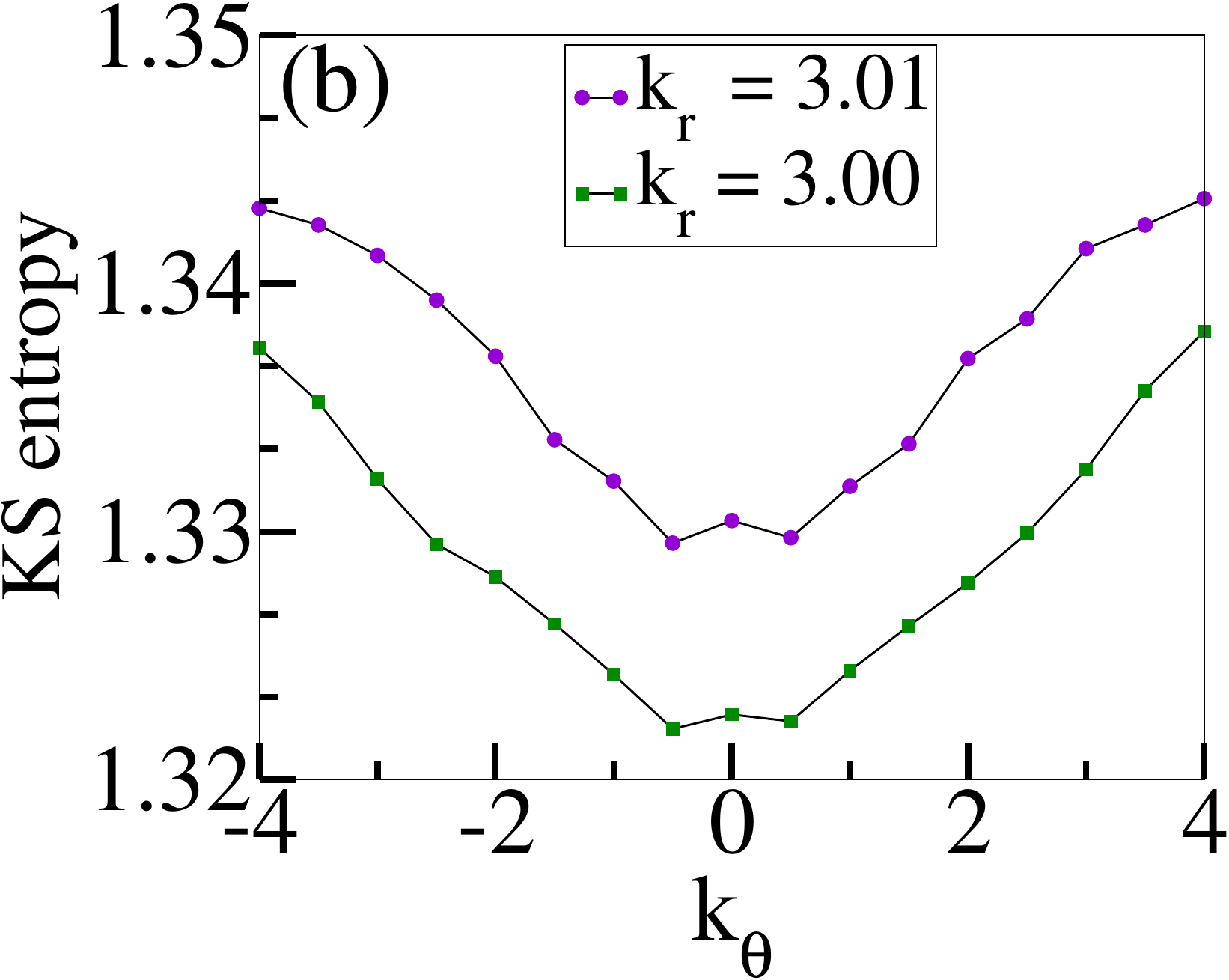}
    \caption{(a) The LLE is averaged ($\bar{\lambda}_+$) over a grid of $200\times 200$ points in the phase-space with each point evolved for 1500 kicks. Here, intervals $k, k' \in (-10,10)$ are divided into a grid of $100\times 100$ points. (b) KSE is also averaged over a grid of $200\times 200$ points in the phase-space with each point evolved for $10^4$ kicks, $k_r = 3.0$ and $k_\theta \in (-3.5, 3.5)$.}\label{fig5}
\end{figure}

We study the effect of stability with the variation of $k_\theta$ for $k_r = 1.0$ by analysing the LLE as shown in Figs.~\ref{fig4} and \ref{ktTwist}. The LLE show excellent agreement with the phase-space structures. However, there are no observable changes apart from \textit{rotation} of the phase-space structures with the variation of $k_\theta$. It should be noted that the parameter $k_\theta$ plays two roles: (1) it induces torsion of the spin in the physical space, and (2) it causes a \textit{rotation} of the phase-space structures, as mentioned earlier. Now, we analytically check the dependence of $k_\theta$ on the LLE. It is given by
\begin{equation}
    \lambda_+ = \lim_{n\rightarrow\infty} \frac{\ln \left(\mu_{+}\left(\mathbf{{X}}_n\right)\right)}{n},
\end{equation}
to check if it shows dependence on $k_\theta$. Here, $\mu_+$ denotes the largest eigenvalue of the product tangent map $\prod_{l=0}^{n-1} \mathcal{T}\left[\mathbf{X}_{l}\right]$. The tangent map $\mathbf{M}$ evaluated at $\mathbf{X}_n$ follows a characteristic polynomial of the form:
\begin{equation}
    \mathbf{M}^3 - T_2 \mathbf{M}^2 + T_1 \mathbf{M} - T_0 = 0,
\end{equation}
with coefficients defined as:
\begin{align}
    T_2 =& k B + \left(1+k' B\right) \cos(kX + k' A) + k' X \sin(kX + k' A),\nonumber  \\ 
    T_1 =& k' B + kY \cos(k' A) + \cos(kX + k' A) + kZ\sin(k' A), \, \text{and} \nonumber \\
    T_0 =& 1,
\end{align}
where $A = Z \cos k X + Y \sin k X$ and $B = Y \cos k X - Z \sin k X$. Since $|\mathbf{X}_n|^2 = 1$, one of the eigenvalues is always 1 (see Appendices B and D of Ref.~\citep{Poggi}). Then, the characteristic polynomial can be rewritten as: 
\begin{equation}
    \left(\mathbf{M}-1\right) \left( \mathbf{M}^2 + B_1 \mathbf{M} + B_2 \right) = 0,
\end{equation}
where, $B_1 = 1- T_2$ and $B_2 = T_0$. Then, the largest eigenvalue is given by
\begin{equation}
    M_+ = \frac{T_2 - 1}{2} + \frac{1}{2} \sqrt{{\left(1-T_2\right)}^2 - 4}.
\end{equation}
By applying the ergodic hypothesis and transitioning from a time average to a phase-space average over the unit sphere, we arrive at:
\begin{align}
    \bar{\lambda}_{+} = \frac{1}{4\pi} \displaystyle{\lim_{n\to\infty}} \left(\frac{1}{n} \int_{0}^{\pi} d\theta \sin\theta \int_0^{2\pi}  d\phi \, \ln \big| M_+ \big|\right),
\end{align}
when, solved up to first-order approximation, yields:
\begin{equation}
    \bar{\lambda}_{+} \approx \ln \left(\frac{k + k'}{2}\right).
\end{equation}
The phase-space averaged LLE $\bar{\lambda}_+$ remains largely constant along the line $k+k' = \textit{const.}$ (see Fig.~\ref{fig5}(a)). Hence, the LLE does not vary with $k_\theta$ up to the linear order of $(k, k')$.

Since, the LLE fails to capture any notable change with respect to $k_\theta$, as shown in Fig.~\ref{fig5}(a). We use the KSE as a more sensitive indicator of chaotic behaviour. The KSE \citep{Zarum1998}, which measures the rate of information loss in a dynamical system, is defined as follows: 
\begin{equation}\label{KSentropy}
    h_{KS} = \lim\limits_{t\to\infty} \frac{1}{t} \sum_{n=1}^{t} \log_{2} l_n,
\end{equation}
where $l_n$ is the changing distance between two nearby points at time $n$. The $\delta \mathbf{X}_n$ are evolved by iterating the \textit{tangent map} $\mathbf{M}$. Fig.~\ref{fig5}(b) presents the KSE averaged over the phase-space for $k_r = 3.0$ and varying values of $k_\theta$.

The results show that the KSE increases with $|k_\theta|$, indicating a growth in chaotic behaviour as $k_\theta$ deviates from zero. The case $k_\theta = 0$ exhibits time-reversal symmetry and corresponds to the minimum level of chaos. The other two cases, $k_\theta = \pm k_r$, are also time-reversal symmetric; however, both show higher levels of chaos compared to the $k_\theta = 0$ for the same value of $k_r$. The chaos continues to increase without showing a sharp transition at $k_\theta = \pm k_r$. There is qualitative but not noteworthy quantitative change in the behaviour from $k_\theta < k_r$ to $k_\theta > k_r$. 

\section{Many-body model with all-to-all Ising interaction}\label{sec:mbfs}
We consider $N$ spin-half particles with all-to-all interactions, with the total spin of the system given by $j = N/2$. Replacing $J_{x,y,z}$ with $\sum_{l=1}^{2j} \sigma_l^{x,y,z}/2$ \citep{wang2004entanglement, milburn1999simulating,sharma2024exactly}, the Floquet operator in Eq. (\ref{U}) is expressed as follows:  
\begin{align}\label{Floquet}
    \begin{split}
        \hat{\mathcal{U}} = &\exp\left(-i \frac{k'}{4j} \sum_{l' < l = 1}^{N} \sigma_{l'}^x \sigma_l^x \right) \exp\left(-i \frac{k}{4j} \sum_{l' < l = 1}^{N} \sigma_{l'}^z \sigma_l^z \right) \\ 
        &\times \exp\left(-i \frac{\pi}{4} \sum_{l=1}^{N} \sigma_l^y\right).
    \end{split}
\end{align}
Since, the transformed parameters $(k_r, k_\theta)$ resolve dynamics into two qualitatively distinct parts. We use $k = k_r + k_\theta$ and $k' = k_r - k_\theta$, anticipating their distinct roles in the quantum dynamics. For our initial states, we utilise the standard $SU(2)$ coherent states \citep{glauber1976superradiant, puri2001mathematical}, which are expressed in the qubit basis as follows:  
\begin{equation}\label{Eq:generalstate}
    |\theta_0, \phi_0\rangle = \otimes^{2j} \left[\cos\left(\frac{\theta_0}{2}\right) | 0\rangle + e^{-i\phi_0}\sin\left(\frac{\theta_0}{2}\right)| 1 \rangle\right].
\end{equation}  
The quantum system under consideration is described by the Floquet operator $\hat{\mathcal{U}}$ with permutation symmetry. The following property \citep{sharma2024exactly}:  
\begin{equation}
    \left[\mathcal{U}, \otimes^{2j}_{l=1}\sigma_l^y\right] = 0,
\end{equation}  
simplifies the quantum dynamics by allowing it to be expressed in a block-diagonal form using a specific basis of states. These basis states are defined \citep{sharma2024exactly} as follows:  
\begin{align}
    |\Phi_q^{\pm}\rangle &= \frac{1}{\sqrt{2}} \left(|W_q\rangle \pm i^{2j - 2q}|\overline{W}_q\rangle\right), \\
    |W_q\rangle &= {\begin{pmatrix} 2j \\ q \end{pmatrix}}^{-\frac{1}{2}} \sum_{\mathcal{P}} {\left( \otimes^q | 1\rangle \otimes^{2j-q}|0\rangle \right)}_{\mathcal{P}},  \\
    |\overline{W}_q\rangle &= {\begin{pmatrix} 2j \\ q \end{pmatrix}}^{-\frac{1}{2}} \sum_{\mathcal{P}} {\left( \otimes^q | 0\rangle \otimes^{2j-q}|1\rangle \right)}_{\mathcal{P}}, 
\end{align}  
where $\mathcal{P}$ denotes the sum over all possible permutations of the qubits. Here, $0 \leq q \leq \frac{2j-1}{2}$ for odd $j$ and $0 \leq q \leq j$ for even $j$, excluding the state $|\Phi_{j}^-\rangle$. Since these states are referred to as parity eigenstates \citep{sharma2024exactly}, given by  
\begin{align}
    \otimes^{2j}_{l=1}\sigma^y_l |\Phi_j^\pm \rangle = \pm |\Phi_j^\pm \rangle,
\end{align}  
significantly simplify the study of the system's quantum dynamics.

\section{Measures of quantum correlations}\label{sec:qm}
For standard QKT, time-averaged quantum correlations seem to show good agreement \citep{neil2016ergodic,madhok_dogra_correlations,ruebeck_entanglement_classical,dogra2019quantum,Bandyopadhyay2004} with the corresponding classical phase-space. On similar lines, we analyse quantum correlations in the light of broken time-reversal symmetry. In this section, we define linear entropy, von Neumann entropy, quantum discord, and concurrence.

The linear entropy for a single-qubit reduced density matrix (RDM) $\rho_1$ \citep{dogra2019quantum,nielsen2010quantum} is given by
\begin{equation}
    S = 1 - \text{tr} \left[\rho_1^2\right].
\end{equation}
This quantity measures the degree of mixing of a given state. The $S = 0$ corresponds to a pure state, whereas $S = 0.5$ corresponds to a maximally mixed state. In the study of entanglement dynamics, linear entropy is easier to analyse and effectively captures qualitative features such as the periodicity of entanglement. Therefore, we focus on linear entropy in the analytical part of this paper.

The von Neumann entropy \citep{neil2016ergodic,Zarum1998} for a single-qubit RDM $\rho_1$, in the log 2 base, is given by
\begin{equation}
    S_{vn} = - \text{tr}\left( \rho_1 \log_2 \rho_1 \right).
\end{equation}
It is commonly used to investigate the generation of entanglement during chaotic dynamics and its correlation with classical chaotic measures, such as the Lyapunov exponent. We employ this entropy in our computations to enable a comparison between our analytical results and numerical findings in the deep quantum regime. Then, we compare the results in the high-spin regime with the LLE \citep{p-spin_poggi,Constantoudis} and KSE \citep{Zarum1998}.

To investigate quantum correlations in addition to entanglement, we calculated the quantum discord \citep{Ollivier2001, Sarandy2005}. The mutual quantum information shared between two qubits is defined as follows:
\begin{equation}
    \mathcal{I}(B:A) = S_{vn}\left(\rho_A\right) + S_{vn}\left(\rho_B\right) - S_{vn}\left(\rho_{AB}\right).
\end{equation}
A quantum measurement on subsystem $A$, represented by a positive-operator valued measure (POVM) $\Pi_i \otimes \mathds{I}_B$, is performed such that the conditional state of $B$ for a given outcome $i$ is given by
\begin{align}
    \begin{split}
        \rho_{B|i} &= \text{Tr}_A (\rho_{AB|i})/p_i, \\ 
        p_i &= \text{Tr}_{A,B} (\rho_{AB|i}),  \\
        \rho_{AB|i} &= \frac{\Pi_i \otimes \mathds{I}_B \, \rho_{AB} \, \Pi_i \otimes \mathds{I}_B}{\text{Tr}(\Pi_i \otimes \mathds{I}_B \, \rho_{AB})}. 
    \end{split}
\end{align}
The conditional entropy is defined as $\tilde{S}_{vn, \lbrace \Pi_i \rbrace}(B|A) = \sum_i p_i S_{vn}(\rho_{B|i})$. The mutual quantum information is then expressed as:
\begin{equation}
    \mathcal{J}(B:A) = \max_{\lbrace \Pi_i \rbrace} \left[ S_{vn} \left(\rho_B \right) - \tilde{S}_{vn, \lbrace \Pi_i \rbrace}(B|A) \right].
\end{equation}
Consequently, the quantum discord is defined by:
\begin{equation}
    \mathcal{D}(B:A) = \mathcal{I}(B:A) - \mathcal{J}(B:A).
\end{equation}

The concurrence serves as a quantifier of entanglement between two qubits. For a given 2-qubit RDM $\rho_{12}$, the concurrence is defined \citep{Hill, Wootters} as follows:
\begin{equation}
    \mathcal{C}\left(\rho_{12}\right) = \max \left( 0, \sqrt{\Lambda_1} - \sqrt{\Lambda_2} - \sqrt{\Lambda_3} - \sqrt{\Lambda_4} \right),
\end{equation}
where $\Lambda_i$'s are the eigenvalues of the non-Hermitian matrix $\rho_{12} \left( \sigma_y \otimes \sigma_y \right) \rho_{12}^* \left( \sigma_y \otimes \sigma_y \right)$ taken in descending order.

\section{Exact Solution for 2-Qubits}\label{sec:2qubit}
The entanglement dynamics of the 2-qubit system have been studied \citep{ruebeck_entanglement_classical} in the standard QKT. In this section, we present an exact solution of the 2-qubit DKT system and find the condition for the periodicity of the entanglement dynamics. Then, we derive the infinite-time averaged linear entropy analytically for a general initial state Eq.~(\ref{Eq:generalstate}) and compare it with the computed long-time-averaged von Neumann entropy. Here, we analyse the time evolution of the linear entropy and its infinite-time average, specifically for initial states: $|\theta_0 = 0, \phi_0 = 0\rangle$ and $|\theta_0 = \pi/2, \phi_0 = -\pi/2\rangle$. 

As mentioned earlier, we use parity eigenstates \citep{sharma2024exactly} as a basis states:
\begin{align}
    \begin{split}
 |\Phi_0^\pm \rangle &= \frac{1}{\sqrt{2}} |00\rangle \mp \frac{1}{\sqrt{2}} |11\rangle \,\, \text{and}  \\ 
 |\Phi_1^+ \rangle &= \frac{1}{\sqrt{2}} |10\rangle +  \frac{1}{\sqrt{2}} |01\rangle.
    \end{split}
\end{align}
The Floquet operator $\hat{\mathcal{U}}$ in this basis takes a block-diagonal form:
\begin{equation}
 \mathcal{U} = \begin{pmatrix}
 \mathcal{U}_+  &0_{2\times 1}\\
        0_{1\times 2}  &\mathcal{U}_-
    \end{pmatrix},
\end{equation}
where the components of the block-diagonal form are given by
\begin{align}
 \mathcal{U}_+ = \begin{pmatrix}
        0  &-e^{-\frac{i k_\theta}{2}}\\
 e^{i \frac{k_\theta}{2}} &0\\
    \end{pmatrix} \, \text{and } \,\,
 \mathcal{U}_- = e^{-\frac{i k_r}{2}}.
\end{align}
The eigenvalues of $\hat{\mathcal{U}}$ are $\lbrace -i, i,  e^{ -\frac{i k_r}{2}} \rbrace$ with corresponding eigenvectors $\left\{{\left[-i e^{- \frac{i k_\theta}{2}},  1, 0 \right]}^T, {\left[-i e^{\frac{i k_\theta}{2}},  1, 0\right]}^T, {\left[  0, 0, 1\right]}^T\right\}$, respectively. The time-evolved Floquet operators for each block are:
\begin{align}
    \begin{split}
 \mathcal{U}_+^n &= \begin{pmatrix}
 \cos(n \pi/2 )  &-\sin(n \pi/2) e^{-  \frac{i k_\theta}{2}}\\
 \sin(n \pi/2 )e^{\frac{i k_\theta}{2}} &\cos(n \pi/2)\\
        \end{pmatrix} \; \text{and} \\
 \mathcal{U}_-^n &= e^{-\frac{i n k_r}{2}}.
    \end{split}
\end{align}
This allows us to evolve any initial state, allowing us to study the entanglement dynamics as the system evolves.
 
We evolve the general initial state using the Floquet operator $\hat{\mathcal{U}}$. The resulting state is given by
\begin{equation}
 |\psi_n\rangle = c_0 |\Phi^+_0\rangle + c_1 |\Phi^+_1\rangle + c_2 |\Phi^-_0\rangle,
\end{equation}
where,
\begin{align}
 c_0 &= \frac{e^{-\frac{i}{2}(k_\theta + 2\phi_0)}}{\sqrt{2}}  \left\{-\sin\left(\frac{n\pi}{2}\right)\sin(\theta_0) \right. \notag \\
    &\left. + e^{\frac{i}{2}k_\theta} \cos\left(\frac{n\pi}{2}\right) \left[\cos(\theta_0)\cos(\phi_0)+i\sin(\phi_0)\right]\right\} \label{Eq:coeff20}, \\
 c_1 &= \frac{e^{-\phi_0}}{\sqrt{2}} \left\{\cos\left(\frac{n\pi}{2}\right)\sin(\theta_0) \right. \notag \\
    &\left. + e^{\frac{i}{2}k_\theta} \sin\left(\frac{n\pi}{2}\right) \left[\cos(\theta_0)\cos(\phi_0)+i\sin(\phi_0)\right]\right\} \label{Eq:coeff21} \text{ and}  \\
 c_2 &= \frac{e^{-\frac{i}{2}(nk_r + 2\phi_0)}}{\sqrt{2}}   \left[\cos(\phi_0)+i\cos(\theta_0)\sin(\phi_0)\right]. \label{Eq:coeff22}
\end{align}
Then, the single-qubit RDM $\rho_1(n)$ is obtained as follows:
\begin{align}
    \rho_1(n) =& \begin{pmatrix}
 \frac{1}{2} + \Re[c_0 c_2^*] &\Re[c_1 c_2^*] + i \Im[c_0 c_1^*]\\
        \Re[c_1 c_2^*] - i \Im[c_0 c_1^*] &\frac{1}{2} - \Re[c_0 c_2^*]
    \end{pmatrix}. \nonumber
\end{align}
Since the eigenvalues of the above single-qubit RDM are $1/2 \pm \sqrt{{\Re[c_0 c_2^*]}^2 + {\Re[c_1 c_2^*]}^2 + {\Im[c_0 c_1^*]}^2}$, the linear entropy is given by
\begin{align}\label{Eq:2qubitgeneral}
    \begin{split}
 S^{(2)}_{(\theta_0, \phi_0)}(n, k_r, k_\theta) =& \frac{1}{2} -2{\Re[c_0 c_2^*]}^2 - 2{\Re[c_1 c_2^*]}^2 \\
        & - 2{\Im[c_0 c_1^*]}^2.
    \end{split}
\end{align}
For the special state $|\theta_0 = 0,\phi_0 = 0\rangle$, it reduces to (see supplementary material \citep{supplementary2025}):
\begin{align}
 S^{(2)}_{(0,0)}\left(n,k_r,k_\theta\right) =& \begin{cases}
 \frac{1}{2}\sin^2\left(\frac{n k_r + k_\theta}{2}\right) & \text{odd } n\\
 \frac{1}{2}\sin^2\left(\frac{n k_r}{2}\right) & \text{even } n.
    \end{cases}
\end{align}
It is zero for even $n$ and when $n k_r/2$ is an integer multiple of $\pi$. This means, when $k_r$ is rational multiple of $\pi$ (modulo $2\pi$), the corresponding period is $2\pi/k_r$. Similarly, for another special state $|\theta_0 = \pi/2,\phi_0 = -\pi/2\rangle$ \citep{dogra2019quantum}, the linear entropy in the Eq.~(\ref{Eq:2qubitgeneral}) reduces to (see supplementary material \citep{supplementary2025}):
\begin{align}
 S^{(2)}_{(\frac{\pi}{2},-\frac{\pi}{2})}\left(n, k_r, k_\theta\right) =& \begin{cases}
 \frac{1}{2} \sin^2\left(\frac{k_\theta}{2}\right)  & \text{odd } n,\\
        0 & \text{even } n.
    \end{cases}
\end{align}
The above expression is always zero for even $n$. Thus, this state has a period of 2.

It can be seen from the Eqns.~(\ref{Eq:coeff20}), (\ref{Eq:coeff21}), (\ref{Eq:coeff22}) and (\ref{Eq:2qubitgeneral}) that the linear entropy is periodic if:
\begin{align}
 k_r = a \pi \,\, \text{ where }\,\, a \in \mathds{Q}.
\end{align}
It should be noted that the above condition gives the periodic nature for any initial state Eq.~(\ref{Eq:generalstate}) and not the actual period which depends on the given initial state. 

The infinite-time averaged linear entropy \citep{dogra2019quantum} is obtained as follows:
\begin{widetext}
    \begin{align}\label{infinite-time averaged-2qubit}
        \begin{split}
            \langle S^{(2)}_{(\theta_0,\phi_0)}\left(k_r,k_\theta\right)\rangle =& \lim_{N\to \infty} \frac{1}{N} \sum_{n=0}^{N-1} S^{(2)}_{(\theta_0,\phi_0)}\left(n, k_r, k_\theta\right) \\
 =& \frac{106 + 8\cos(2\theta_0) + 14 \cos(4\theta_0) - 4\cos(2\theta_0 - 4\phi_0) + \cos(4\theta_0 - 4\phi_0) + 6\cos(4\phi_0) + \cos(4\theta_0 + 4\phi_0)}{1024} \\
            &+\frac{32 \cos(2k_\theta)\sin^2(\theta_0) \left[\cos(2\phi_0)(3 + \cos(2\theta_0)) - 2\sin^2(\theta_0)\right] - 4\cos(2\theta_0 + 4\phi_0)}{1024}   \\
            &+\frac{1}{128} {\left[3 + \cos(2\theta_0) + 2\cos(2\phi_0)\sin^2(\theta_0)\right]}^2 + \frac{1}{16} \sin(2k_\theta)\sin(\theta_0) \sin(2\theta_0) \sin(2\phi_0).
        \end{split}
    \end{align}  
\end{widetext}
Although the periodicity of the linear entropy is determined solely by $k_r$, the infinite-time averaged linear entropy does not depend on the chaos parameter $k_r$. It saturates to $1/4$ for the initial state $|\theta_0 = 0, \phi_0 = 0\rangle$ regardless of the kick strength $k_\theta$. Whereas for the initial state $|\theta_0 = \pi/2, \phi_0 = -\pi/2\rangle$, it is $\sin^2\left(k_\theta/2\right)/4$.

\begin{figure}
    \includegraphics[width=\linewidth]{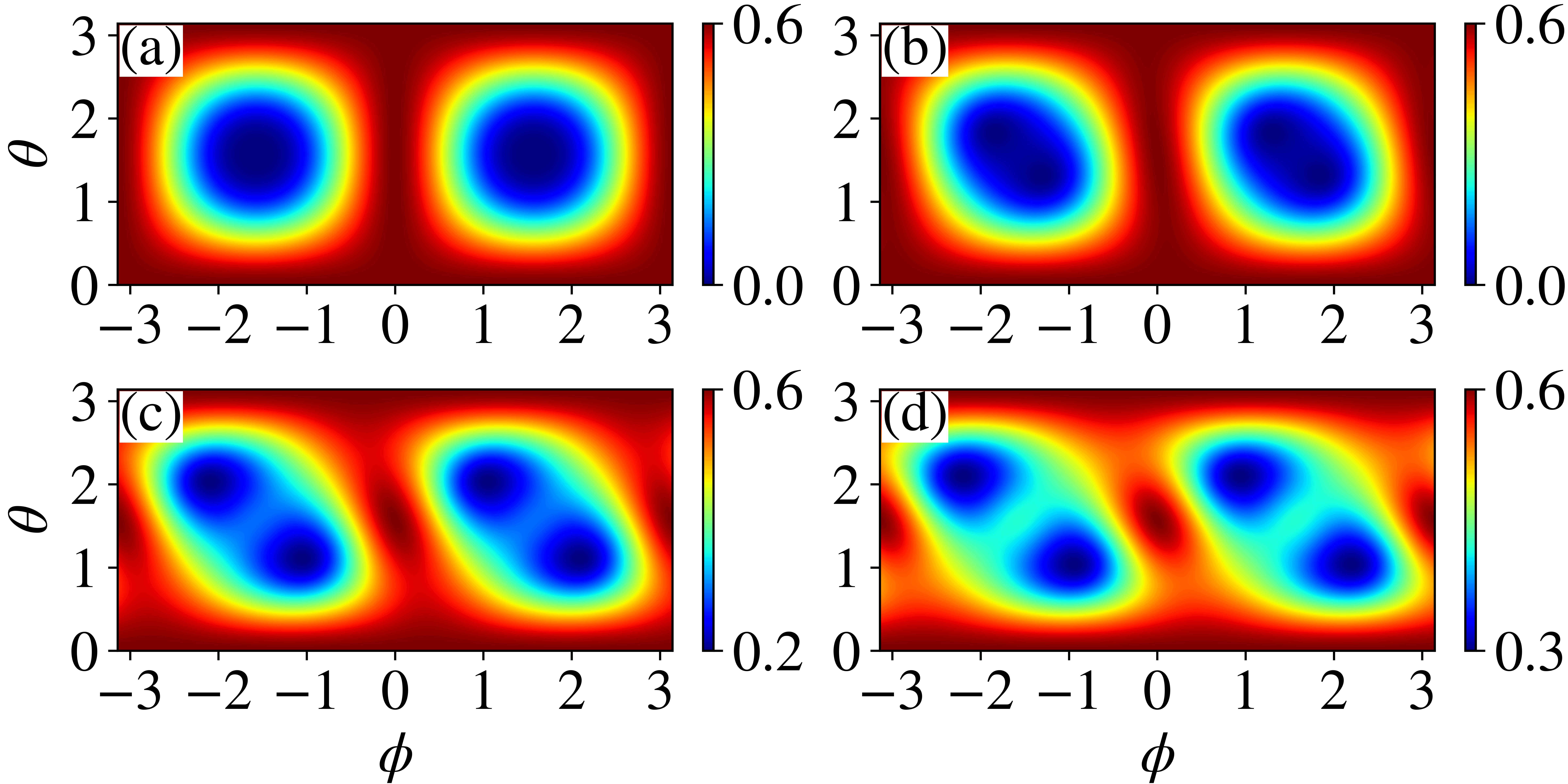}
    \caption{Long-time-averaged von Neumann entropy for single-qubit RDM $\rho_1(n)$, $n=1000$ and a grid of $200\times 200$ initial coherent states. Here, $j = 1$, $k_r = 1$, (a) $k_\theta = 0$, (b) $k_\theta = 0.25$, (c) $k_\theta = 0.75$ and (d) $k_\theta = 1$.}\label{j1}
\end{figure}

The infinite-time averaged linear entropy produces the same qualitative results as the computed long-time-averaged von Neumann entropy (see Fig.~\ref{j1}). Since we numerically analyse von Neumann entropy with quantum discord and concurrence in the semi-classical regime, we plot von Neumann entropy to compare results with the deep quantum regime. The entanglement increases significantly with $k_\theta$ as shown in Fig.~\ref{j1}. The results are consistent with phase-space structures, with minimum entanglement corresponding to the regular region. This includes an observation of $k_\theta$ being responsible for twisting the phase-space region around trivial fixed points. Interestingly, the two low-valued time-averaged entanglement (blue colour) regions separate further along the $Z= - X$ line with the increase in the value of $k_\theta$ from zero to 1 for $k_r = 1$.

\section{Exact Solution for 3-Qubits}\label{sec:3qubit}
The 3-qubit QKT system is an exactly solvable model \citep{dogra2019quantum} that shows ergodicity and thermalisation signatures. This section presents an exact solution for the 3-qubit DKT system. The detailed analytical calculations for the entanglement dynamics and the infinite-time averaged linear entropy for the general initial state are given in the supplementary material \citep{supplementary2025}. Since we are specifically interested in two initial spin-coherent states: the state $|\theta_0 = 0, \phi_0 = 0\rangle$ and the state $|\theta_0 = \pi/2, \phi_0 = - \pi/2\rangle$, we provide their detailed analysis in the supplementary material \citep{supplementary2025}. Moreover, we study the periodicity of entanglement dynamics and analyse the relationship between different measures of quantum correlations.

We consider the following basis states for the 3-qubit system:
\begin{align}
 |\Phi^\pm_0\rangle &= \frac{1}{\sqrt{2}} \left(|000\rangle \pm |111\rangle\right),  |\Phi^\pm_1\rangle = \frac{1}{\sqrt{2}} \left(|W\rangle \pm i |\overline{W}\rangle\right), \notag \\
 |W\rangle &= \frac{1}{\sqrt{3}} \sum_{\mathcal{P}} |001\rangle_{\mathcal{P}} \text{ and } |\overline{W}\rangle = \frac{1}{\sqrt{3}} \sum_{\mathcal{P}} |110\rangle_{\mathcal{P}},
\end{align}
where $\sum_{\mathcal{P}}$ denotes summation over all possible permutations. The Floquet operator in these basis states is given by
\begin{align}
    \begin{split}
 \mathcal{U} = \begin{pmatrix}
 \mathcal{U}_+    &0\\
            0   &\mathcal{U}_-
        \end{pmatrix} ,
 \mathcal{U}_{\pm} = \pm e^{\mp i\frac{\pi}{4}} e^{-\frac{i}{3}k_r} \begin{pmatrix}
            \alpha   &\mp \beta^*\\
            \pm \beta   &\alpha^*
        \end{pmatrix}, 
    \end{split}
\end{align}
where
\begin{align}
        \alpha =& \frac{1}{2} \sin\left(\frac{2k_r}{3}\right) + \frac{i}{4} \left[3\cos\left(\frac{2k_\theta}{3}\right) - \cos\left(\frac{2k_r}{3}\right)\right] , \nonumber \\
        \beta =& \frac{\sqrt{3}}{4} \left[ \cos\left(\frac{2k_r}{3}\right) + \cos\left(\frac{2k_\theta}{3}\right) + 2i \sin\left(\frac{2k_\theta}{3}\right)\right].\nonumber
\end{align}
The eigenvalues of the Floquet operator are $-e^{\frac{3i\pi}{4}} \left\{ \Re[\alpha] \mp C/2, -i\Re[\alpha] \mp i C/2 \right\}$, with corresponding eigenvectors given by ${(2\beta)}^{-1}\left\{ {\left[-i \text{Im}[\alpha] \mp C, 1, 0, 0\right]}^T, {\left[0, 0, -i \text{Im}[\alpha] \mp C, 1\right]}^T\right\}$ respectively. Here, $C = \sqrt{2\Re[\alpha^2] - 2{|\beta|}^2 - 2}$. By expressing $\mathcal{U}_+$ as a rotation $e^{-i\gamma \vec{\sigma}\cdot \hat{\eta}}$ by an angle $\gamma$ about an axis $\hat{\eta} = \sin\theta \cos\chi \hat{x} + \sin\theta \sin\chi \hat{y} + \cos\theta \hat{z}$ and comparing terms, we obtain:
\begin{align}\label{gamma3}
    \begin{split}
 \cos\gamma =& \frac{1}{2}\sin\left(\frac{2k_r}{3}\right), \; \sin\theta = \frac{\sqrt{3}}{2\sin\gamma}\;\; \text{ and } \\
 \tan\chi =& -\frac{\cos\left(\dfrac{2k_r}{3}\right) + \cos\left(\dfrac{2k_\theta}{3}\right)}{2\sin\left(\dfrac{2k_\theta}{3}\right)}.
    \end{split}
\end{align}
The time-evolved Floquet operator is given by
\begin{align}
 \mathcal{U}^n_{\pm} =& {\left(\pm 1\right)}^n e^{\mp i n \frac{\pi}{4}} e^{-\frac{i}{3}n k_r} \begin{pmatrix}
        \alpha_n   &\mp \beta^*_n\\
        \pm \beta_n   &\alpha^*_n
    \end{pmatrix}, \\ 
    \alpha_n =& \cos(n\gamma) + \frac{i}{4} \frac{\sin (n\gamma)}{\sin\gamma}\left[3\cos\left(\frac{2k_\theta}{3}\right) - \cos\left(\frac{2k_r}{3}\right)\right], \\
    \beta_n =& \frac{\sqrt{3}}{4} \frac{\sin (n\gamma)}{\sin\gamma} \left[\cos\left(\frac{2k_r}{3}\right) + \cos\left(\frac{2k_\theta}{3}\right) + 2i\sin\left(\frac{2k_\theta}{3}\right)\right].
\end{align}
The standard QKT with $k \neq 0$ and $k' = 0$ can be recovered by setting $k_\theta = k_r$ and $2k_r = \kappa_0$. In this case, we recover to Eqs. (13) and (14) of Ref. \citep{dogra2019quantum}. 
 
The general initial state Eq.~(\ref{Eq:generalstate}) is evolved using the Floquet operator $\hat{\mathcal{U}}$ to obtain:
\begin{align}
 |\psi_n\rangle = c_0' |\Phi_0^+\rangle + c_1' |\Phi_1^+\rangle + c_2' |\Phi_0^-\rangle + c_3' |\Phi_1^-\rangle,
\end{align}
where,
\begin{align}
c_0' &= t_1 t_2 t_4 \left[2\alpha_n t_6 - \sqrt{3}\beta_n^* t_7 + i\alpha_n t_9 \right], \\
c_1' &= t_1 t_2 t_4 \left[2\beta_n t_6 + \sqrt{3}\alpha_n^* t_7 + i\beta_n t_9 \right], \\
c_2' &= t_1 t_3 t_5 \left[2\alpha_n t_6 + \sqrt{3}\beta_n^* t_7 - i\alpha_n t_{10} \right], \\
c_3' &= -t_1 t_3 t_5 \left[2\beta_n t_6 - \sqrt{3}\alpha_n^* t_7 - i\beta_n t_{10} \right],
\end{align}
and, $t_1 = \frac{1}{2\sqrt{2}}$, 
$t_2 = e^{-\frac{i}{4}(n\pi + 6\phi_0)}$, 
$t_3 = e^{\frac{i}{4}(5n\pi - 6\phi_0)}$, 
$t_4 = \cos\left(\frac{\theta_0 + \phi_0}{2}\right) - i\sin\left(\frac{\theta_0 - \phi_0}{2}\right)$, 
$t_5 = \cos\left(\frac{\theta_0 - \phi_0}{2}\right) + i\sin\left(\frac{\theta_0 + \phi_0}{2}\right)$, 
$t_6 = \cos(\theta_0)\cos(\phi_0)$, 
$t_7 = \sin(\theta_0)$, 
$t_8 = \sin(\phi_0)$, 
$t_9 = \sin(\theta_0) + 2\sin(\phi_0)$, and 
$t_{10} = \sin(\theta_0) - 2\sin(\phi_0)$. The single-qubit RDM $\rho_1(n)$ associated with the above state is given by
\begin{align}
        \rho_1(n) =& \begin{pmatrix}
 \frac{1}{2} + p_1'  &p_{12}' \\
 {p_{12}'}^*   &\frac{1}{2} - p_1'
        \end{pmatrix},  \\
 p_1' =& \Re[c_1' {c_3'}^*] + \frac{1}{3}\Re[c_2' {c_4'}^*] \\ 
 p_{12}' =& -\frac{i}{3} (c_2' + c_4'){(c_2' - c_4')}^* + \frac{\sqrt{3}}{6} (c_1' + c_3') {(c_2' + c_4')}^* \notag \\
        &-\frac{1}{2\sqrt{3}} (c_2' - c_4'){(c_1' - c_3')}^*,
\end{align}
The eigenvalues of the above single-qubit RDM are $1/2 \pm \sqrt{{p_1'}^2 + {|p_{12}'|}^2}$. Then, the linear entropy is given by
\begin{align}\label{Eq:3qubitlinear}
 S^{(3)}_{(\theta_0,\phi_0)}(n, k_r, k_\theta) =& \frac{1}{2} - 2{p_1'}^2 - 2 {|p_{12}'|}^2.
\end{align}

The infinite-time averaged linear entropy for the special state $|\theta_0 = 0, \phi_0 = 0\rangle$ \citep{dogra2019quantum} is given by (see supplementary material \citep{supplementary2025})
\begin{widetext}
    \begin{align} \label{infinite-time averaged-3qubit-00}
        \begin{split}
            \langle S^{(3)}_{(0,0)} \left(k_r, k_\theta\right) \rangle =& \frac{1}{64} {\left[7+\cos\left(\dfrac{4k_r}{3}\right)\right]}^{-2} \left\lbrace 1026 + 13 \cos\left(\dfrac{8k_r}{3}\right) + \left[304 - 52\cos\left(\dfrac{4k_\theta}{3}\right)\right] \cos\left(\dfrac{4k_r}{3}\right) - 112\cos\left(\dfrac{4k_\theta}{3}\right) \right. \\
            & \left. +\, 8 \cos\left(\dfrac{2k_r}{3}\right) \cos\left(\dfrac{2k_\theta}{3}\right) \left[-2 + 9 \cos \left(\dfrac{4k_\theta}{3}\right) + \cos\left(\dfrac{4k_r}{3}\right) \right] - 27\cos\left(\dfrac{8k_\theta}{3}\right) \right\rbrace.
        \end{split}
    \end{align}
\end{widetext}
For the case of $k_\theta = k_r$ and $2k_r = \kappa_0$, the above expression reduces to Eq. (29) of Ref. \citep{dogra2019quantum}. It can be seen that $\langle S^{(3)}_{(0,0)} \left(k_r, k_r\right) \rangle \geq \langle S^{(3)}_{(0,0)} \left(k_r, 0 \right) \rangle$.

The infinite-time averaged linear entropy for another special state $|\theta_0 = \pi/2, \phi_0 = -\pi/2\rangle$ is given by (see supplementary material \citep{supplementary2025})
\begin{widetext}
    \begin{align}\label{infinite-time averaged-3qubit_piby2}
        \begin{split}
            \langle S^{(3)}_{(\frac{\pi}{2},-\frac{\pi}{2})} \left(k_r,k_\theta\right) \rangle =& \frac{1}{32} {\left[7+\cos\left(\dfrac{4k_r}{3}\right)\right]}^{-2} \left\lbrace 410 + 5 \cos\left(\dfrac{8k_r}{3}\right) + 4\left[28 - 9\cos\left(\dfrac{4k_\theta}{3}\right)\right]\cos\left(\dfrac{4k_r}{3}\right)-144\cos\left(\dfrac{4k_\theta}{3}\right) \right. \\
            & \left. +\, 8 \cos\left(\dfrac{2k_r}{3}\right) \cos\left(\dfrac{2k_\theta}{3}\right) \left[10 + 9\cos\left(\dfrac{4k_\theta}{3}\right) + \cos\left(\dfrac{4k_r}{3}\right) \right] - 27\cos\left(\dfrac{8k_\theta}{3}\right) \right\rbrace.
        \end{split}
    \end{align}
\end{widetext}
For the case of $k_\theta = k_r$ and $2k_r = \kappa_0$, the above expression reduces to Eq. (40) from the Ref. \citep{dogra2019quantum}. For this state also the inequality $\langle S^{(3)}_{(\pi/2, -\pi/2)} \left(k_r, k_r\right) \rangle \geq \langle S^{(3)}_{(\pi/2,-\pi/2)} \left(k_r, 0 \right) \rangle$ holds true.

\begin{figure}
    \includegraphics[width=\linewidth]{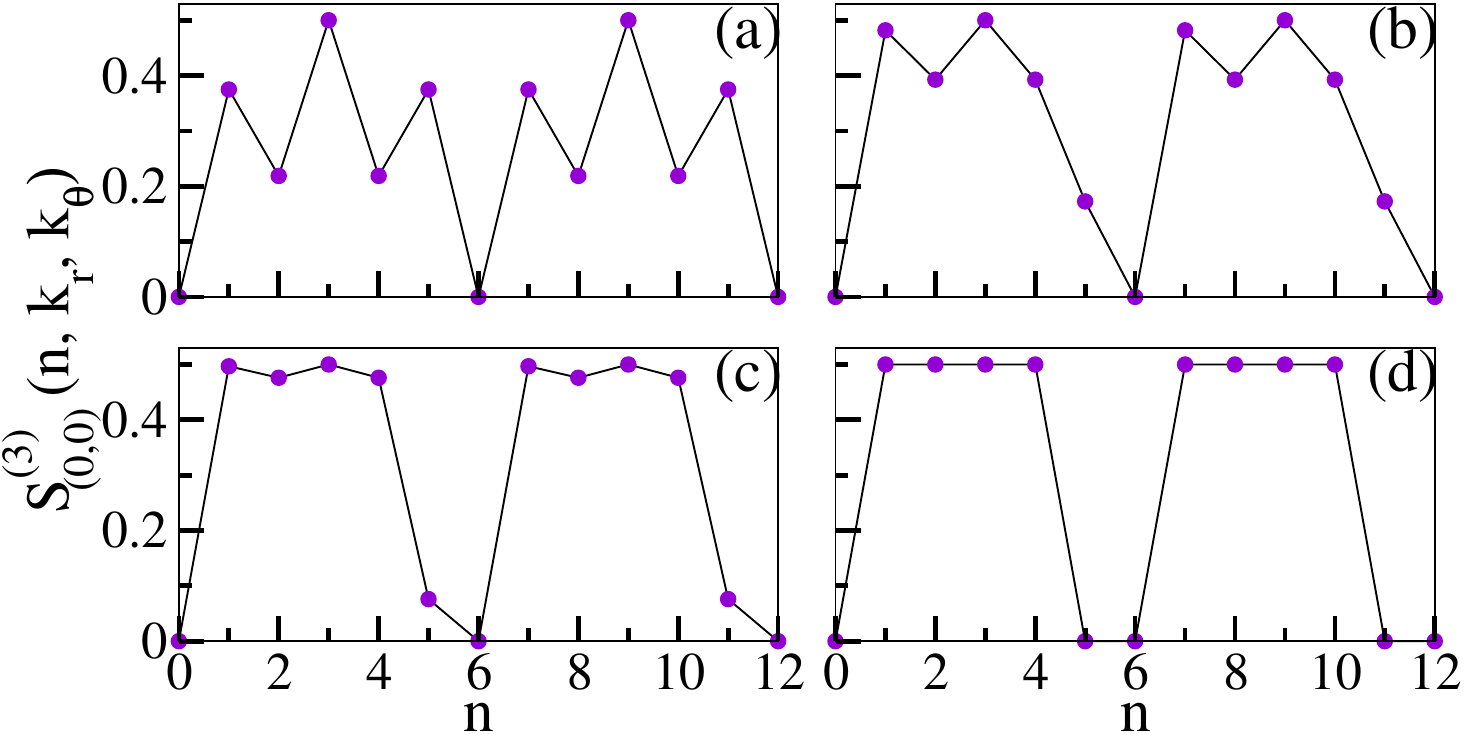}
    \caption{Linear entropy is plotted as a function of discrete time $n$ for the 3-qubit state $|\theta_0 = 0, \phi_0 = 0\rangle$. Here, $k_r = 3\pi/4$, (a) $k_\theta = 0$, (b) $k_\theta = 1$, (c) $k_\theta = 1.5$ and (d) $k_\theta = 3\pi/4$.}\label{fig6}
\end{figure}

\begin{figure}
    \includegraphics[width=\linewidth]{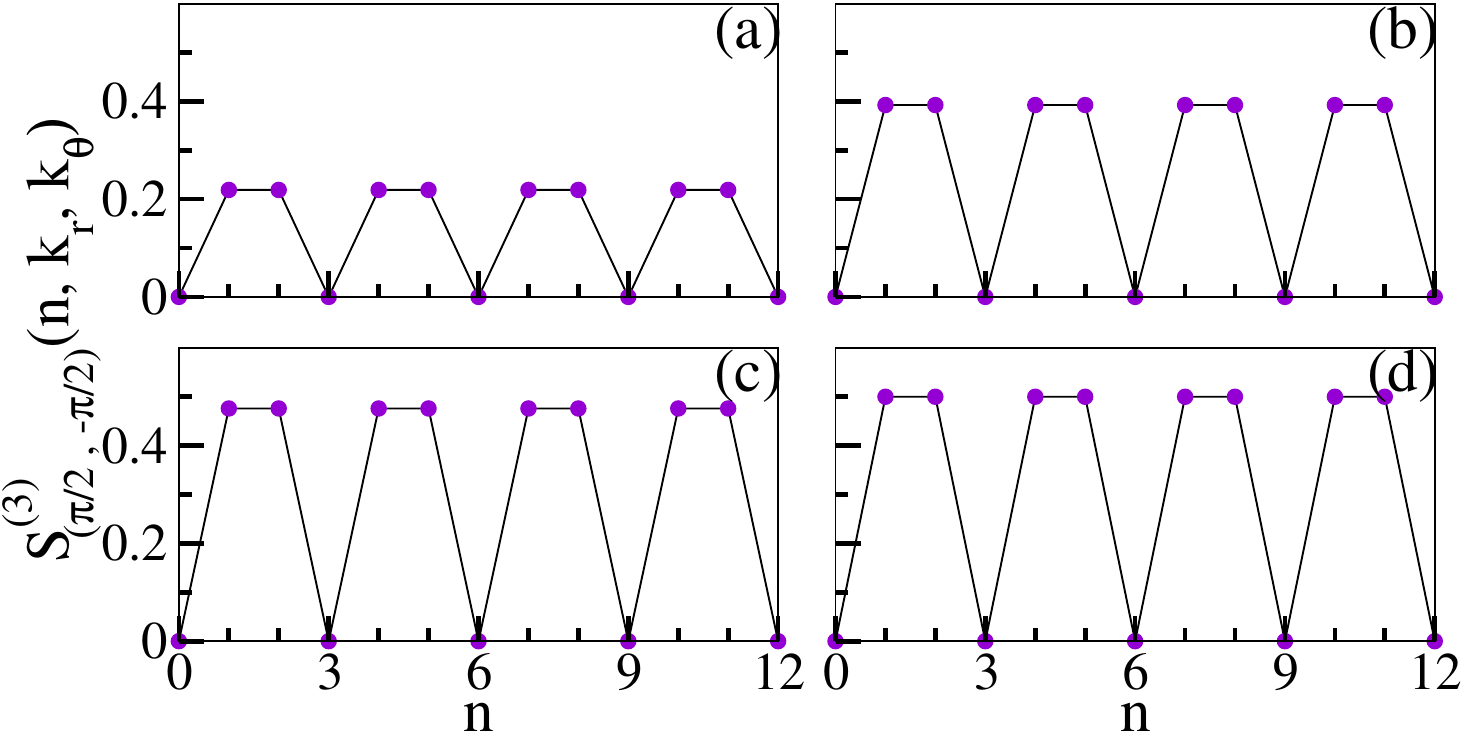}
    \caption{Linear entropy is plotted as a function of discrete time $n$ for the 3-qubit state $|\theta_0 = \pi/2,\phi_0 = -\pi/2\rangle$. Here, $k_r = 3\pi/4$, (a) $k_\theta = 0$, (b) $k_\theta = 1$, (c) $k_\theta = 1.5$ and (d) $k_\theta = 3\pi/4$.}\label{fig7}
\end{figure}

\begin{figure}
    \includegraphics[width=\linewidth]{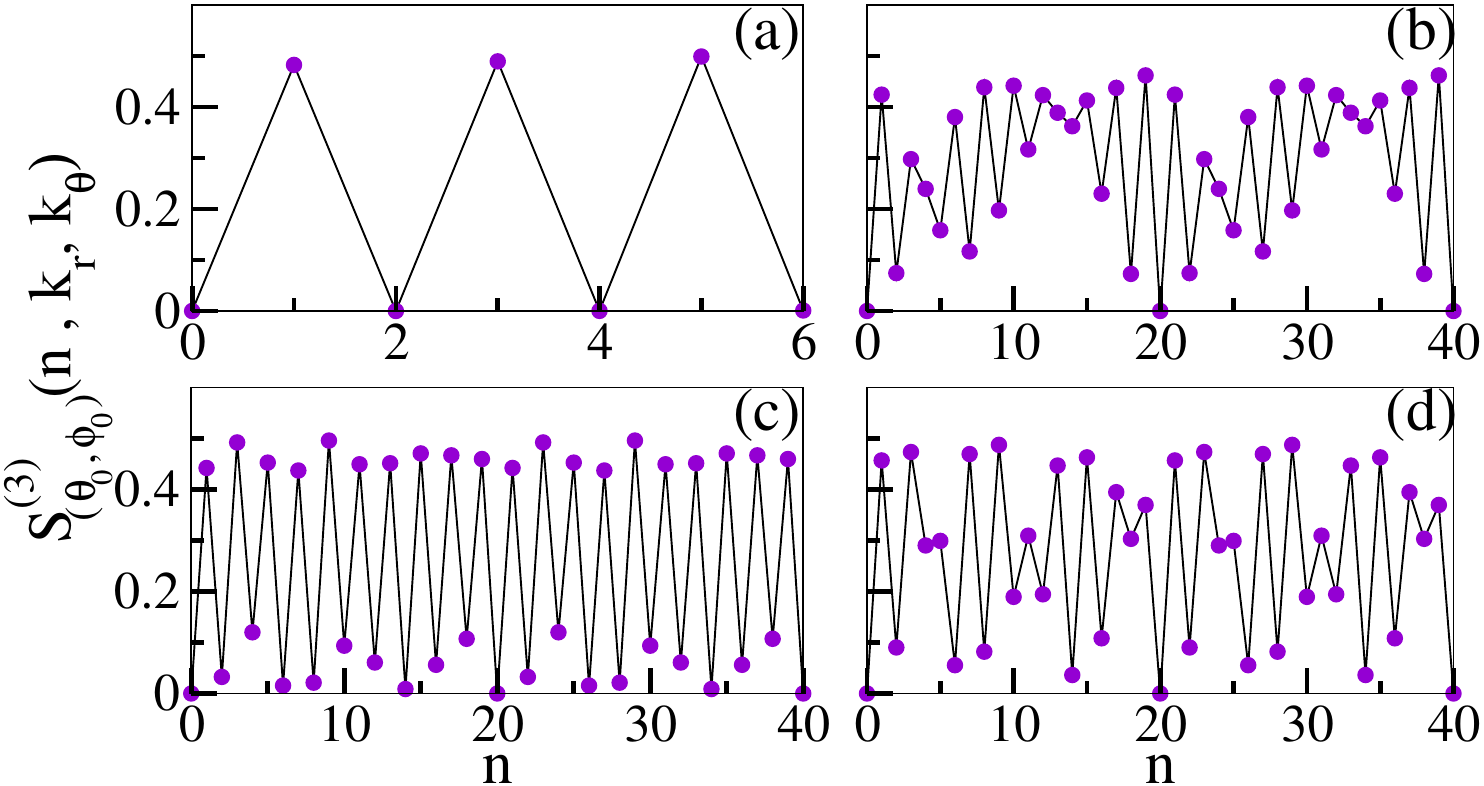}
    \caption{Linear entropy is plotted as a function of discrete time $n$ for $j=1.5$ and $k_\theta = 0$. The choice $a = 9/20 \implies k_r = 3\pi/2 - 3\sin^{-1}\left[2 \sin\left(\pi/20\right)\right]/2$ is used for sub-figures: (a) $|\theta_0 = 0.2,\phi_0 = 1.3\rangle$ and (b) $|\theta_0 = 1.9,\phi_0 = 0.72\rangle$. The choice $a = 7/20 \implies k_r = 3\pi/2 - 3\sin^{-1}\left[2 \sin\left(3\pi/20\right)\right]/2$ is used for sub-figures: (c) $|\theta_0 = 0.35,\phi_0 = 0.6\rangle$ and (d) $|\theta_0 = 0.86,\phi_0 = 0.45\rangle$.}\label{3qubitGeneral}
\end{figure}

The linear entropy given by Eq.~(\ref{Eq:3qubitlinear}) is polynomial in $\gamma$ through $\alpha_n$ and $\beta_n$. Therefore, the linear entropy is periodic when $\gamma = a\pi$ with $a \in \mathds{Q}$ satisfying:
\begin{align}\label{periodicEq}
 \frac{1}{2}\sin\left(\frac{2k_r}{3}\right) = \cos(a\pi) \,\,\text{and} \,\, \frac{1}{3} \leq a \leq \frac{2}{3}.
\end{align}
The lowest value of $a = \frac{1}{3}$ gives $k_r = 3\pi/4$. The periodicity can be confirmed in the computations of the linear entropy for $k_r = 3\pi/4$ as shown in Figs.~\ref{fig6} and \ref{fig7}. The results further show that the periodicity remains unaffected by $k_\theta$. Here, the periodicity condition for the standard QKT with $\kappa_0 = 3\pi/2$ \citep{dogra2019quantum} is a special case of our model. This analysis shows that for countably infinite values of $k_r$, the linear entropy is periodic, and for each such $k_r$, non-countably infinite values of $k_\theta$ also show periodic behaviour.
Our computations also show that if the entanglement dynamics is periodic for a particular initial state, then it is periodic for every initial state (see Fig.~\ref{3qubitGeneral}) and every state satisfies the same periodicity condition given by Eq. (\ref{periodicEq}).

\subsection{Measures of quantum correlations in deep quantum regime}
This subsection presents computational results regarding various measures of quantum correlations for a 3-qubit system. We particularly focus on the kick strength $k_\theta$ as it decides the temporal symmetry.

\begin{figure}
    \includegraphics[width=\linewidth]{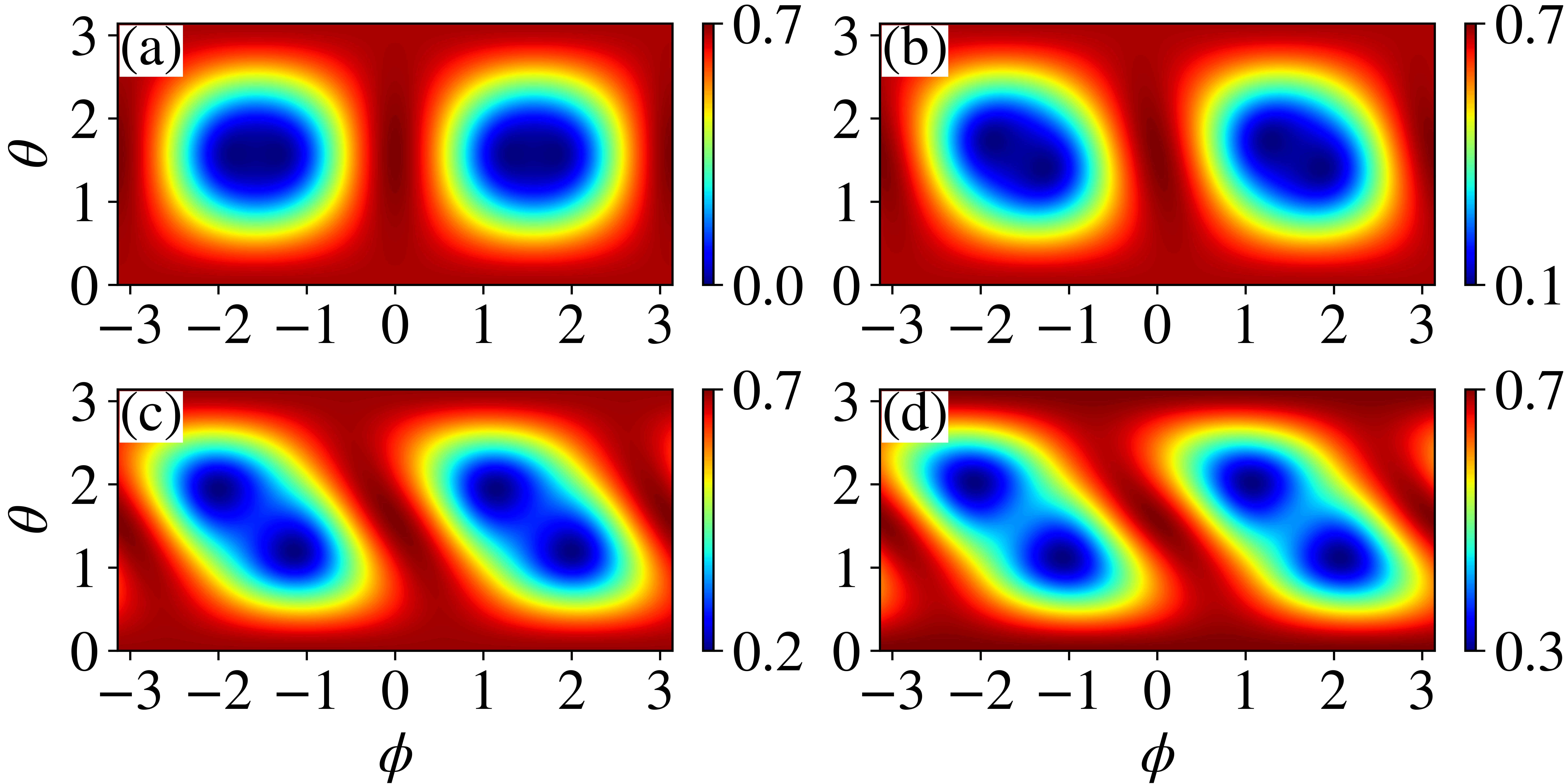}
    \caption{Long-time-averaged von Neumann entropy for single-qubit RDM $\rho_1(n)$, $n=1000$ and a grid of $200\times 200$ initial coherent states. Here, $j = 1.5$, $k_r = 1$, (a) $k_\theta = 0$, (b) $k_\theta = 0.25$, (c) $k_\theta = 0.75$ and (d) $k_\theta = 1$.}\label{fig8}
\end{figure}
\begin{figure}
    \includegraphics[width=\linewidth]{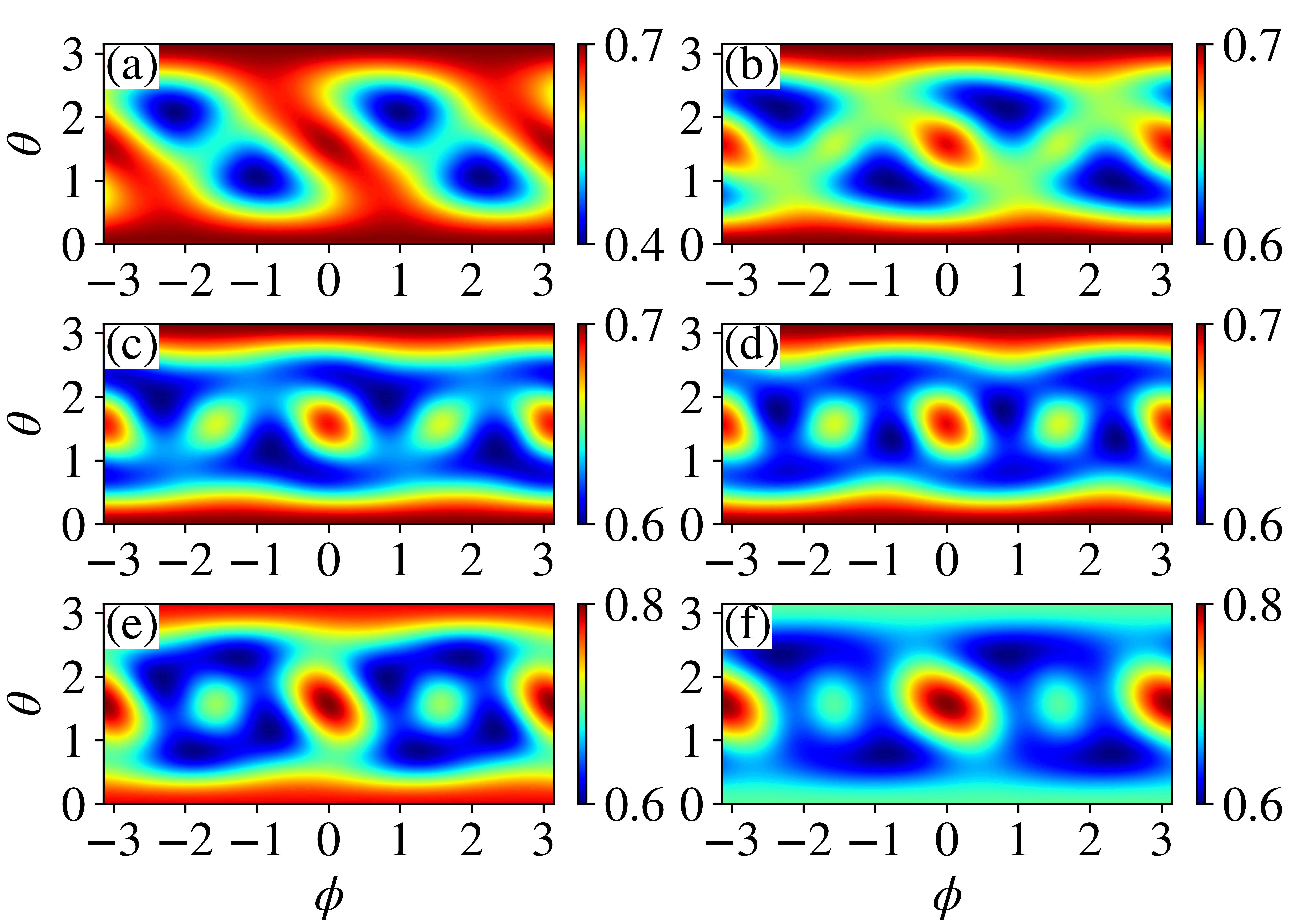}
    \caption{Long-time-averaged von Neumann entropy for single-qubit RDM $\rho_1(n)$, $n=1000$ and a grid of $200\times 200$ initial coherent states. Here, $j = 1.5$, $k_r = 1$, (a) $k_\theta = 1.25$, (b) $k_\theta = 1.75$, (c) $k_\theta = 3.0$ and (d) $k_\theta = 3.75$.}\label{fig8_2}
\end{figure}
The long-time-averaged von Neumann entropy for the 3-qubit system produces similar results as that of the 2-qubit system as shown in Fig.~\ref{fig8}. However, the two blue regions indicating low entanglement get closer in the case of $k_\theta = 1$ in the 3-qubits compared to the 2-qubit system.
Note that the $k_\theta$ is a weak chaos parameter. It increases chaos by twisting the phase-space structures. However, it is not responsible for bifurcation in the classical dynamics. We analyse the semi-classical regime in Sec. \ref{sec:hs} to clarify this issue.

The long-time-averaged von Neumann entropy for the second quadrant $(k > 0, k' < 0)$ is illustrated in Figs.~\ref{fig8_2}. We observe a twisting effect in the coarse-grained phase-space structures. The entanglement show deviation from the coarse-grained phase-space for large values of $k_\theta$.

\begin{figure}
    \includegraphics[width=\linewidth]{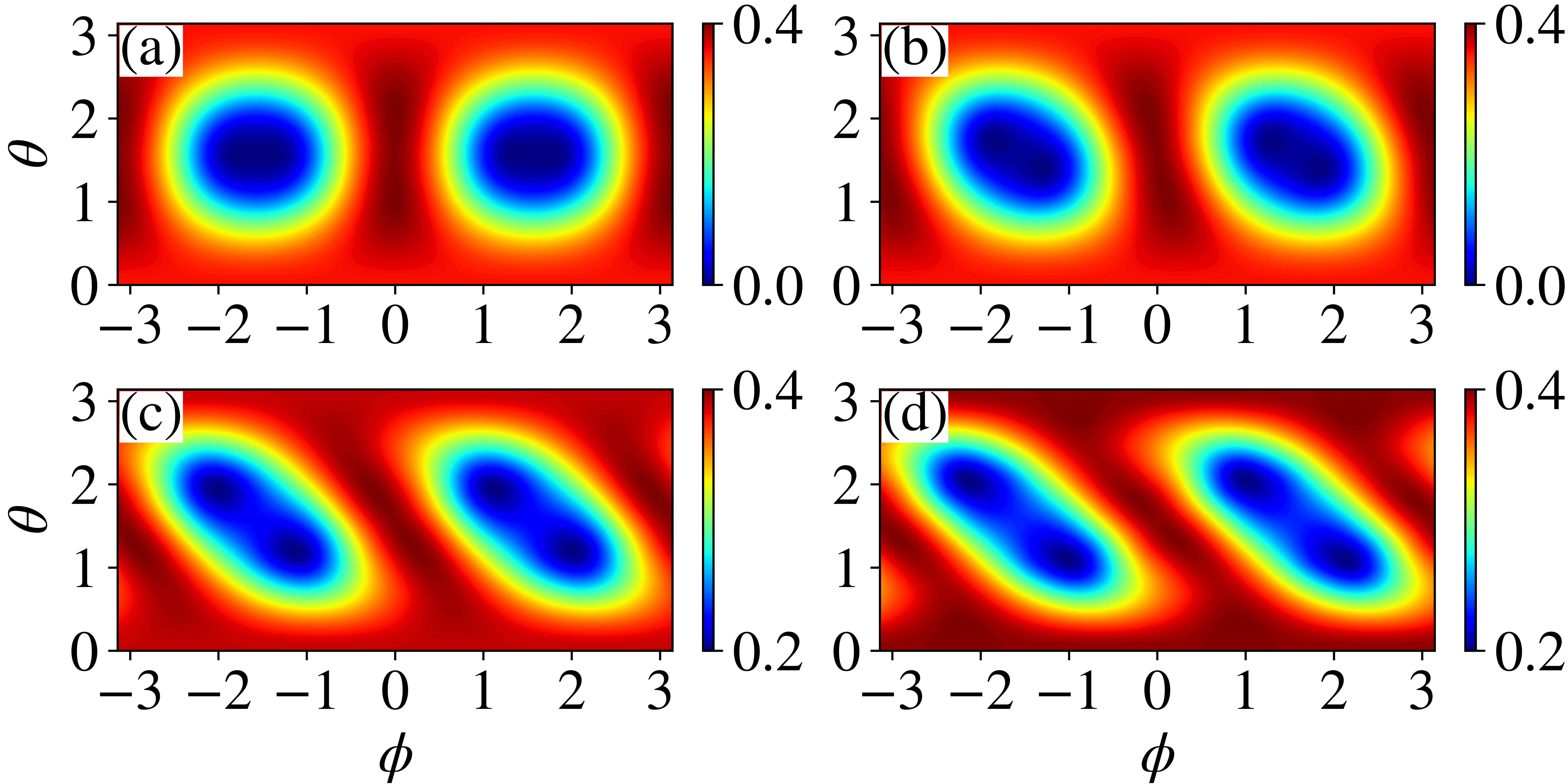}
    \caption{Long-time-averaged quantum discord for the RDM $\rho_{12}(n)$, $n=1000$ and a grid of $200\times 200$ initial coherent states. Here, $j = 1.5$, $k_r = 1$, (a) $k_\theta = 0$, (b) $k_\theta = 0.25$, (c) $k_\theta = 0.75$ and (d) $k_\theta = 1$.}\label{fig9}
\end{figure}
The long-time averaged quantum discord shows that the quantum correlations increase with $k_\theta$ for states $|\theta_0 = \pi/2, \phi_0 = \pm \pi/2\rangle$ and $|\theta_0 = 0, \phi_0 = 0\rangle$ as shown in Fig.~\ref{fig9}. Whereas, the state $|\theta_0 = \pi/2, \phi_0 = 0\rangle$ remains unchanged. The minimum value of the long-time averaged quantum discord for $k_\theta = 0$ is close to zero. However, for $k_\theta = 1$, it is approximately 0.2 with the both cases having 0.4 maximum long-time averaged value of the quantum correlations. It further indicates that, on an average, the long-time averaged quantum correlations increase with $k_\theta$. We observe excellent agreement between the quantum discord and the von Neumann entropy.

\begin{figure}
    \includegraphics[width=\linewidth]{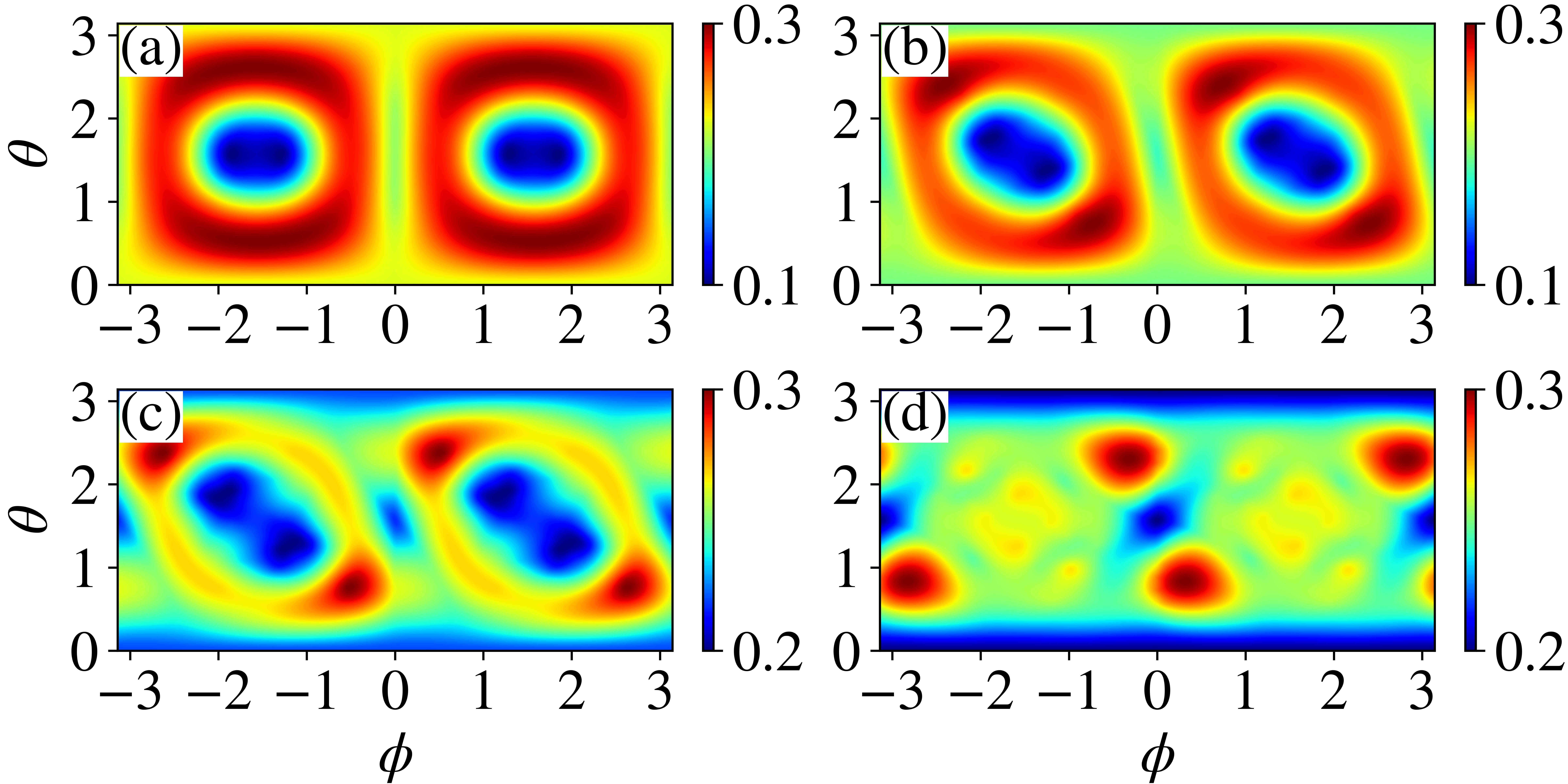}
    \caption{Long-time-averaged concurrence for the RDM $\rho_{12}(n)$, $n=1000$ and a grid of $200\times 200$ initial coherent states. Here, $j = 1.5$, $k_r = 1$, (a) $k_\theta = 0$, (b) $k_\theta = 0.25$, (c) $k_\theta = 0.75$ and (d) $k_\theta = 1$.}\label{fig10}
\end{figure}
Although the long-time averaged concurrence captures features of formation of pairs of blue regions in the case $(k_r = 1, k_\theta = 0.75)$, the blue coloured subregions become yellow for $k_\theta = 1$ leaving high valued regions outside them unchanged (see Fig.~\ref{fig10}). In contrast to earlier two measures, the long-time averaged concurrence of states $|\theta_0 = \pi/2, \phi_0 = 0\rangle$ and $|\theta_0 = 0, \phi_0 = 0\rangle$ increases with slower rate with $k_\theta$. The long-time averaged concurrence, thus, have distinct features and does not fully agree with the earlier two measures of quantum correlations.

\section{Exact Solution for 4-Qubits}\label{sec:4qubit}
The 4-qubit kicked top represents the smallest quantum system showing all-to-all interactions \citep{dogra2019quantum}, which differ from nearest-neighbour interactions. Despite the non-integrability of this system, an exact solution can be derived. Similar to the earlier section, we present an exact solution for the 4-qubit DKT system, followed by the entanglement dynamics and periodicity condition. The detailed analytical calculations for the entanglement dynamics and the infinite-time averaged linear entropy for the general initial state are given in the supplementary material \citep{supplementary2025}.

Following the approach used in previous studies \citep{dogra2019quantum}, we use basis states in which the Floquet operator takes on a block-diagonal form given by
\begin{align}
 |\Phi^\pm_0\rangle &= \frac{1}{\sqrt{2}} \left(|0000\rangle \pm |1111\rangle\right),
 |\Phi^\pm_1\rangle = \frac{1}{\sqrt{2}} \left(|W\rangle \mp |\overline{W}\rangle\right) \text{ and} \notag \\
 |\Phi^+_2\rangle &= \frac{1}{\sqrt{6}} \sum_{\mathcal{P}} |0011\rangle_{\mathcal{P}},
\end{align}
where $|W\rangle = \frac{1}{2} \sum_{\mathcal{P}} |0001\rangle_{\mathcal{P}}$, $|\overline{W}\rangle = \frac{1}{2} \sum_{\mathcal{P}} |1110\rangle_{\mathcal{P}}$, and $\sum_{\mathcal{P}}$ denotes the sum over all possible permutations. In this basis, the $n$-th power of the Floquet operator is given by
\begin{align}
 \mathcal{U} = \begin{pmatrix}
 -1 &0 &0\\
        0 &\mathcal{U}_+ &0\\
        0 &0 &\mathcal{U}_-
    \end{pmatrix},
\end{align}
where, 
\begin{align}
    \begin{split}
 \mathcal{U}_+ &= e^{-\frac{i}{2}(k_r + \pi)} \begin{pmatrix}
            \alpha' &i{\beta'}^*\\
 i\beta' &{\alpha'}^*
        \end{pmatrix}  \; \text{and} \\
 \mathcal{U}_- &= \begin{pmatrix}
            0   &e^{-\frac{3}{4}i(k_r - k_\theta)}\\
 -e^{-\frac{3}{4}i(k_r + k_\theta)}    &0
        \end{pmatrix},
    \end{split}
\end{align}
further, $\alpha'$ and $\beta'$ are given by
\begin{align}
    \alpha' =& \frac{1}{2} \sin(k_r) + \frac{i}{4} \left[3\cos(k_\theta) - \cos(k_r)\right]   \, \text{and} \nonumber \\
    \beta' =& \frac{\sqrt{3}}{4} \left[ \cos(k_r) + \cos(k_\theta) + 2i \sin(k_\theta)\right]. \nonumber    
\end{align}
Utilizing a method analogous to the 3-qubit system, we derive the time-evolved Floquet operator as follows:
\begin{align}
    \begin{split}
 \mathcal{U}_+^n &= e^{-\frac{i}{2}n (k_r + \pi)} \begin{pmatrix}
            \alpha'_n &i{\beta'_n}^*\\
 i\beta'_n &{\alpha'_n}^*
        \end{pmatrix} \,\, \text{and} \\
 \mathcal{U}_-^n &= e^{-\frac{3i}{4} n k_r} \begin{pmatrix}
 \cos\left(\frac{n\pi}{2}\right) &-\sin\left(\frac{n\pi}{2}\right)e^{-\frac{3i}{4} k_\theta}\\
 \sin\left(\frac{n\pi}{2}\right) e^{\frac{3i}{4} k_\theta} &\cos\left(\frac{n\pi}{2}\right)
        \end{pmatrix},
    \end{split}
\end{align}
where $\alpha'_n$ and $\beta'_n$ are given by
\begin{align}
    \alpha'_n =& \cos(n \gamma) + \frac{i}{4} \frac{\sin(n\gamma)}{\sin(\gamma)} \left[3\cos(k_\theta) - \cos(k_r)\right], \\
    \beta'_n =& \frac{\sqrt{3}}{4} \frac{\sin(n\gamma)}{\sin(\gamma)} \left[ \cos(k_r) + \cos(k_\theta) + 2i \sin(k_\theta)\right],
\end{align}
and $\cos \gamma = \frac{1}{2}\sin k_r$. In the special case where $k_\theta = k_r$ and $2k_r = \kappa_0$, we recover the 4-qubit system discussed in Ref. \citep{dogra2019quantum}.

We consider the general initial state Eq.~(\ref{Eq:generalstate}) and evolve using the Floquet operator to get the following state:
\begin{align}
 |\psi_n\rangle = c_0'' |\Phi_0^+\rangle + c_1'' |\Phi_1^+\rangle + c_2'' |\Phi_2^+\rangle + c_3'' |\Phi_0^-\rangle + c_4'' |\Phi_1^-\rangle, \nonumber 
\end{align}
where the coefficients are defined as follows:
\begin{align}
c_0'' &= \frac{t_1}{2\sqrt{2}} \left[2\alpha_n (t_2 t_3 + i\; t_4) + i\; \sqrt{3} t_5 \beta_n^* t_6 \right], \\
c_1'' &= \frac{t_7 t_8 t_{9}}{\sqrt{2}} \left[t_{10} t_{11} + i\; t_{12} \right], \\
c_2'' &= \frac{t_1}{2\sqrt{2}} \left[2i \beta_n (t_2 t_3 + i\; t_4) + \sqrt{3} t_5 \alpha_n^* t_6 \right], \\
c_3'' &= \frac{t_{13}}{\sqrt{2}}  \left[t_{14} t_{15} (t_2 t_3 - t_4) - t_5 t_{16} t_{9} (t_{11} + t_{10} t_{12}) \right], \\
c_4'' &= \frac{t_{17}}{\sqrt{2}}  \left[t_{14} t_{16} (t_3 - t_{18} t_4) + t_7 t_{15} t_{9} (t_{11} + t_{10} t_{12}) \right],
\end{align}
and $t_1 = e^{-\frac{i}{2}(nk_r + n\pi + 8\phi_0)}$, $t_2 = e^{4i\phi_0}$, $t_3 = \cos^4\left(\frac{\theta_0}{2}\right)$, $t_4 = \sin^4\left(\frac{\theta_0}{2}\right)$, $t_5 = e^{2i\phi_0}$, $t_6 = \sin^2(\theta_0)$, $t_7 = e^{-2i\phi_0}$, $t_8 = (-1)^n$, $t_{9} = \sin(\theta_0)$, $t_{10} = \cos(\theta_0)$, $t_{11} = \cos(\phi_0)$, $t_{12} = \sin(\phi_0)$, $t_{13} = e^{-\frac{i}{4}(3nk_r + 3k_\theta + 16\phi_0)}$, $t_{14} = e^{\frac{3i}{4}k_\theta}$, $t_{15} = \cos\left(\frac{n\pi}{2}\right)$, $t_{16} = \sin\left(\frac{n\pi}{2}\right)$, $t_{17} = e^{-\frac{3i}{4}nk_r}$, and $t_{18} = e^{-4i\phi_0}$. The single-qubit RDM $\rho_1(n)$ corresponding to the above state is expressed as follows:
\begin{align}
    \rho_1(n) = \begin{pmatrix}
 p_{11}'' &p_{12}'' \\
 {p_{12}''}^*   &1- p_{11}''
    \end{pmatrix},
\end{align}
where, 
\begin{align}
 p_{11}'' =& \frac{1}{2}\left(1+2\,\Re[c''_0 {c''_3}^*]+2\,\Re[c''_1 {c''_4}^*]\right) \\
 p_{12}'' =& \frac{1}{4}(c_1''-c_4'') {(c_3''-c_0'')}^* + \frac{1}{4}(c_0''+c_3'') {(c_1''+c_4'')}^* \notag \\
    &+ \frac{\sqrt{3}}{4}c_2'' {(c_4''-c_1'')}^* + \frac{\sqrt{3}}{4}{c_2''}^* (c_4''+c_1'').
\end{align}
Using its eigenvalues $1/2\pm \sqrt{{|p_{11}''|}^2 + {|p_{12}''|}^2}$, the linear entropy is obtained as follows:
\begin{align}
 S^{(4)}_{(\theta_0,\phi_0)}(n, k_r, k_\theta) &= \frac{1}{2} - 2 {\left(\Re[c''_0 {c''_2}^*] + \Re[c''_1 {c''_4}^*]\right)}^2 \notag \\
    & - 2 {|p_{12}''|}^2.
\end{align}

The infinite-time averaged linear entropy for the special state $|\theta_0 = 0, \phi_0 = 0\rangle$ is given by (see supplementary material \citep{supplementary2025})
\begin{align}\label{infinite-time averaged-4qubit-00}
    \langle S^{(4)}_{(0,0)} \left(k_r,k_\theta\right) \rangle =\frac{160+25\cos(2k_r)-9\cos(2k_\theta)}{64\left[7+\cos(2k_r)\right]}.
\end{align}
Similar to the 3-qubit case, we find that the inequality $\langle S^{(4)}_{(0,0)}(k_r, k_\theta = k_r) \rangle \geq \langle S^{(4)}_{(0,0)} (k_r, k_\theta = 0) \rangle$ holds true also for the 4-qubit system. The maximum value of the infinite-time averaged linear entropy for $k_\theta = k_r$ is $\frac{3}{8}$, whereas for $k_\theta = 0$, it is $\frac{11}{32}$. For the case $k_\theta = k_r$ and $2k_r = \kappa_0$, we can recover the result presented in Eq. (55) of \citep{dogra2019quantum}.

The infinite-time averaged linear entropy for the another special state $|\theta_0 = \pi/2, \phi_0 = -\pi/2\rangle$ is given by (see supplementary material \citep{supplementary2025})
\begin{align}\label{infinite-time averaged-4qubit-piby2}
    \langle S^{(4)}_{(\frac{\pi}{2},-\frac{\pi}{2})} \left(k_r, k_\theta\right) \rangle = \frac{3}{8} -  \frac{{\left[\cos(k_r) + 3\cos(k_\theta)\right]}^2}{16 {\left[7+\cos(2k_r)\right]}}.
\end{align}
For the case $k_\theta = k_r$ and $2k_r = \kappa_0$, we recover the expression provided in Eq. (59) of \citep{dogra2019quantum}.
For $k_\theta = k_r$ its maximum value is $3/8$ at $k_r = \pi/2$. For $k_\theta = 0$, it is $11/32$ at $k_r = \pi$. Our results show that similar to the 3-qubit system the 4-qubit system also exhibits higher entanglement for the case $k_\theta = k_r$ than the case $k_\theta = 0$.

The linear entropy is periodic if there exist $a, b \in \mathds{Q}$, such that
\begin{align}\label{4qubitPeriocity}
    \begin{split}
 \frac{1}{2}\sin(k_r) = \cos(a\pi), \; \frac{k_r}{4} = b\pi \;\text{ and }\; \frac{1}{3} \leq a \leq \frac{2}{3}.
    \end{split}
\end{align}
Refer supplementary material for the detailed derivation \citep{supplementary2025}. Our computations reveal $k_\theta$ independence of the periodicity of entanglement dynamics (see Figs.~\ref{fig11} and \ref{fig12}). Similar to the 3-qubit system, the 4-qubit system also reveals the initial state independence of the periodicity of entanglement dynamics (see Fig.~\ref{4qubitGeneral}).
\begin{figure}
    \includegraphics[width=\linewidth]{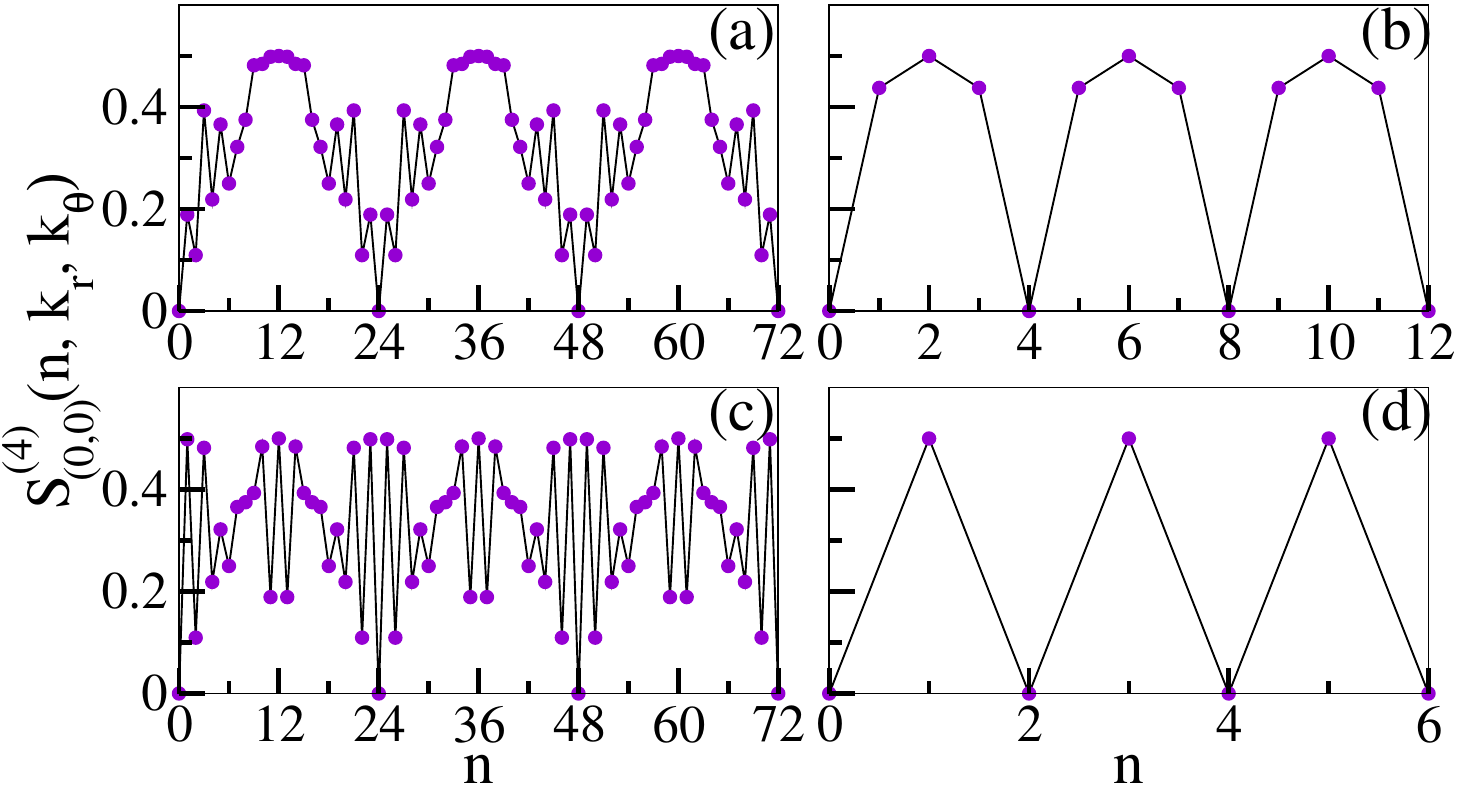}
    \caption{Linear entropy is plotted as a function of discrete time $n$ for the 4-qubit state $|\theta_0 = 0, \phi_0 = 0\rangle$ and $k_\theta = 0$. Here, (a) $k_r = \pi/2$ and (c) $k_r = 3\pi/2$ has a period of 24, (b) $k_r = \pi$ has a period of 4, and (d) $k_r = 2\pi$ has a period of 2.}\label{4qubitPeriodic}
\end{figure}
\begin{figure}
    \includegraphics[width=\linewidth]{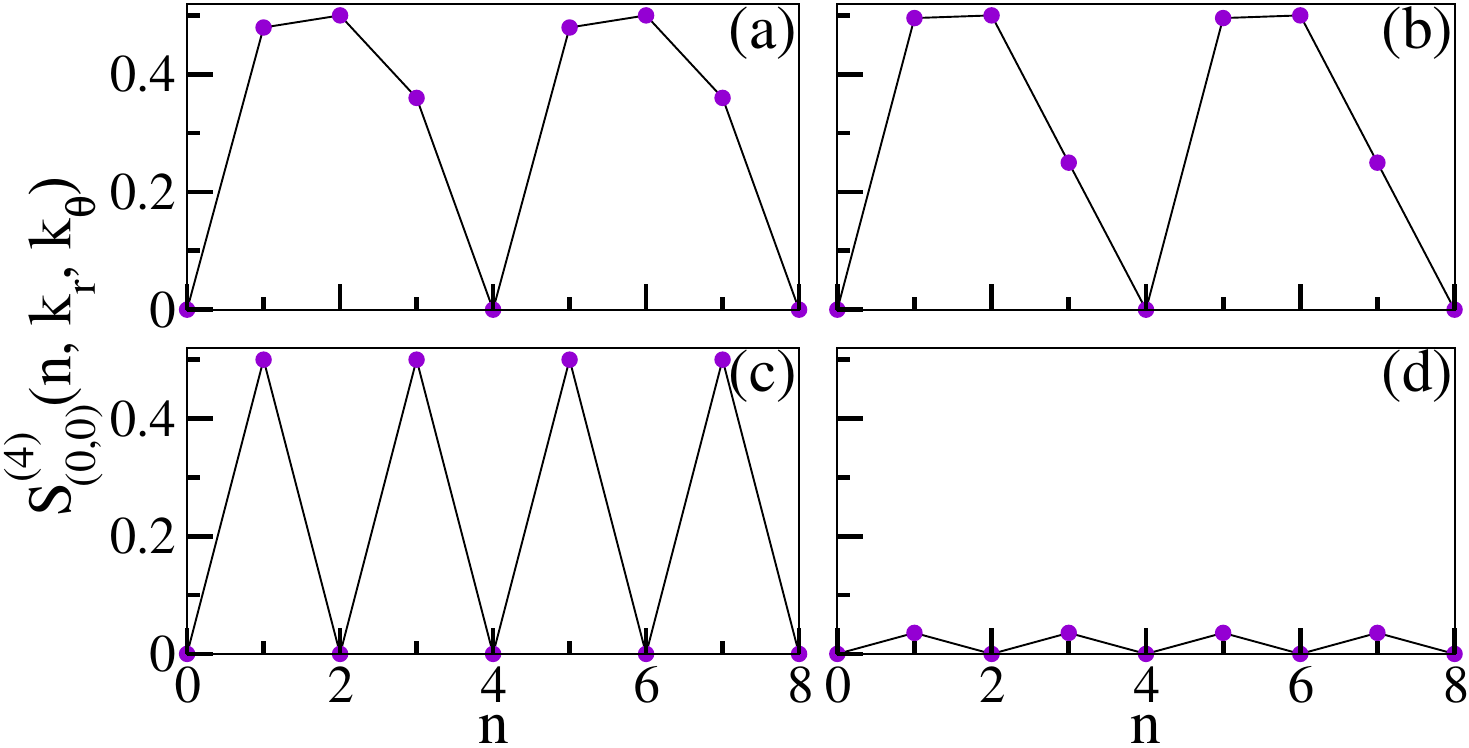}
    \caption{Linear entropy is plotted as a function of discrete time $n$ for the 4-qubit initial coherent state $|\theta_0 = 0, \phi_0 = 0\rangle$. Here, (a) $k_r = \pi, k_\theta = 0.2\pi$, (b) $k_r = \pi, k_\theta = 0.4\pi$, (c) $k_r = 2\pi, k_\theta = 0.2\pi$, and (d) $k_r = 4\pi, k_\theta = 0.2\pi$.}\label{fig11}
\end{figure}
\begin{figure}
    \includegraphics[width=0.65\linewidth]{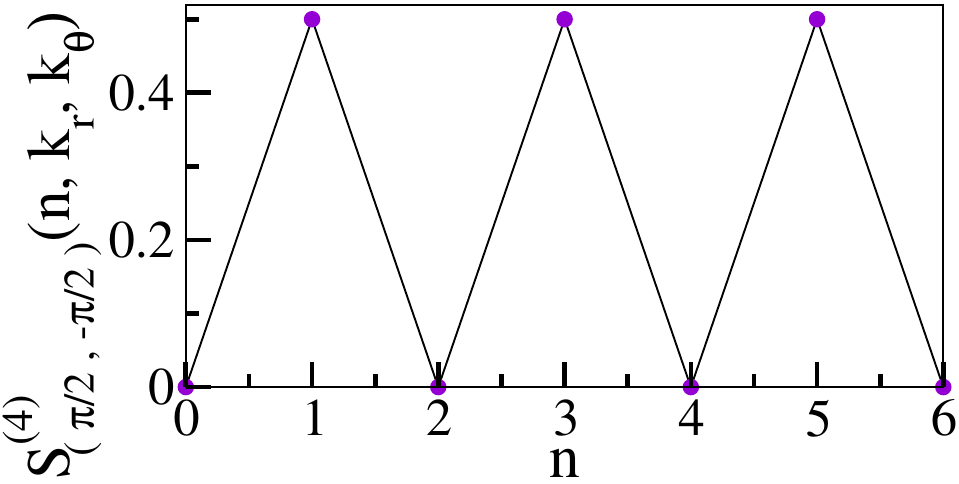}
    \caption{Linear entropy is plotted as a function of discrete time $n$ corresponding to the 4-qubit initial coherent state $|\theta_0 = \pi/2,\phi_0 = -\pi/2\rangle$ for $k_r = \pi$ and $k_\theta \in \left(-k_r,k_r\right)$.}\label{fig12}
\end{figure}
\begin{figure}
    \includegraphics[width=\linewidth]{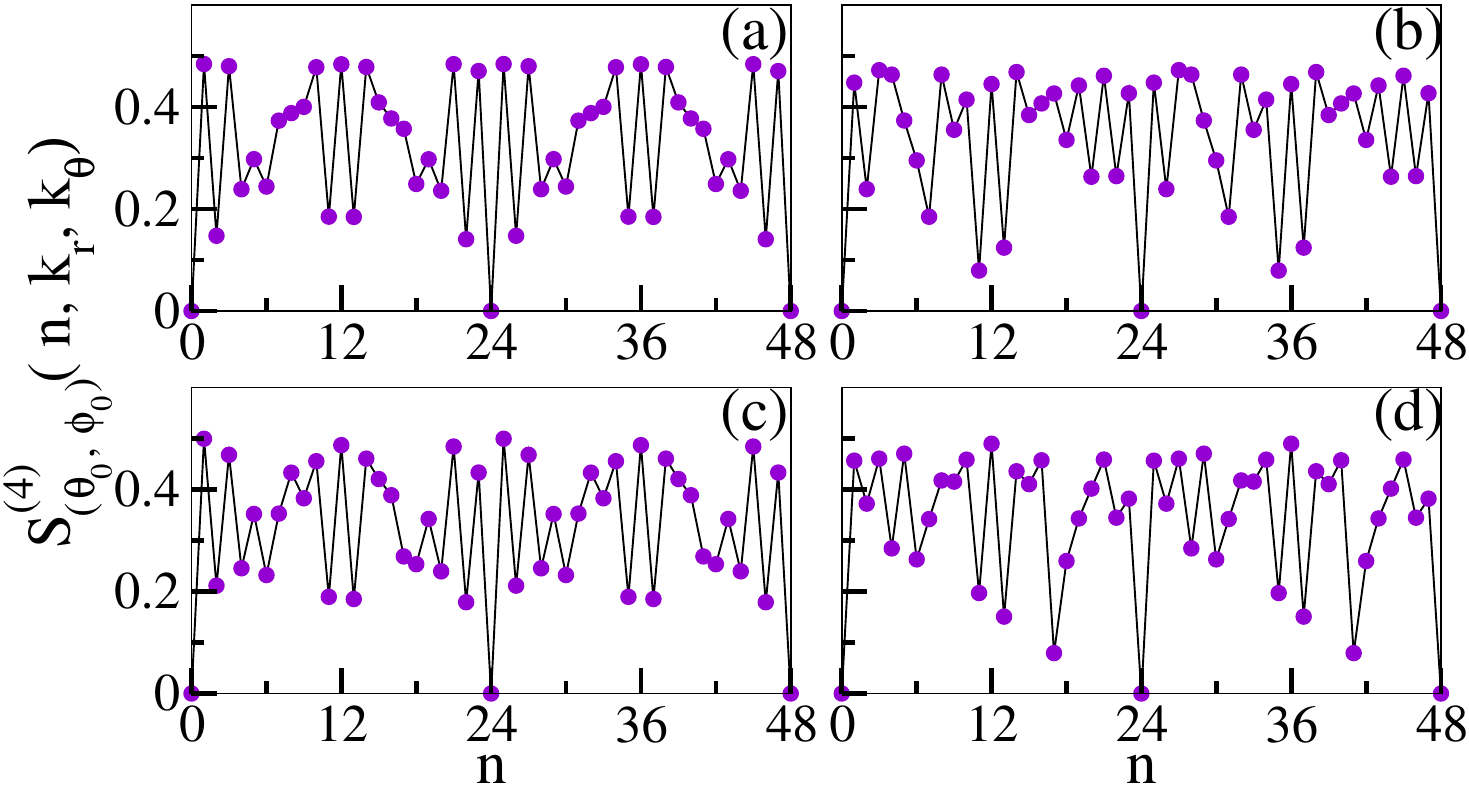}
    \caption{Linear entropy is plotted as a function of discrete time $n$ for 4-qubit RDM $\rho_1(n)$ with $k_r=3\pi/2$ and $k_\theta = 0$. Here, the initial coherent states are: (a) $|\theta_0 = 0.2,\phi_0 = 1.3\rangle$, (b) $|\theta_0 = 1.9,\phi_0 = 0.72\rangle$, (c) $|\theta_0 = 0.35,\phi_0 = 0.6\rangle$ and (d) $|\theta_0 = 0.86,\phi_0 = 0.45\rangle$.}\label{4qubitGeneral}
\end{figure}

\begin{figure}
    \includegraphics[width=\linewidth]{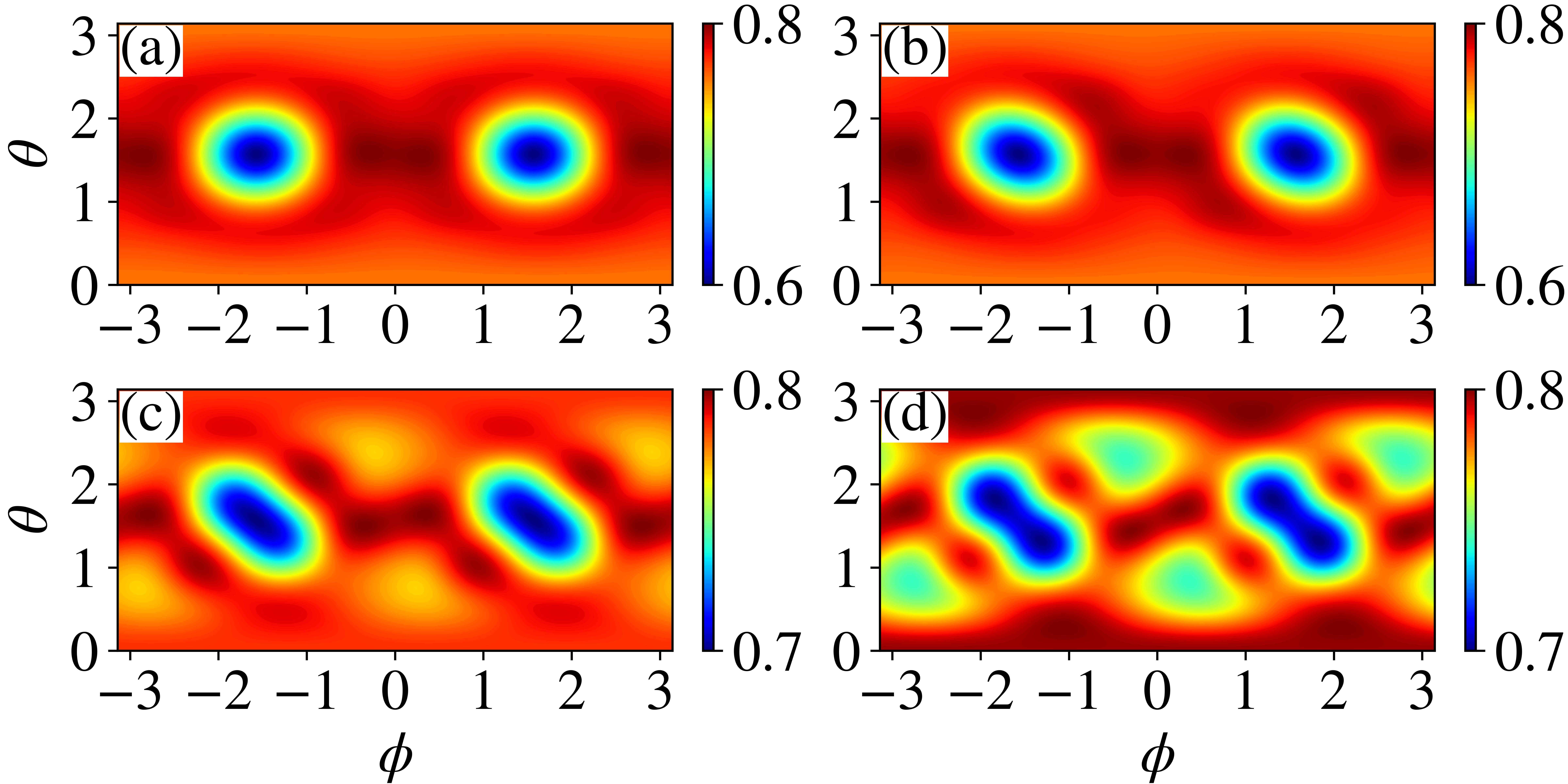}
    \caption{Long-time-averaged von Neumann entropy for single-qubit RDM $\rho_1(n)$, $n=1000$ and a grid of $200\times 200$ initial coherent states. Here, $j = 2$, $k_r = 1$, (a) $k_\theta = 0$, (b) $k_\theta = 0.25$, (c) $k_\theta = 0.75$ and (d) $k_\theta = 1$.}\label{vn4qubit}
\end{figure}

Our analysis of the long-time averaged von Neumann entropy reveals that the 4-qubit system follows the trend set by the 2-qubit and 3-qubit systems. Here, we observe the formation of low-valued long-time averaged entanglement subregions (see Fig.~\ref{vn4qubit}). For the case $(k_r = 1, k_\theta = 1)$, the low-valued long-time averaged entanglement regions do not fully split into further two. The results, thus, show consistency with the 2-qubit and 3-qubit systems.

\section{High-spin system}\label{sec:hs}
In this section, we analyse the average quantum correlations as a function of $k_r$ and $k_\theta$ over the entire phase-space. Additionally, we study a special initial coherent state, $|\theta_0 = \pi/2, \phi_0 = -\pi/2\rangle$, since its corresponding point in the classical dynamics bifurcates at $k_r = 1$. Notably, this point is a saddle point in phase-space, and the behaviour of average quantum correlations for this state reveals an interesting dependence on $j$. These quantum correlations in the high-spin limit are then compared with those from the deep quantum regime and the classical phase-space. To explore the semi-classical regime, we consider system of qubits as high as $j=500.5$.

\subsection{von Neumann entropy}
In classical dynamics, it was noted that $k_r$ is a strong chaos parameter and is responsible for bifurcations. The $k_\theta$, on the other hand, is a weak chaos parameter. It only shrinks the regular region by twisting the phase-space structures without causing bifurcations. In the 2-qubit system, however, we observe that the low-valued regions of long-time averaged entanglement further divide into two subregions with an increase in $k_\theta$ (see Fig.~\ref{j1}). The distinction between these two subdivided low-valued regions of the long-time averaged entanglement becomes less pronounced as $j$ is increased from one to two (see Figs.~\ref{j1}, \ref{fig8}, and \ref{vn4qubit}).
\begin{figure}[h!]
    \includegraphics[width=\linewidth]{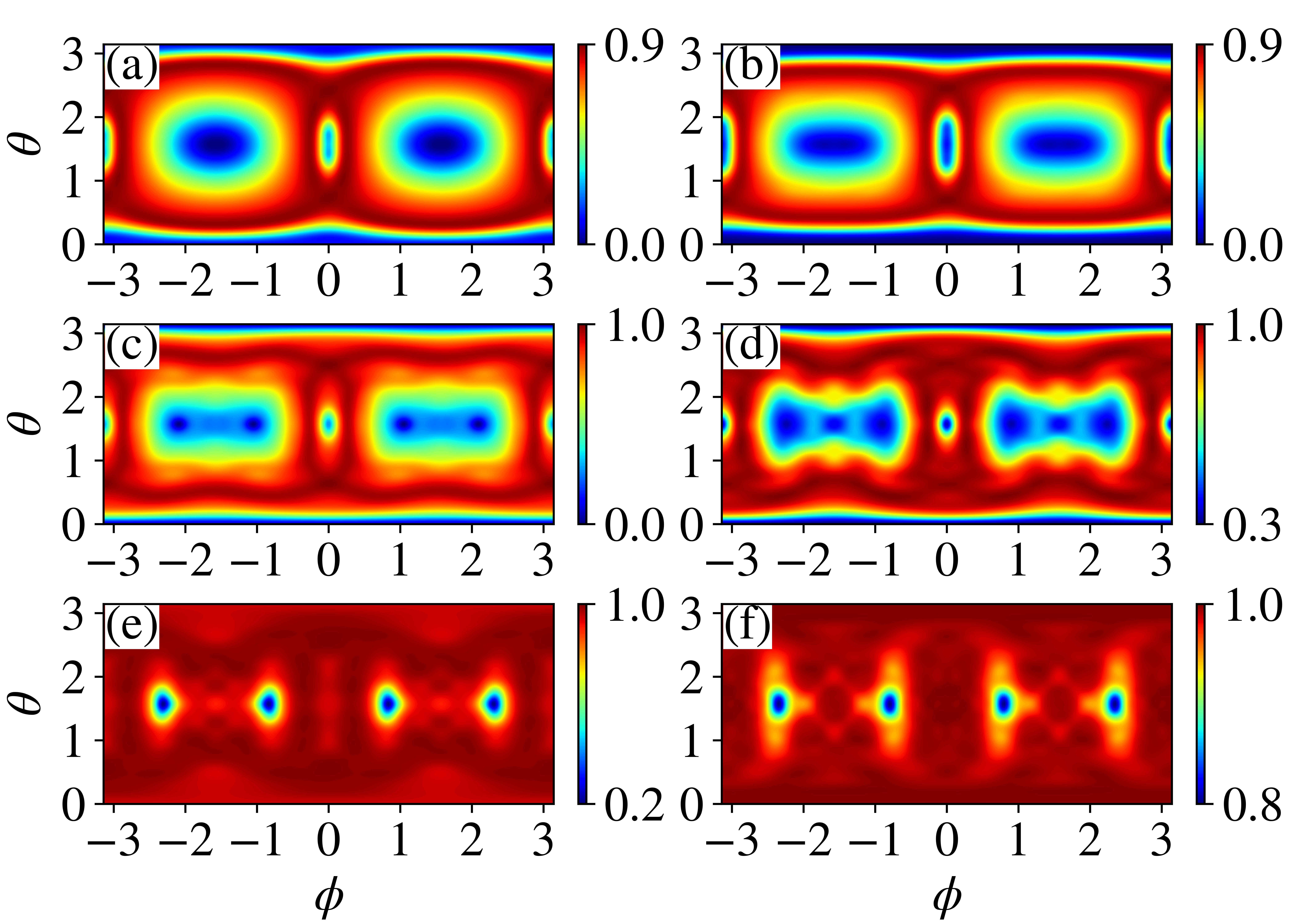}
    \caption{Long-time-averaged von Neumann entropy for a single-qubit RDM $\rho_1(n)$ and a grid of $200\times 200$ initial coherent states. Here, $n=1000$, $j = 50.5$, $k_\theta = 0$. (a) $k_r = 0.75$, (b) $k_r = 1$, (c) $k_r = 1.25$, (d) $k_r = 1.5$, (e) $k_r = 1.75$ and (f) $k_r = 2$.}\label{fig13}
\end{figure}
\begin{figure}[h!]
    \includegraphics[width=\linewidth]{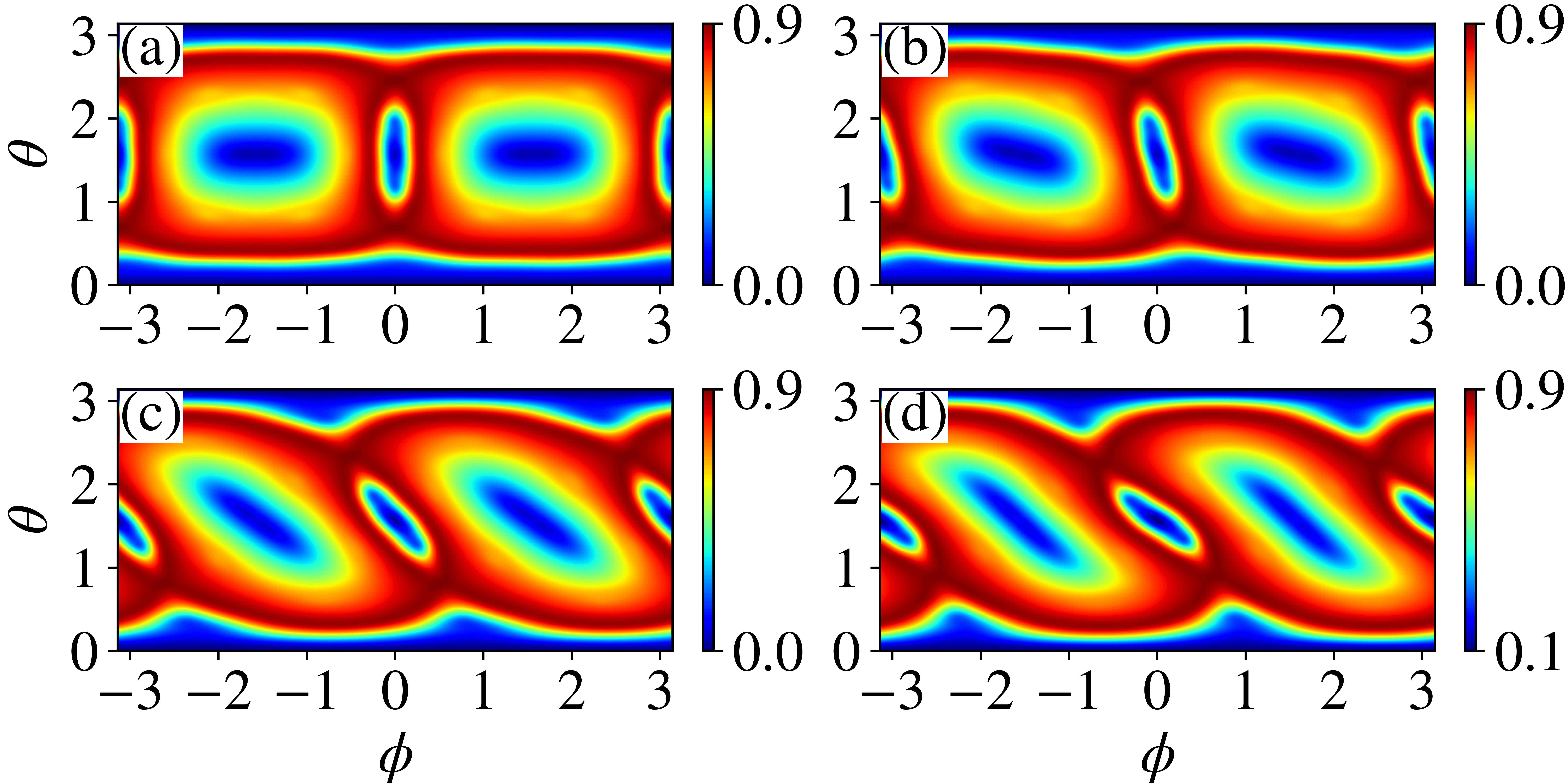}
    \caption{Long-time-averaged von Neumann entropy for a single-qubit RDM $\rho_1(n)$ and a grid of $200\times 200$ initial coherent states. Here, $n=1000$, $j = 75.5$, $k_r = 1$, (a) $k_\theta = 0$, (b) $k_\theta = 0.25$, (c) $k_\theta = 0.75$ and (d) $k_\theta = 1$.}\label{fig14}
\end{figure}
\begin{figure}[h!]
    \includegraphics[width=\linewidth]{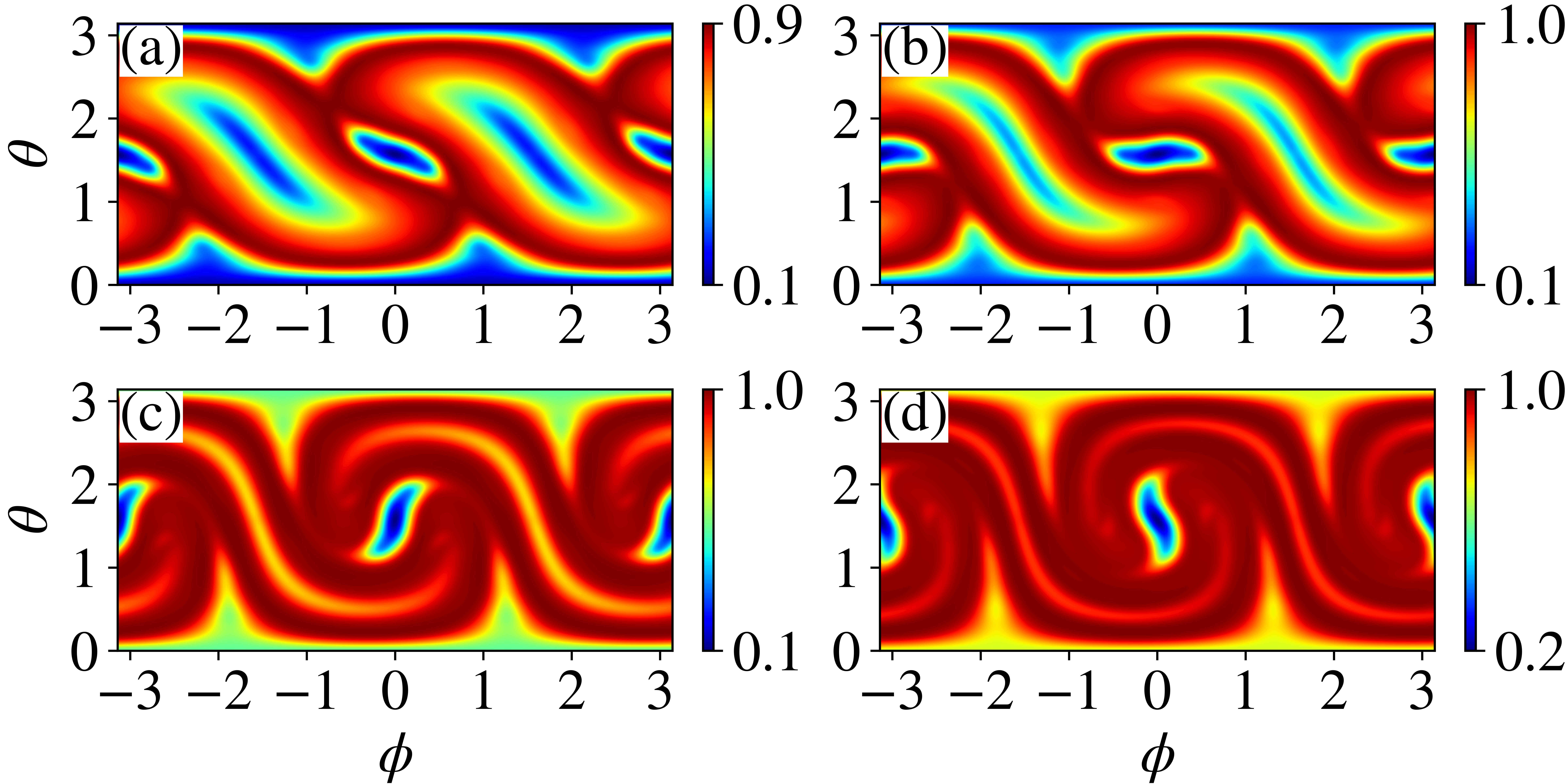}
    \caption{Long-time-averaged von Neumann entropy for a single-qubit RDM $\rho_1(n)$ and a grid of $200\times 200$ initial coherent states. Here, $n=1000$, $j = 75.5$, $k_r = 1$, (a) $k_\theta = 1.25$, (b) $k_\theta = 1.75$, (c) $k_\theta = 3.0$ and (d) $k_\theta = 3.75$.}\label{fig14_2}
\end{figure}

The long-time averaged von Neumann entropy reveals fine-grained phase-space structures (see Fig.~\ref{fig13}). The low-valued long-time averaged entanglement of states $|\theta_0 = \pi/2, \phi_0 = \pm \pi/2 \rangle$ corresponds to the trivial fixed points for $k_r = 0.75$, as shown in Fig.~\ref{fig13}(a). States $|\theta_0 = 0, \phi_0 = 0 \rangle$ and $|\theta_0 = \pi/2, \phi_0 = 0 \rangle$ show lower entropy compared to the deep quantum regime (see Figs.~\ref{j1}, \ref{fig8}, and \ref{vn4qubit}). Interestingly, their long-time averaged entanglement decreases for $k_r = 1.0$ (see Fig.~\ref{fig13}(b)), remains close to zero for $k_r = 1.25$ (see Fig.~\ref{fig13}(c)), and for $k_r > 1.5$, it increases rapidly. States belonging to the boundaries of the period-4 cycle have high long-time averaged entanglement and saturate to maximum (see Fig.~\ref{fig13}(c)). The formation of low-valued regions of long-time averaged entanglement corresponds to stable fixed points, and high long-time averaged entanglement corresponds to the chaotic boundaries of the period-4 cycles. For $k_r = 1.5$, the long-time averaged entanglement closely follows fine-grained phase-space structures, as shown in Fig.~\ref{fig13}(d). However, points $(\theta_0 = \pi/2, \phi_0 = \pm \pi/2)$ are chaotic in the classical dynamics (see Fig.~\ref{ppkr}(d)). In contrast, the corresponding states have lower value of the long-time averaged entanglement. The results of Figs.~\ref{fig13}(e) and \ref{fig13}(f) are consistent with the corresponding phase-space structures.

\begin{figure}
    \includegraphics[width=\linewidth]{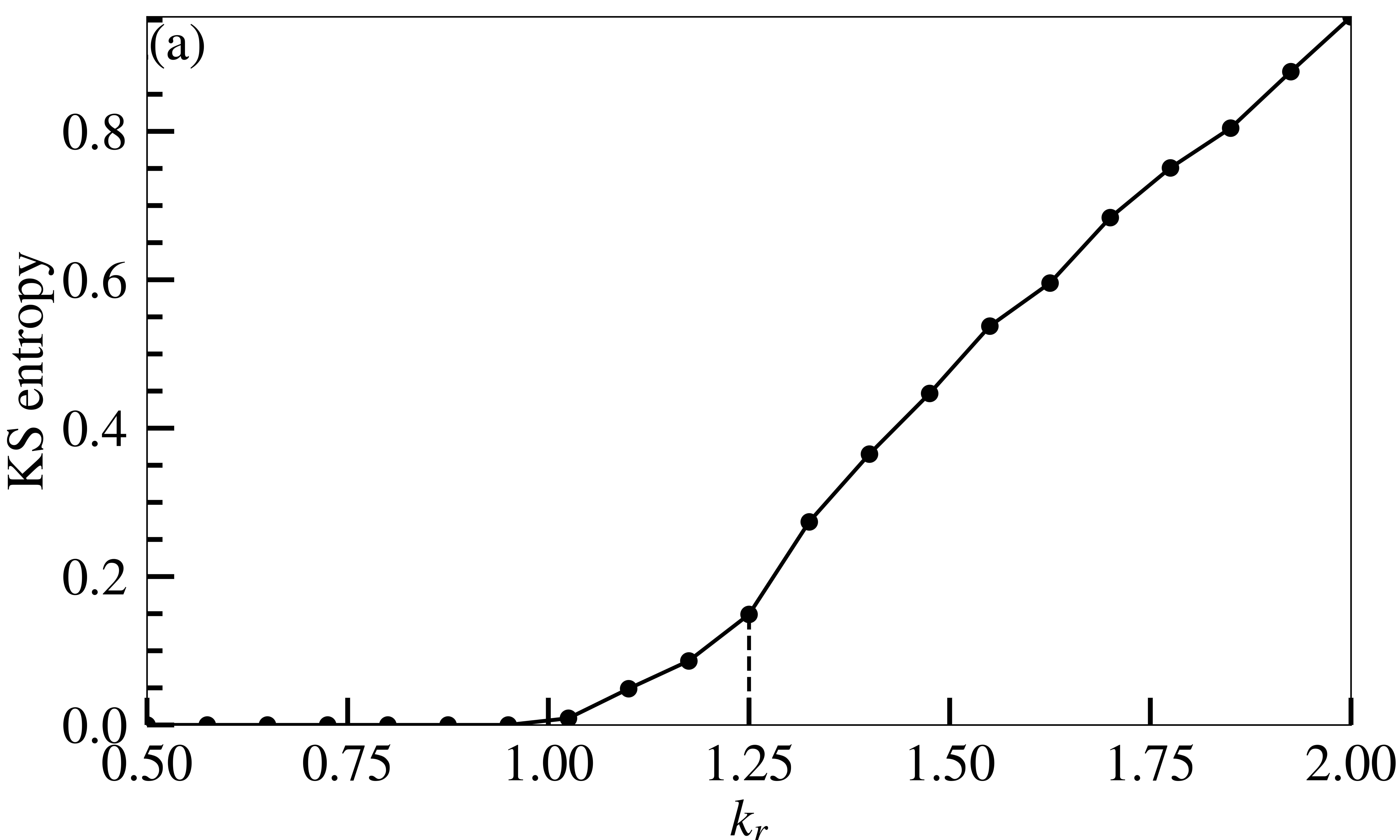}
    \includegraphics[width=\linewidth]{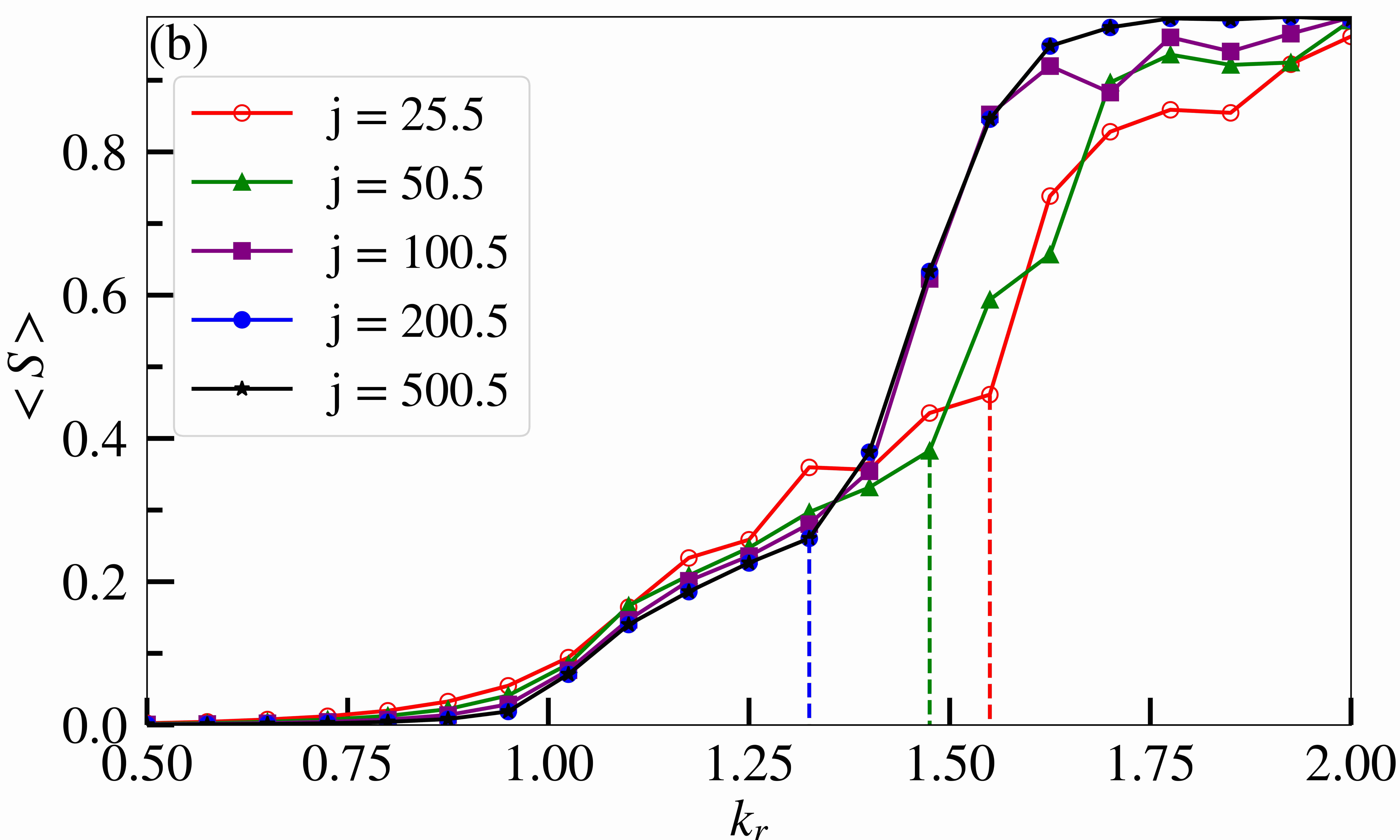}
    \includegraphics[width=\linewidth]{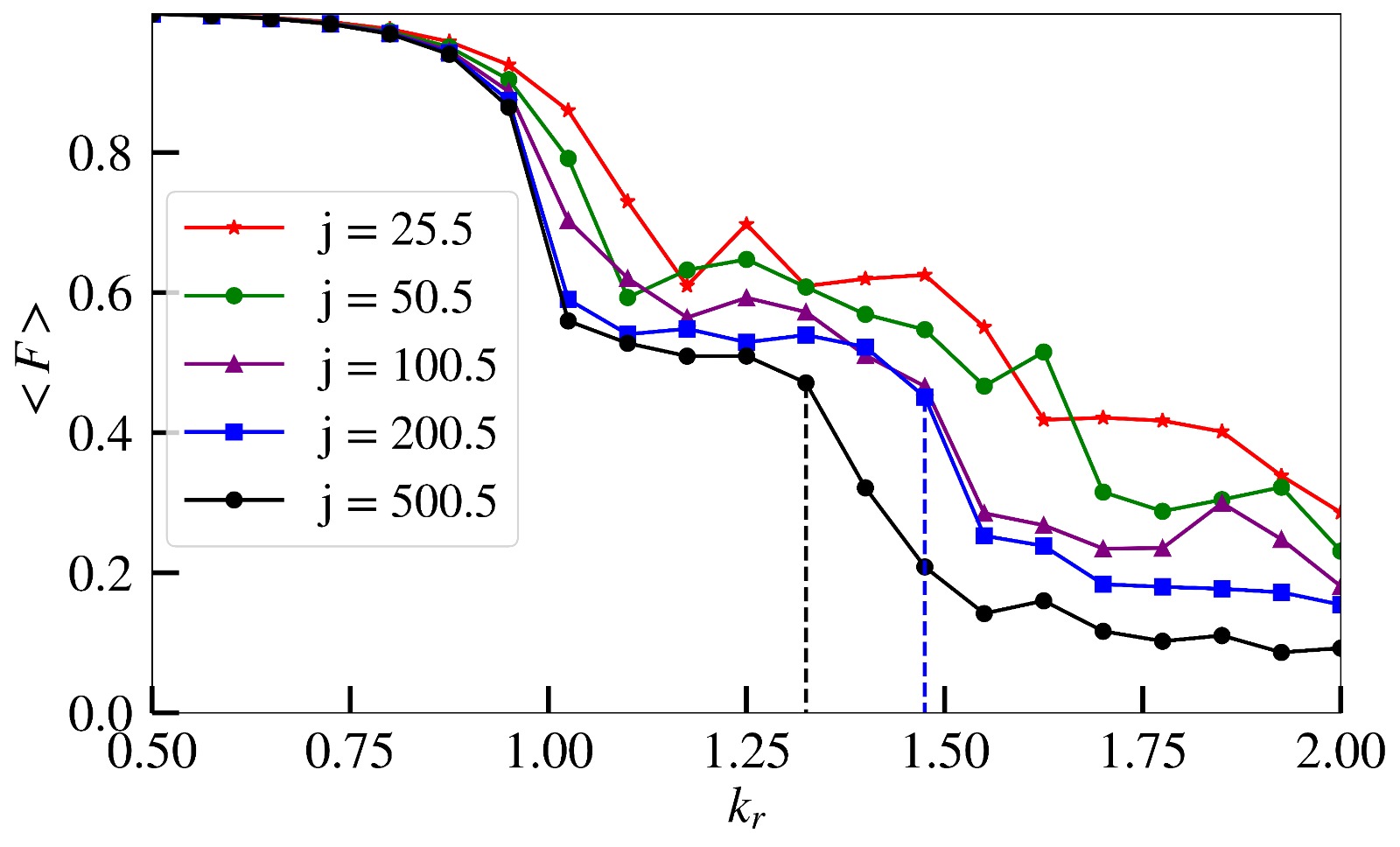}
    \caption{(a) KSE versus $k_r$ for the point $(\theta_0 = \pi/2, \phi_0 = -\pi/2)$ evolved for $10^5$ kicks keeping $k_\theta = 0$. (b) Long-time-averaged von Neumann entropy plotted versus $k_r$ for the initial coherent state $|\theta_0 = \pi/2, \phi_0 = -\pi/2\rangle$, $n = 1000$ and $k_\theta = 0$. (c) Long-time-averaged fidelity plotted versus $k_r$ for the initial coherent state $|\theta_0 = \pi/2, \phi_0 = -\pi/2\rangle$, $T = 1000$ and $k_\theta = 0$.}\label{y-1}
\end{figure}

The KSE indicates the formation of a homoclinic point (intersection of unstable and stable manifolds) at $k_r = 1$ before transitioning to chaos. This is shown by the slow growth of KSE for $k_r \in (1.0, 1.25)$ (see Fig.~\ref{y-1}(a)). Beyond $k_r = 1.25$, the point $(\theta_0 = \pi/2, \phi_0 = - \pi/2)$ becomes chaotic. The quantum dynamics of the corresponding state splits into three regions, as shown in Fig.~\ref{y-1}(b). The first region, $0\leq k_r < 1$, has an extremely low value of the long-time averaged entanglement. The second region starts at $k_r = 1$ onwards until it reaches a value that depends on the total spin-$j$. At this value of $k_r$, a transition occurs, and the long-time averaged entanglement quickly saturates to maximum. For $j = 25.5$, $j = 50.5$, and $j = 200.5$, the transition point occurs at $k_r = 1.55$, $k_r = 1.475$, and $k_r = 1.325$, respectively. Beyond $j = 200.5$, the shift in the transition point becomes extremely slow. This is evident from the overlap between the $j = 200.5$ and $j = 500.5$ plots. In the limit $j\to\infty$, the transition point gradually approaches its classical counterpart. Before transitioning to maximum, the long-time averaged entanglement grows linearly in the second region for the state $|\theta_0 = \pi/2, \phi_0 = - \pi/2 \rangle$. This indicates the formation of a partially stable state.

To further analyse the stability of the state $|\theta_0 = \pi/2, \phi_0 = - \pi/2 \rangle$, we compute the long-time averaged fidelity (see Fig.~\ref{y-1}(c)). The fidelity $F(\rho, \sigma) = \tr (\sqrt{\sqrt{\rho} \sigma \sqrt{\rho}})$ quantifies the closeness between two density matrices $\rho$ and $\sigma$ \citep{nielsen00}, and is used to measure how closely the evolved state resembles the initial state. The long-time averaged fidelity is given by
\begin{align}
    \langle F \rangle = \frac{1}{T}\sum_{t=0}^{T-1} F(\rho_1(0), \rho_1(t)).
\end{align}
Here,
\begin{align}
    \rho_1(0) &= \tr_{\neq 1}\left(|\psi_0\rangle \langle \psi_0|\right), \\
     \rho_1(t) &= \tr_{\neq 1}\left(\mathcal{U}^t|\psi_0\rangle \langle \psi_0|{\mathcal{U}^\dagger}^t\right),
\end{align}
and $|\psi_0 \rangle = |\theta_0,\phi_0\rangle$ is defined in the Eq.~(\ref{Eq:generalstate}). It is comparatively low in chaotic regions and high in regular ones, thus providing insight into the stability of the corresponding state. Similar to the long-time averaged von Neumann entropy, the fidelity splits into three distinct regions. For $k_r < 1$, the fidelity is close to 1, indicating the stable nature of the initial state. A sharp drop occurs at near $k_r = 1$, indicating decrease in the stability. The distinction between the second and third region becomes more pronounced for the larger values of $j$. The emergence of a plateau for $j \geq 200.5$ signals the presence of a \textit{metastable} state (see Fig.~\ref{y-1}(c)). This plateau narrows as $j$ increases. The sharp decline in fidelity at the onset of the third region marks a sudden rise in instability.

The classical dynamics of the initial point $(\theta_0 = \pi/2, \phi_0 = \pm \pi/2)$ also divides into three distinct parts. The homoclinic point appears before the onset of chaos. Correspondingly, the quantum dynamics exhibit three regions: low entanglement matches regular behaviour, while entanglement saturation signals chaos. Interestingly, the transition points between these regions depend on $j$. In the limit $j \to \infty$, these quantum transition points converge to their classical counterparts. This convergence shows that partial stability (homoclinic point) in classical dynamics emerges as metastability in the semi-classical dynamics. This metastability is characterised by the linear growth in average entanglement (see Fig.~\ref{y-1}(b)) and the emergence of plateau in average fidelity (see Fig.~\ref{y-1}(c)).

\begin{figure}
    \includegraphics[width=\linewidth]{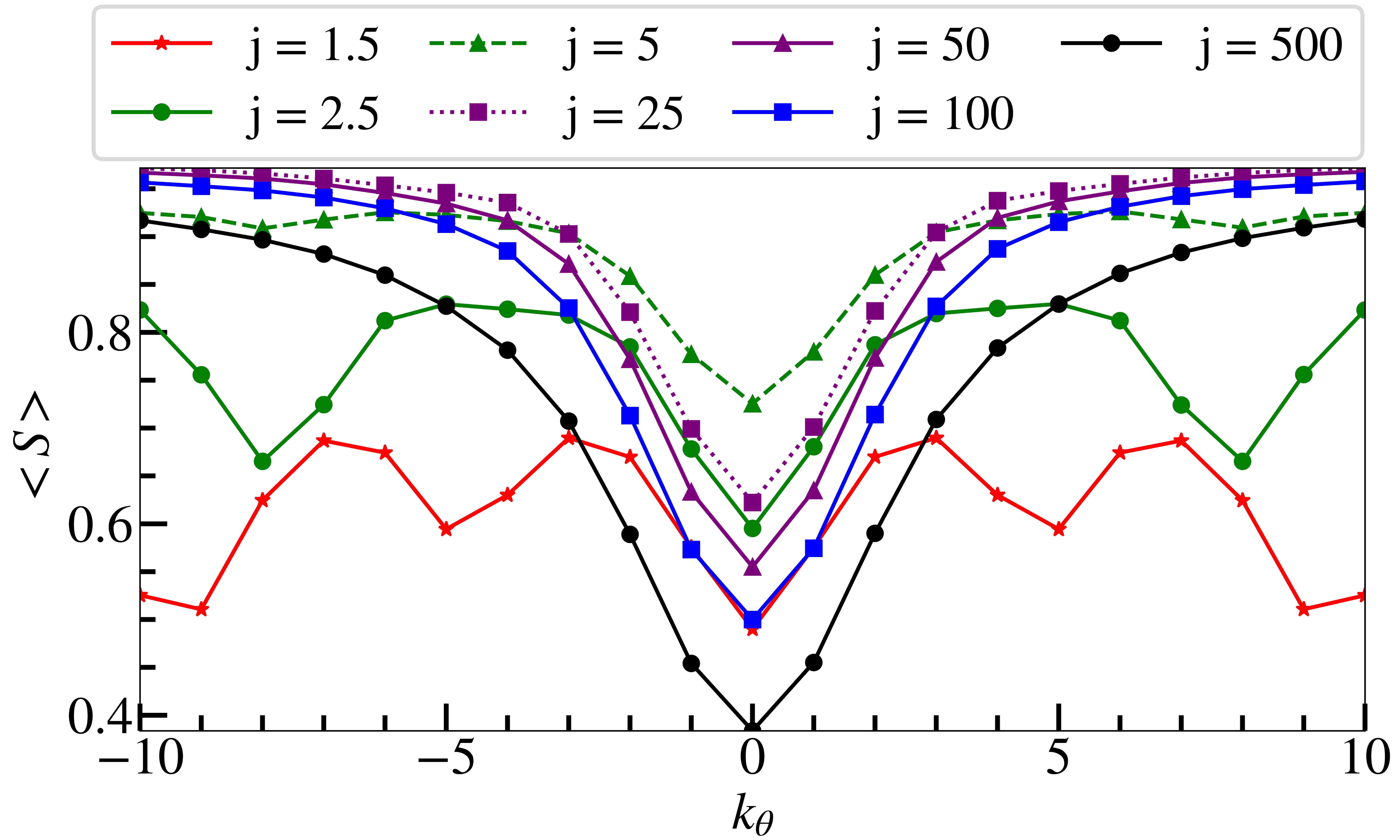}
    \caption{Long-time-averaged von Neumann entropy for the single-qubit RDM $\rho_1(n)$ and a grid of $200\times 200$ initial coherent states. Here, $n=1000$ and $k_r = 1$.}\label{fig14_3}
\end{figure}

As noted earlier, the role of $k_r$ is analogous to that of $k$ in the standard QKT. In contrast, $k_\theta$ is a unique feature of our model. The long-time averaged entanglement shows excellent agreement with the classical phase-space structures for $k_\theta < k_r$ (see Figs.~\ref{ppktheta}, \ref{fig4} and \ref{fig14}). In contrast, when $k_\theta > k_r$, regions of high long-time averaged entanglement grow rapidly with $k_\theta$ (see Fig.~\ref{fig14_2}) compared to its classical counterpart (see Figs.~\ref{ppkthetaTwist}, \ref{ktTwist} and \ref{fig5}(b)).

Building on these observations, we analyse how long-time averaged entanglement depends on $j$ and $k_\theta$. It can be observed that $\langle S_{j = 500}(k_\theta)\rangle < \langle S_{j = 100}(k_\theta)\rangle < \langle S_{j = 50}(k_\theta)\rangle < \langle S_{j = 25}(k_\theta)\rangle$, as shown in Fig.~\ref{fig14_3}. For these cases, the $\langle S_{j > 25}(k_\theta = 0)\rangle$ is minimum and grows rapidly with $|k_\theta |$. This behaviour arises due to the stretching caused by $k_\theta$ that shrinks the width of fixed orbits. Consequently, states associated with these fixed orbits get higher long-time averaged entanglement with more twist $k_\theta$ due to the spread of wave packets. The deep quantum regime deviates from the above trend, specifically, for large values of $k_\theta$.

Recall that the two low-valued regions of the long-time averaged entanglement further split into two, as $k_\theta$ increases from zero to one in the deep quantum regime (see Figs.~\ref{j1}, \ref{fig8}, and \ref{vn4qubit}). An excellent agreement is observed between low-spin and high-spin systems for the case $(k_r = 1, k_\theta = 0)$, as shown in Figs.~\ref{j1}(a), \ref{fig8}(a), \ref{vn4qubit}(a), and \ref{fig14}(a). However, for $k_\theta = k_r = 1$, the distinction between two low-valued subregions of the long-time averaged entanglement gets less pronounced with a increase in $j$ (see Figs.~\ref{j1}(d), \ref{fig8}(d), \ref{vn4qubit}(d), and \ref{fig14}(d)).

\begin{figure}[h!]
    \includegraphics[width=\linewidth]{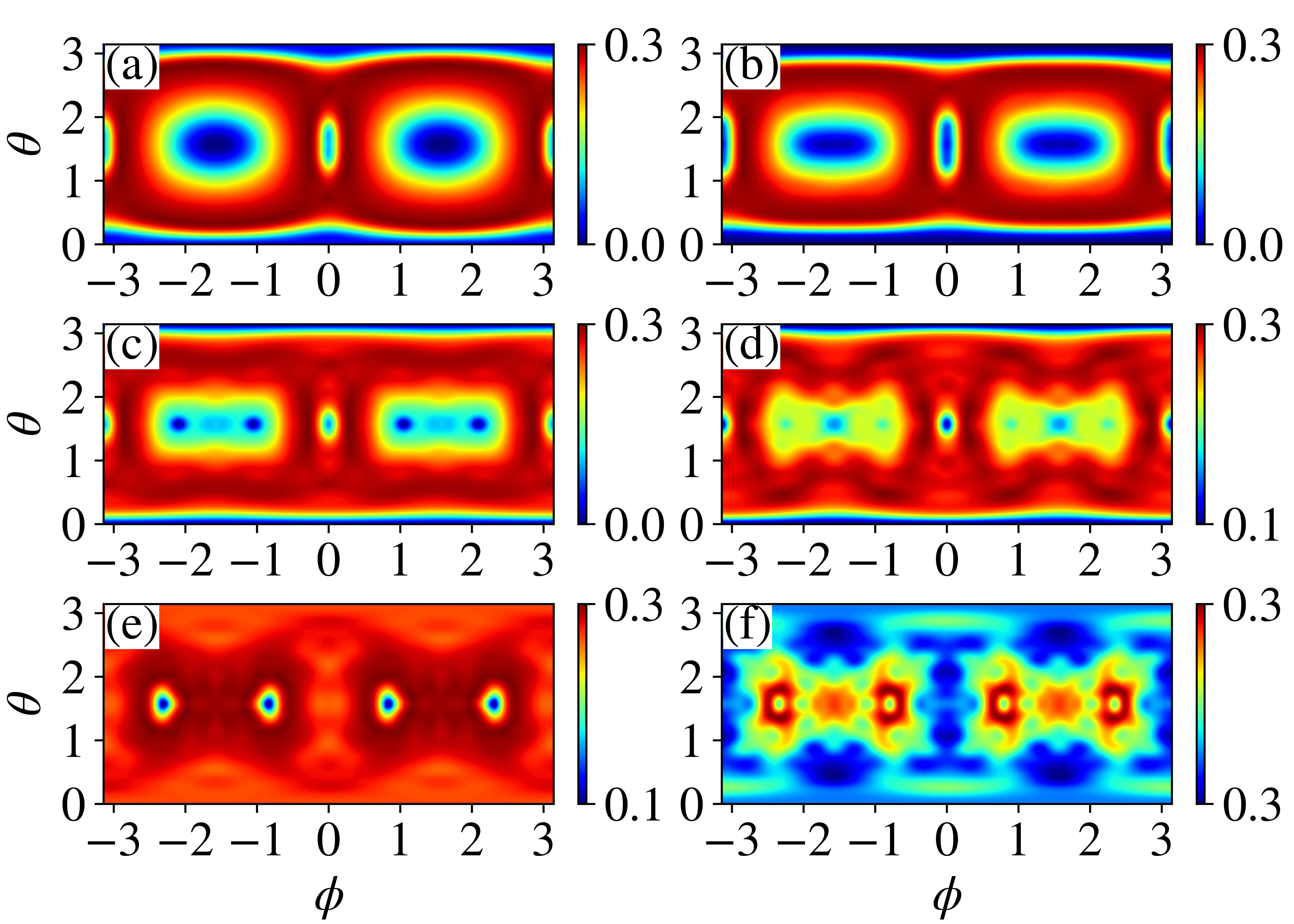}
    \caption{Long-time-averaged quantum discord for a 2-qubit RDM $\rho_{12}(n)$, $n=1000$ and a grid of $200\times 200$ initial coherent states. Here, $j = 50.5$, $k_\theta = 0$, (a) $k_r = 0.75$, (b) $k_r = 1$, (c) $k_r = 1.25$, (d) $k_r = 1.5$, (e) $k_r = 1.75$ and (f) $k_r = 2$.}\label{fig15}
\end{figure}
\begin{figure}[h!]
    \includegraphics[width=\linewidth]{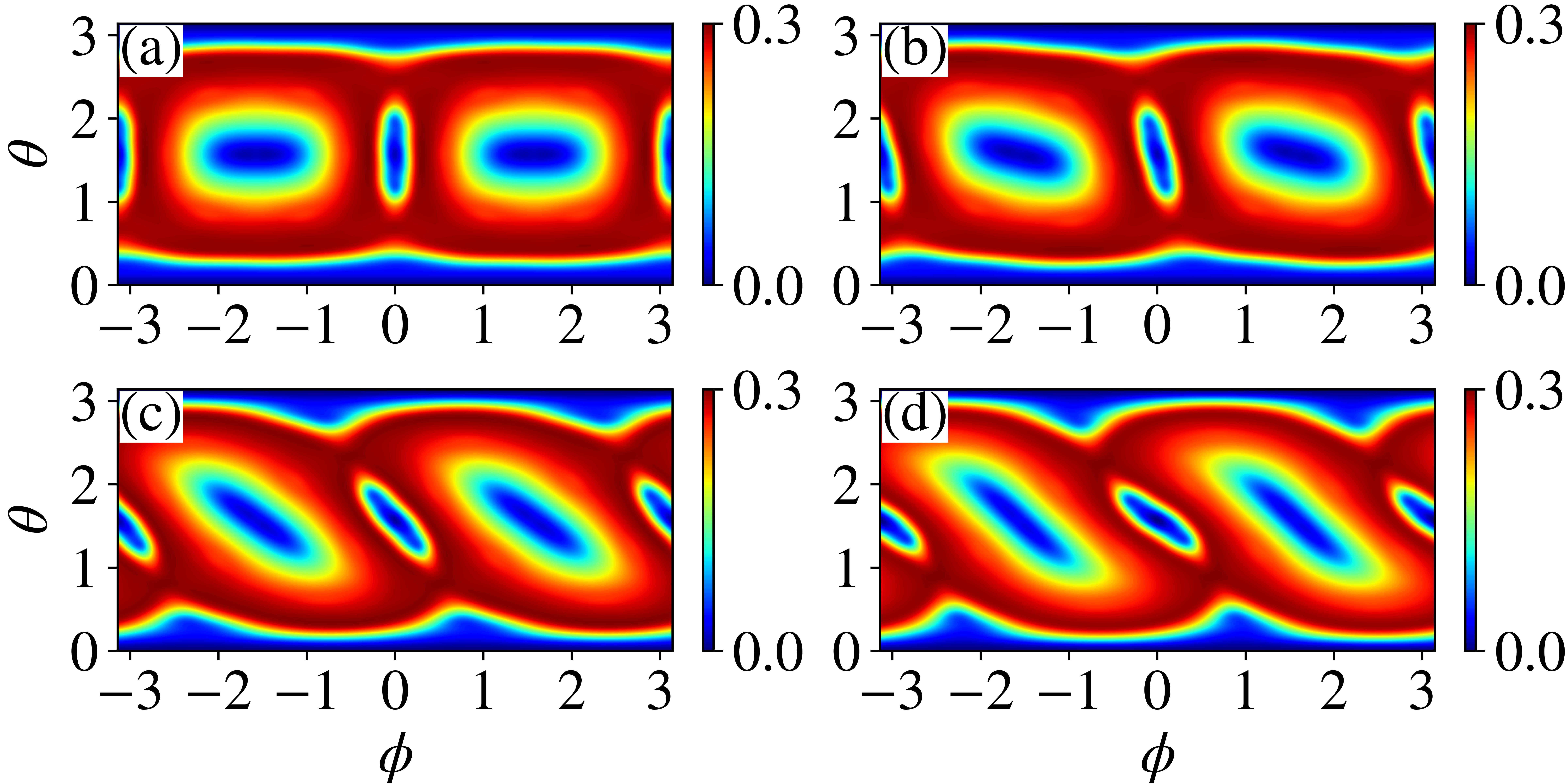}
    \caption{Long-time-averaged quantum discord for 2-qubit RDM $\rho_{12}(n)$, $n=1000$ and a grid of $200\times 200$ initial coherent states. Here, $j = 75.5$, $k_r = 1$, (a) $k_\theta = 0$, (b) $k_\theta = 0.25$, (c) $k_\theta = 0.75$ and (d) $k_\theta = 1$.}\label{fig16}
\end{figure}
\begin{figure}[h!]
    \includegraphics[width=\linewidth]{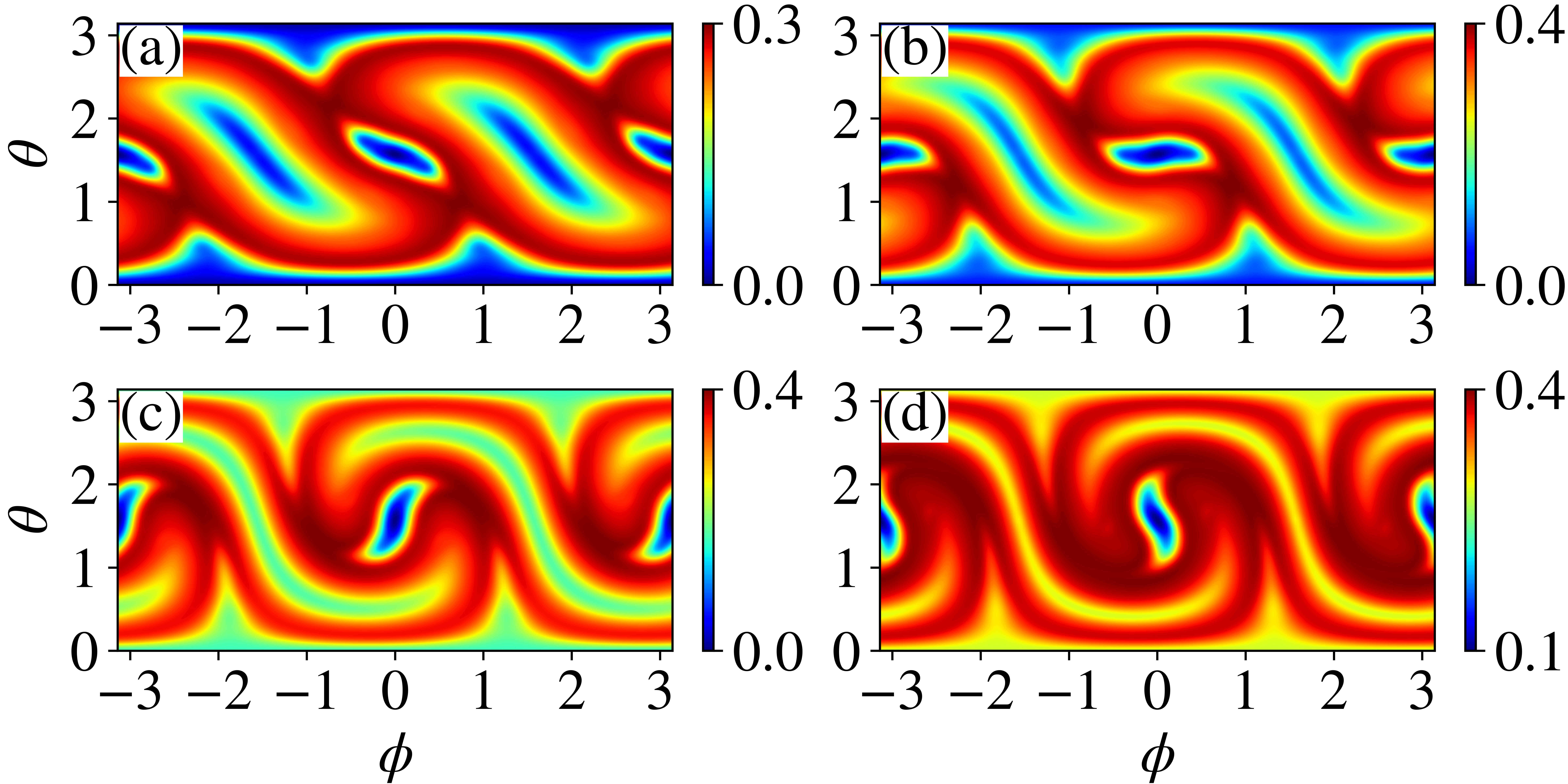}
    \caption{Long-time-averaged quantum discord for 2-qubit RDM $\rho_{12}(n)$, $n=1000$ and a grid of $200\times 200$ initial coherent states. Here, $j = 75.5$, $k_r = 1$, (a) $k_\theta = 1.25$, (b) $k_\theta = 1.75$, (c) $k_\theta = 3.0$ and (d) $k_\theta = 3.75$.}\label{fig16_2}
\end{figure}
\subsection{Quantum discord}
In this subsection, we will discuss the long-time averaged quantum discord as a function of $k_r$ and $k_\theta$ in the high-spin regime. Then, these results are compared with the corresponding long-time averaged von Neumann entropy and KSE. 

The long-time averaged quantum discord computed in Figs.~\ref{fig15}(a), \ref{fig15}(b), \ref{fig15}(c), \ref{fig15}(e), \ref{fig16}, and \ref{fig16_2} show excellent agreement with that of the long-time averaged von Neumann entropy computed in Figs.~\ref{fig13}(a), \ref{fig13}(b), \ref{fig13}(c), \ref{fig13}(e), \ref{fig14}, and \ref{fig14_2} respectively. The low value of the long-time averaged quantum discord (see \ref{fig15}(d)) for states $|\theta_0 = \pi/2, \phi_0 = \pm \pi/2\rangle$ further supports results of the von Neumann entropy (see Fig.~\ref{fig13}(d)). However, states associated with bifurcated orbits have relatively higher long-time averaged quantum discord, as shown in Fig.~\ref{fig15}(d). Soon for $k_r = 2$, the long-time averages of the quantum discord for all initial coherent states get saturated to maximum (see \ref{fig15}(f)). Since the quantum discord may qualitatively differ with the KSE, it can be seen as a slightly weaker signature of chaos.

\subsection{Concurrence}
In this subsection, we will discuss the long-time averaged concurrence as a function of $k_r$ and $k_\theta$ in the high-spin regime. Then, these results are compared with the corresponding long-time averaged von Neumann entropy, quantum discord and KSE. We also compare results obtained with the long-time averaged concurrence in the deep quantum regime.

\begin{figure}[h!]
    \includegraphics[width=\linewidth]{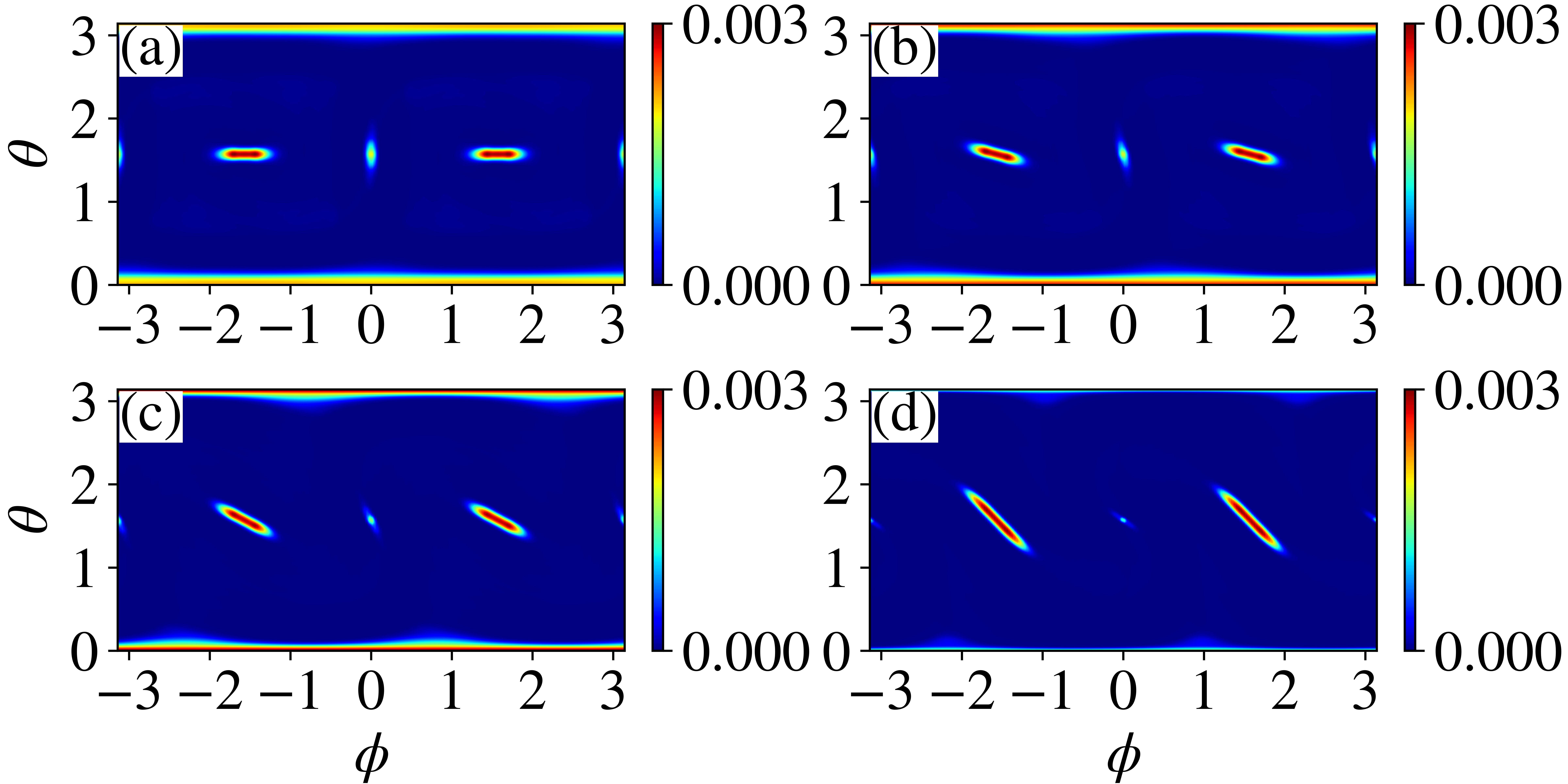}
    \caption{Long-time-averaged concurrence for 2-qubit RDM $\rho_{12}(n)$, $n=1000$ and a grid of $200\times 200$ initial coherent states. Here, $j= 75.5$, $k_r = 1$, (a) $k_\theta = 0$, (b) $k_\theta = 0.25$, (c) $k_\theta = 0.75$ and (d) $k_\theta = 1$.}\label{fig17}
\end{figure}

\begin{figure}[h!]
    \includegraphics[width=\linewidth]{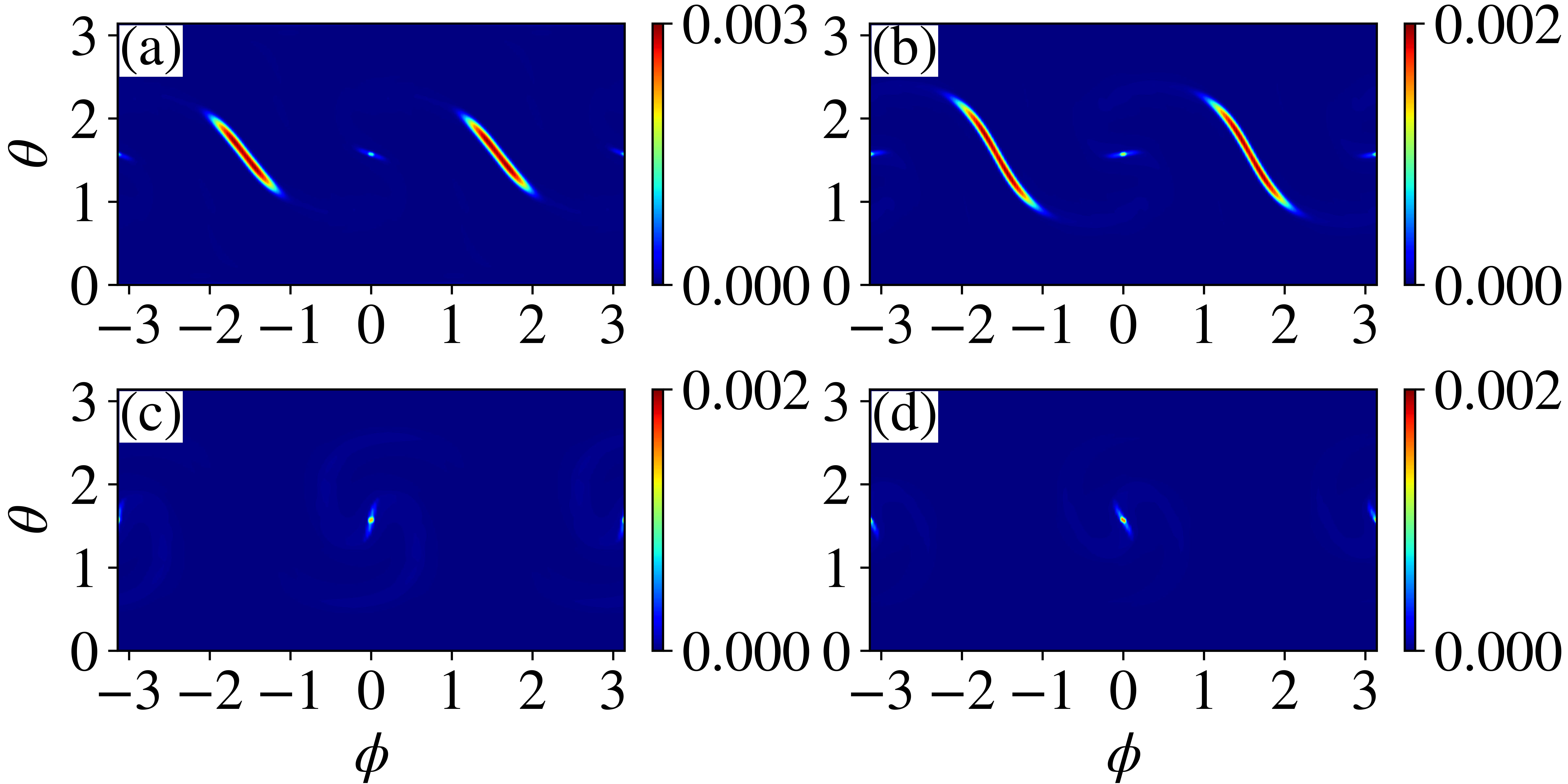}
    \caption{Long-time-averaged concurrence for 2-qubit RDM $\rho_{12}(n)$, $n=1000$ and a grid of $200\times 200$ initial coherent states. Here, $j= 75.5$, $k_r = 1$, (a) $k_\theta = 1.25$, (b) $k_\theta = 1.75$, (c) $k_\theta = 3.0$ and (d) $k_\theta = 3.75$.}\label{fig18}
\end{figure}

It can be seen from the Figs.~\ref{fig10} and \ref{fig17} that the long-time averaged concurrence decays with $j$, indicating multipartite nature of the entanglement. States associated with the trivial fixed points have relatively higher values compared to that for other states (see Fig.~\ref{fig17}). For $k_\theta \leq k_r$, the rate of decay for the long-time averaged concurrence is greater for states $|\theta_0 = \pi/2, \phi_0 = 0 \rangle$ and $|\theta_0 = 0, \phi_0 = 0 \rangle$ than that for states $|\theta_0 = \pi/2, \phi_0 = \pm \pi/2 \rangle$ with increase in $k_\theta $ (see Fig.~\ref{fig17}). The converse is true for $k_\theta > k_r$ (see Fig.~\ref{fig18}). Concurrence, therefore, complements with the results of von Neumann entropy in the high-spin regime and show partial deviations in the deep quantum regime.

\section{Results}\label{sec:r}
The classical dynamics of DKT presents a rich phase-space structure, with one part governed by $k_r$ being \textit{equivalent} to the kicking strength of the standard kicked top. The other part, governed by $k_\theta$, is a unique feature of our model. It allows us to study the dynamics in a system with broken time-reversal symmetry. It can be further divided into two parts. For $k_\theta < k_r$, the phase-space structure \textit{rotates} with the same angular speed around the trivial fixed points (see Fig.~\ref{ppktheta}). Whereas for $k_\theta > k_r$, however, the phase-space structure \textit{rotates} at an angular speed that depends on its distance from the trivial fixed points (see Fig.~\ref{ppkthetaTwist}). Although this qualitative change occurs at the time-reversal symmetric case $k_\theta = \pm k_r$, the KSE keeps increasing without any noteworthy change at these points. We also obtained the time-reversal operator for the case $k_\theta = 0$.

The paper presents exact solutions for 2- to 4-qubit systems and provides analytical expressions for the Floquet operator, its eigenvalues and the entanglement dynamics. 

We give a criterion for the periodicity of entanglement in systems with two to four qubits. Although the infinite-time averaged entanglement depends on $k_\theta$, the periodicity of entanglement dynamics does not. We numerically verify that the periodicity of entanglement dynamics is independent of the initial coherent state. Nevertheless, the \textit{period} depends on the initial state. Since the periodicity depends only on $k_r$, it remains unaffected by the time-reversal symmetry.

In the deep quantum regime, the long-time averaged quantum correlations produce coarse-grained phase-space structures. Here, the von Neumann entropy shows good agreement with the quantum discord and shows partial agreement with the concurrence. It also shows good agreement with the LLE and KSE. In this regime, our study reveals an interesting phenomenon that does not occur in the corresponding classical counterpart. The regions with low values of the long-time averaged entanglement further split into two subregions as $k_\theta$ increases from zero onwards.

In the semi-classical regime, for $k_\theta < k_r$, the long-time averaged von Neumann entropy shows excellent agreement with the LLE and KSE. For $k_\theta > k_r$, however, the results depend strongly on the values of $j$. For extremely high values of $j$, the long-time averaged von Neumann entropy shows good agreement but deviates as $j$ decreases. This happens because of quantum effects associated with the twisting of classical phase-space structures. In contrast to the long-time averaged von Neumann entropy, the long-time averaged quantum discord and the long-time averaged concurrence show slightly weaker agreement with the KSE. When averaged over all initial states, the long-time averaged entanglement is minimum for $k_\theta = 0$. This result aligns with that of the corresponding KSE. The quantum correlations give signatures of bifurcations for states associated with the homoclinic point. The total spin $j$ dependent metastability is observed in the semi-classical regime corresponding to the (partially stable) homoclinic point. 

We show that if the standard kicked top (DKT with $k' = 0$) is in the chaotic regime for a given kick strength $k$, then chaos can be minimised by setting $k' = -k$, that is $k_r = 0$. In the corresponding QKT, the aperiodic dynamics ($k \neq j \pi/2$) can be tuned to be periodic by setting $k'$ such that $k_r = (k+k')/2 = j\pi/2$. This, in turn, minimises the average entanglement of the quantum system.

\section{Discussion}\label{sec:d}
The time-reversal symmetric cases $k_\theta = 0$ and $k_\theta = k_r$, both show linear level repulsion. For a particular $k_r$, both these cases have similar structures related by a \textit{rotation} around the trivial fixed points. The role of $k_r$ in the DKT is similar to that of the kick strength $k$ in the standard kicked top. The $k_\theta$ increases chaos without introducing bifurcations, which is an interesting feature of our model. 

Our analysis shows that the case $k_\theta = k_r$ for which there exists a time-reversal symmetry in the quantum dynamics does show qualitative changes in the corresponding classical system. However, these observed changes are only qualitative and could be quantified in the future. These results provide deeper insights into the quantum-classical correspondence.

In the semi-classical limit, we observe that each of the two low-entropy regions corresponding to the trivial fixed points further splits into two as $k_\theta$ increases from zero onwards. This split is prominent in the deep quantum regime. This bifurcation-like phenomenon is absent in the corresponding classical dynamics. The analysis of the state $|\theta_0=\pi/2,\phi_0=-\pi/2\rangle$ reveals the correspondence between partially stable fixed point in the classical dynamics and a metastable state in the associated semi-classical regime. This metastability becomes increasingly pronounced with larger values of $j$. The value of the kick strength $k_r$ at which this metastability breaks down decreases as $j$ increases.

Finally, we give a mechanism to minimise classical and quantum chaos in the standard kicked top systems by introducing an appropriate second kick $k'$, applied perpendicular to both: the free precession and the first kick $k$. Specifically, any aperiodic quantum dynamics in QKT (DKT with $k' = 0$) can be made periodic and thus can minimise the entanglement generation by using a suitable choice of the kick strength $k'$. Following the broad range of applications and experimental realisations of the QKT \citep{neil2016ergodic,krithika2022quantum,krithika2023nmr,chaudhary_quantum_signatures,Meier2019}, our model can be used to control chaos in the classical systems and entanglement growth in the quantum system. 
Moreover, the chaos resulting from $k_\theta$ can be observed experimentally.

\section{Acknowledgments}
The authors are grateful to the Department of Science and Technology (DST) for their generous financial support, making this research possible through sanctioned Project No. SR/FST/PSI/2017/5(C) to the Department of Physics of VNIT, Nagpur.

\bibliographystyle{apsrev4-2}
\bibliography{ref_main,ref_supp}


\clearpage

\makeatletter
\let\addcontentsline\addcontentslineOriginal
\makeatother

\renewcommand{\thesection}{S\Roman{section}}
\renewcommand{\thesubsection}{S\Roman{section}.\Alph{subsection}}
\renewcommand{\theequation}{S\arabic{equation}}
\renewcommand{\thefigure}{S\arabic{figure}}
\renewcommand{\thetable}{S\arabic{table}}

\setcounter{section}{0}
\setcounter{subsection}{0}
\setcounter{equation}{0}
\setcounter{figure}{0}
\setcounter{table}{0}


\newcommand{\suppsection}[2][]{%
  \refstepcounter{section}%
  \phantomsection%
  \section*{\texorpdfstring{S\Roman{section}\quad #2}{S\Roman{section} #2}}%
  \addcontentsline{toc}{section}{S\Roman{section}\quad #2}%
  \ifstrempty{#1}{}{%
    \label{#1}%
  }%
}

\newcommand{\suppsubsection}[2][]{%
  \refstepcounter{subsection}%
  \phantomsection%
  \subsection*{\texorpdfstring{S\Roman{section}.\Alph{subsection}\quad #2}{S\Roman{section}.\Alph{subsection} #2}}%
  \addcontentsline{toc}{subsection}{S\Roman{section}.\Alph{subsection}\quad #2}%
  \ifstrempty{#1}{}{%
    \label{#1}%
  }%
}

\clearpage
\onecolumngrid

\begin{center}
\textbf{\large Supplementary Material for\\ 
``\textit{The double kicked top: classical, quantum and semiclassical analysis}''}
\end{center}

\tableofcontents

\suppsection[]{Introduction}
This supplementary material provides detailed analytical calculations of the classical map and its tangent map (see Secs. \ref{supp:sec:CM} and \ref{supp:sec:tangenet}) for our model from the main text. In Sec. \ref{supp:sec:timeOperator}, we derive non-conventional time-reversal operators for the cases discussed in the main text. Then we study the entanglement dynamics in the deep quantum regime. To be precise, we derive expressions for the eigenvalues, eigenvectors, and the time-evolved linear entropy for an arbitrary initial spin-coherent state for 2-, 3- and 4-qubit systems (see Secs. \ref{supp:sec:2qubit}, \ref{supp:sec:3qubit} and \ref{supp:sec:4qubit}). In addition, we obtain an infinite-time average linear entropy for any general initial spin-coherent state $|\theta_0, \phi_0\rangle$ for the case of 2-qubits. Whereas, in the case of 3- and 4-qubits, we obtain an infinite-time average linear entropy for two initial states $|\theta_0 = 0, \phi_0 = 0 \rangle$ and $|\theta_0 = \pi/2, \phi_0 = -\pi/2 \rangle$.

\suppsection[supp:sec:CM]{Classical Map}
In this section, we first derive the following transformations:
\begin{equation}
    J_l' = \mathcal{U}^\dagger J_l \mathcal{U},
\end{equation}
and then obtain the classical map by taking limit $j\to\infty$. Here, $l=x,y,z$ and the Floquet operator is given by
\begin{equation}\label{Uop}
    \mathcal{U} = \exp\left(- i \frac{k'}{2j}J_x^2\right)\exp\left(- i \frac{k}{2j}J_z^2\right)\exp\left(- i p J_y\right).
\end{equation}
The derivation is organized into four subsections: (A) transformations of $J_l$ around $x$-axis, (B) transformations of $J_l$ around $z$-axis, (C) transformations of $J_l$ around $y$-axis, and finally (D) the full derivation combining these transformations.

\suppsubsection[A]{Transformation around \textit{x}-axis}
The transformation induced by the operator $\exp(i \frac{k'}{2j}J_x^2)$ is evaluated as follows. Since $\exp(i \frac{k'}{2j}J_x^2)$ commutes with $J_x$, we have:
\begin{equation}\label{xx}
    \exp(i \frac{k'}{2j}J_x^2) J_x \exp(-i \frac{k'}{2j}J_x^2)  = J_x.
\end{equation}
To transform $J_y$, we use the Baker-Campbell-Hausdorff formula \citep{sakurai1994modern}:
\begin{equation}
    \exp(i \frac{k'}{2j}J_x^2) J_y \exp(-i \frac{k'}{2j}J_x^2) = \sum_{n=0}^{\infty} \frac{1}{n!} \left(i\frac{k'}{2j}\right)^n \big[ J_x^2, J_y \big]_n,
\end{equation}
where $\big[A, B\big]_n$ represents the nested commutator defined recursively as $\big[A, B\big]_0 = B$ and $\big[A, B\big]_n = \big[A, \big[A, B\big]_{n-1}\big]$. Expanding the first few terms, we find:
\begin{align}
    \big[J_x^2, J_y\big]_0 =& J_y \\
    \big[J_x^2, J_y\big]_1 =& \big[J_x^2, J_y\big] = J_x \big[J_x, J_y\big] + \big[J_x, J_y\big] J_x = 2i J_z J_x + J_y, \\
    \big[J_x^2, J_y\big]_2 =& \big[J_x^2, \big[J_x^2, J_y\big]\big] = 2^2 J_y J_x^2 + 4i J_z J_x + J_y, \\
    \big[J_x^2, J_y\big]_3 =& \big[J_x^2, \big[J_x^2, \big[J_x^2, J_y\big]\big]\big] = 2^3i J_z J_x^3 + 12 J_y J_x^2 -6i J_z J_x + J_y, \\ 
    \big[J_x^2, J_y\big]_4 =& \big[J_x^2,\big[J_x^2, \big[J_x^2, \big[J_x^2, J_y\big]\big]\big]\big] = 2^4 J_y J_x^4 + 30i J_z J_x^3 + 24 J_y J_x^2 + 8i J_z J_x + J_y, \\
    \big[J_x^2, J_y\big]_5 =& \big[J_x^2,\big[J_x^2,\big[J_x^2, \big[J_x^2, \big[J_x^2, J_y\big]\big]\big]\big]\big] = 2^5 i J_z J_x^5 + 76 J_y J_x^4 + 78 i J_z J_x^3 + 40 J_y J_x^2 + 10 i J_z J_x + J_y.
\end{align}
While applying commutation relations, we set $\hbar = 1$. Combining and re-arranging all these terms, we get
\begin{equation}
    e^{i \frac{k'}{2j}J_x^2} J_y e^{-i \frac{k'}{2j}J_x^2}  = J_y \left(1 - \frac{{\left( \frac{k'}{j} \right)}^2}{2!}J_x^2 + \frac{{\left( \frac{k'}{j} \right)}^4}{4!}J_x^4 \ldots\right) - J_z \left(\frac{k'}{j} J_x - \frac{{\left(\frac{k'}{j}\right)}^3}{3!}J_x^3 + \frac{{\left( \frac{k'}{j} \right)}^5}{5!}J_x^5 + \ldots\right).
\end{equation}
In the large $j$ limit, we get the following equation:
\begin{equation}\label{yx}
    e^{i \frac{k'}{2j}J_x^2} J_y e^{-i \frac{k'}{2j}J_x^2} = J_y \cos(\frac{k'}{j}J_x) - J_z \sin(\frac{k'}{j}J_x).
\end{equation}
Similarly, for $J_z$, we have
\begin{equation}\label{zx}
    e^{i \frac{k'}{2j}J_x^2} J_z e^{-i \frac{k'}{2j}J_x^2} = J_z \cos(\frac{k'}{j}J_x) + J_y \sin(\frac{k'}{j}J_x).
\end{equation}

\suppsubsection[B]{Transformation around \textit{z}-axis}
In this subsection, we analyze the transformation around the \textit{z}-axis using the \(|j, m\rangle\) basis. The approach involves evaluating the transformation properties of the creation operator \(J_+\) and annihilation operator \(J_-\). These results are subsequently utilized to determine the transformations of the spin components \(J_x\) and \(J_y\).

The transformation of the operator \(J_+\) is evaluated as follows:
\begin{equation}
    \langle j, m | e^{i \frac{k}{2j}J_z^2} J_+ e^{-i \frac{k}{2j}J_z^2}  | j, n  \rangle = e^{i \frac{k}{2j} \left(m^2 - n^2\right)} C_{j m, n+1 }\delta_{m, n+1},
\end{equation}
where the right-hand side is nonzero only if \(m = n+1\). Therefore, the expression simplifies to
\begin{equation}
    \langle j, m | e^{i \frac{k}{2j}J_z^2} J_+ e^{-i \frac{k}{2j}J_z^2} | j, n  \rangle = e^{i \frac{k}{j} \left(n +\frac{1}{2}\right)} C_{j n+1}.
\end{equation}
This implies the operator identity:
\begin{equation}
    e^{i \frac{k}{2j}J_z^2} J_+ e^{-i \frac{k}{2j}J_z^2} = J_+ e^{i \frac{k}{j} \left( J_z +\frac{1}{2}\right)}.
\end{equation}
Similarly, for the annihilation operator \(J_-\), we obtain:
\begin{align}
    \langle j, m | e^{i \frac{k}{2j}J_z^2} J_- e^{-i \frac{k}{2j}J_z^2} | j, n  \rangle = e^{i \frac{k}{j} \left( - n +\frac{1}{2}\right)} C_{j n-1} \implies e^{i \frac{k}{2j}J_z^2} J_- e^{-i \frac{k}{2j}J_z^2} = J_- e^{i \frac{k}{j} \left( - J_z +\frac{1}{2}\right)}.
\end{align}
Using the standard definitions of the spin components, \(J_x \coloneqq \frac{1}{2}\left(J_+ + J_-\right)\) and \(J_y \coloneqq \frac{1}{2i}\left(J_+ - J_-\right)\), we derive their transformations:
\begin{align}
    e^{i \frac{k}{2j}J_z^2} J_x e^{-i \frac{k}{2j}J_z^2} =& \frac{1}{2}\left[ J_+ e^{i \frac{k}{j} \left( J_z +\frac{1}{2}\right) } +  J_- e^{ -i \frac{k}{j} \left( J_z - \frac{1}{2}\right) }\right], \\
    e^{i \frac{k}{2j}J_z^2} J_y e^{-i \frac{k}{2j}J_z^2} =& \frac{1}{2i}\left[ J_+ e^{i \frac{k}{j} \left( J_z +\frac{1}{2}\right) } -  J_- e^{ -i \frac{k}{j} \left( J_z - \frac{1}{2}\right) }\right].
\end{align}
In the large \(j\)-limit, the transformations simplify to:
\begin{align}
    e^{i \frac{k}{2j}J_z^2} J_x e^{-i \frac{k}{2j}J_z^2} =& J_x \cos\left(\frac{k}{j} J_z\right) - J_y \sin\left(\frac{k}{j} J_z\right), \label{xz} \\
    e^{i \frac{k}{2j}J_z^2} J_y e^{-i \frac{k}{2j}J_z^2} =& J_y \cos\left(\frac{k}{j} J_z\right) + J_x \sin\left(\frac{k}{j} J_z\right), \label{yz}\\
    e^{i \frac{k}{2j}J_z^2} J_z e^{-i \frac{k}{2j}J_z^2} =& J_z. \label{zz}
\end{align}

\suppsubsection[C]{Transformation around \textit{y}-axis}
In this subsection, we present the derivation of the transformations of the angular momentum operators $J_x$, $J_y$, and $J_z$ under a rotation around the \textit{y}-axis. Using the BCH formula, the transformed operators are obtained as follows. The transformation of $J_x$ under a rotation by an angle $p$ around the \textit{y}-axis is expressed as:
\begin{equation}
    e^{i p J_y} J_x e^{-i p J_y}.
\end{equation}
Using the BCH formula, this can be expanded as:
\begin{align}
    e^{i p J_y} J_x e^{-i p J_y} &= \sum_{n=0}^{\infty} \frac{(i p)^n}{n!} \underbrace{[J_y, J_x]_n}_{\text{nested commutators}} \nonumber \\
    &= J_x + (i p) [J_y, J_x] + \frac{(i p)^2}{2!} [J_y, [J_y, J_x]] + \frac{(i p)^3}{3!} [J_y, [J_y, [J_y, J_x]]] + \dots. \nonumber
\end{align}
Evaluating the commutators using the angular momentum algebra \([J_y, J_x] = i J_z\) and \([J_y, J_z] = -i J_x\), we find:
\begin{align}
    e^{i p J_y} J_x e^{-i p J_y} &= J_x + p J_z - \frac{p^2}{2!} J_x - \frac{p^3}{3!} J_z + \dots \nonumber \\
    &= J_x \cos(p) + J_z \sin(p). \label{eq:Jx_transformation}
\end{align}
The operator $J_y$ commutes with the unitary operator $e^{i p J_y}$, since $[J_y, J_y] = 0$. Thus, the transformation of $J_y$ is trivial:
\begin{equation}\label{eq:Jy_transformation}
    e^{i p J_y} J_y e^{-i p J_y} = J_y.
\end{equation}
The transformation of $J_z$ under a rotation around the \textit{y}-axis is given by:
\begin{equation}
    e^{i p J_y} J_z e^{-i p J_y} = \sum_{n=0}^{\infty} \frac{(i p)^n}{n!} [J_y, J_z]_n.
\end{equation}
Expanding this transformation using the BCH formula:
\begin{align}
    e^{i p J_y} J_z e^{-i p J_y} &= J_z + (i p) [J_y, J_z] + \frac{(i p)^2}{2!} [J_y, [J_y, J_z]] + \frac{(i p)^3}{3!} [J_y, [J_y, [J_y, J_z]]] + \dots. \nonumber
\end{align}
Using the commutation relations, we find:
\begin{align}
    e^{i p J_y} J_z e^{-i p J_y} &= J_z - p J_x - \frac{p^2}{2!} J_z + \frac{p^3}{3!} J_x + \dots \nonumber \\
    &= J_z \cos(p) - J_x \sin(p). \label{eq:Jz_transformation}
\end{align}

\suppsubsection[D]{Derivation of the final classical map}
To facilitate the derivation, we first establish a general result for any polynomial $f(J_l)$ under the transformation
\begin{equation} \label{eq:general_transformation}
    e^{i g(J_m)} f(J_l) e^{-i g(J_m)} = f\left(e^{i g(J_m)} J_l e^{-i g(J_m)}\right),
\end{equation}
where $g(J_m)$ is any function of the angular momentum operators $\bm{J}$. To verify this, consider $f(J_x) = \sin\left(\frac{k'}{j}J_x\right)$. Using the Taylor series expansion of the sine function, we write:
\begin{equation}
    \sin\left(\frac{k'}{j}J_x\right) = \sum_{l=0}^\infty \frac{(-1)^l}{(2l+1)!} \left(\frac{k'}{j} J_x\right)^{2l+1}.
\end{equation}
Applying the transformation $e^{i g(J_m)} \cdot e^{-i g(J_m)}$ term by term yields:
\begin{align}
    e^{i g(J_m)} \sin\left(\frac{k'}{j}J_x\right) e^{-i g(J_m)} 
    &= \sum_{l=0}^\infty \frac{(-1)^l}{(2l+1)!} \left[ e^{i g(J_m)} \left(\frac{k'}{j} J_x\right)^{2l+1} e^{-i g(J_m)} \right] \\
    &= \sum_{l=0}^\infty \frac{(-1)^l}{(2l+1)!} \left[ \frac{k'}{j} \left(e^{i g(J_m)} J_x e^{-i g(J_m)}\right) \right]^{2l+1} \\
    &= \sin\left[\frac{k'}{j} \left(e^{i g(J_m)} J_x e^{-i g(J_m)}\right)\right].
\end{align}
The transformation of $J_x$ is given by:
\begin{align}\label{eq:jx_prime}
    J_x' &= \mathcal{U}^\dagger J_x \mathcal{U} \notag \\
    &= e^{i p J_y} \left[ e^{i \frac{k}{2j} J_z^2} \left(e^{i \frac{k'}{2j} J_x^2} J_x e^{-i \frac{k'}{2j} J_x^2}\right) e^{-i \frac{k}{2j} J_z^2} \right] e^{-i p J_y}.
\end{align}
Using Eq.~\eqref{xx}, the inner transformation becomes:
\begin{equation}
    e^{i \frac{k'}{2j} J_x^2} J_x e^{-i \frac{k'}{2j} J_x^2} = J_x.
\end{equation}
For the outer bracket, we use the established commutation relations and Eq.~\eqref{eq:general_transformation}. Applying $e^{i \frac{k}{2j} J_z^2}$ yields:
\begin{equation}
    e^{i \frac{k}{2j} J_z^2} J_x e^{-i \frac{k}{2j} J_z^2} = J_x \cos\left(\frac{k}{j} J_z\right) - J_y \sin\left(\frac{k}{j} J_z\right).
\end{equation}
Finally, applying the rotation $e^{i p J_y}$, we get
\begin{align}
    J_x' &= \left(e^{i p J_y} J_x e^{-i p J_y}\right) \cos\left[\frac{k}{j} \left(e^{i p J_y} J_z e^{-i p J_y}\right)\right] - J_y \sin\left[\frac{k}{j} \left(e^{i p J_y} J_z e^{-i p J_y}\right)\right].\label{eq:Jx_prime}
\end{align}
Similarly, the transformation equations for the variables $J_y$ and $J_z$, expressed as:
\begin{align}
    J_y' =& \left[J_y \cos\left(\frac{k}{j} J_x \right) - J_z \sin\left(\frac{k}{j} J_x \right)\right] 
    \cos\left\{\frac{k'}{j}\left[J_z \cos\left(\frac{k}{j} J_x \right) + J_y \sin\left(\frac{k}{j} J_x \right) \right]\right\} \notag \\
    &\quad + J_x \sin\left\{\frac{k'}{j}\left[J_z \cos\left(\frac{k}{j} J_x \right) + J_y \sin\left(\frac{k}{j} J_x \right) \right]\right\}, \label{eq:Jy_prime} \\
    J_z' =& - J_x \cos\left\{\frac{k'}{j}\left[J_z \cos\left(\frac{k}{j} J_x \right) + J_y \sin\left(\frac{k}{j} J_x \right) \right]\right\} \notag \\
    &\quad + \left[J_y \cos\left(\frac{k}{j} J_x \right) - J_z \sin\left(\frac{k}{j} J_x \right)\right] 
    \sin\left\{\frac{k'}{j}\left[J_z \cos\left(\frac{k}{j} J_x \right) + J_y \sin\left(\frac{k}{j} J_x \right) \right]\right\}. \label{eq:Jz_prime}
\end{align}
Now, we introduce the normalized variables $X = \dfrac{J_x}{j}, \quad Y = \dfrac{J_y}{j}, \quad Z = \dfrac{J_z}{j}$. In the limit $j \to \infty$, the transformations in Eqs.~\eqref{eq:Jx_prime}, \eqref{eq:Jy_prime} and \eqref{eq:Jz_prime} converge to a classical map for the variables $X$, $Y$, and $Z$. The resulting classical map is given by:
\begin{align}
    X' &= Z \cos(k X) + Y \sin(k X), \label{eq:X_prime} \\
    Y' &= \left[Y \cos(k X) - Z \sin(k X)\right] 
    \cos\left[k'\left(Z \cos(k X) + Y \sin(k X)\right)\right] + X \sin\left[k'\left(Z \cos(k X) + Y \sin(k X)\right)\right], \label{eq:Y_prime}\\
    Z' &= - X \cos\left[k'\left(Z \cos(k X) + Y \sin(k X)\right)\right] + \left[Y \cos(k X) - Z \sin(k X)\right] 
    \sin\left[k'\left(Z \cos(k X) + Y \sin(k X)\right)\right]. \label{eq:Z_prime}
\end{align}
The corresponding phase space is plotted for various values of $k$ and $k'$ in Figs. (1), (3) and (4) in the main text. The Eqns.~(\ref{eq:X_prime}), (\ref{eq:Y_prime}) and (\ref{eq:Z_prime}) are presented in the main text.

\suppsection[]{Derivation of the Tangent map}\label{supp:sec:tangenet}
In this section, we give the explicit form of the tangent map  for the classical map $\mathbf{F}$ evaluated at a point $\mathbf{X} \coloneqq \left(X, Y, Z\right)$. The tangent map $\mathbf{M}\left(\mathbf{X}\right) \coloneqq \frac{\partial \mathbf{F}}{\partial \mathbf{X}} $ is given as follows:
\begin{equation}\label{eq:matrixM}
    \mathbf{M}\left(\mathbf{X}_n\right) = 
    \begin{pmatrix}
        \mathbf{M}_{11} &\mathbf{M}_{12}  & \mathbf{M}_{13}  \\
        \mathbf{M}_{21} & \mathbf{M}_{22} & \mathbf{M}_{23} \\
        \mathbf{M}_{31} & \mathbf{M}_{32} & \mathbf{M}_{33}
    \end{pmatrix},
\end{equation}
where the elements $\mathbf{M}_{ij}$ are found explicitly as follows:

\begin{alignat}{2}
    \mathbf{M}_{11} =& k Y \cos(k X) - k Z \sin(k X), \\[10pt]
    \mathbf{M}_{12} =& \sin(k X), \\[10pt]
    \mathbf{M}_{13} =& \cos(k X), \\[10pt]
    \mathbf{M}_{21} =& -k k' \left[ Y \cos(k X) - Z \sin(k X) \right]^2 \sin\left[ k' Y \sin(k X) + k' Z \cos(k X) \right] \notag \\
    & \quad + k k' X \left[ Y \cos(k X) - Z \sin(k X) \right] \cos\left[ k' Y \sin(k X) + k' Z \cos(k X) \right]  \\
    & \quad - k \left[ Y \sin(k X) + Z \cos(k X) \right] \cos\left[ k' Y \sin(k X) + k' Z \cos(k X) \right]  + \sin\left[ k' Y \sin(k X) + k' Z \cos(k X) \right], \notag \\[10pt]
    \mathbf{M}_{22} = & \cos(k X) \cos\left[ k' Y \sin(k X) + k' Z \cos(k X) \right]  + k' X \sin(k X) \cos\left[ k' Y \sin(k X) + k' Z \cos(k X) \right] \notag \\
    & \quad + k' \sin(k X) \left[ Z \sin(k X) - Y \cos(k X) \right] \sin\left[ k' Y \sin(k X) + k' Z \cos(k X) \right], \\[10pt]
    \mathbf{M}_{23} = & k' X \cos(k X) \cos\left[ k' Y \sin(k X) + k' Z \cos(k X) \right] - \sin(k X) \cos\left[ k' Y \sin(k X) + k' Z \cos(k X) \right] \notag \\
    & \quad - k' \cos(k X) \left[ Y \cos(k X) - Z \sin(k X) \right] \sin\left[ k' Y \sin(k X) + k' Z \cos(k X) \right], \\[10pt]
    \mathbf{M}_{31} = & \cos\left[ k' Y \sin(k X) + k' Z \cos(k X) \right] \Big[ k k' Y^2 \cos^2(k X) - k k' Y Z \sin(2 k X) + k k' Z^2 \sin^2(k X) - 1 \Big] \notag \\
    & \quad + k \left[ \cos(k X) (k' X Y - Z) - \sin(k X) (k' X Z + Y) \right] \sin\left[ k' Y \sin(k X) + k' Z \cos(k X) \right], \\[10pt]
    \mathbf{M}_{32} = & k' \sin(k X) \left[ Y \cos(k X) - Z \sin(k X) \right] \cos\left[ k' Y \sin(k X) + k' Z \cos(k X) \right] \notag \\
    & \quad + \cos(k X) \sin\left[ k' Y \sin(k X) + k' Z \cos(k X) \right]  + k' X \sin(k X) \sin\left[ k' Y \sin(k X) + k' Z \cos(k X) \right], \\[10pt]
    \mathbf{M}_{33} = & k' \cos(k X) \left[ Y \cos(k X) - Z \sin(k X) \right] \cos\left[ k' Y \sin(k X) + k' Z \cos(k X) \right] \notag \\
    & \quad + k' X \cos(k X) \sin\left[ k' Y \sin(k X) + k' Z \cos(k X) \right] - \sin(k X) \sin\left[ k' Y \sin(k X) + k' Z \cos(k X) \right].
\end{alignat}
This tangent map is used to compute the largest Lyapunov exponent and the Kolmogorov-Sinai entropy in the main text.

\suppsection[]{Derivation of the time-reversal operator}\label{supp:sec:timeOperator}
In this section, we find the non-conventional time-reversal operator $T=U K$ for our model with $K$ being a conjugation operator. Here, we find unitary operator $U$, one for each of two cases: $k'=k\neq 0$ and $(k'=0, k\neq 0)$ or equivalently $(k=0,k'\neq 0)$. A model is said to be time-reversal symmetric if the floquet operator $\mathbf{F}$ obeys the following transformation (see Secs. 2.11 and 2.12 of Ref. \citep{Haake1991Quantum}):
\begin{equation}\label{TRevOp}
    T \mathbf{F} T^{-1} = \mathbf{F}^\dagger,
\end{equation}
where the Floquet operator is given by 
\begin{equation}
    \mathbf{F} = \exp\left(- i \frac{k'}{2j}J_x^2\right)\exp\left(- i \frac{k}{2j}J_z^2\right)\exp\left(- i p J_y\right).
\end{equation}
Applying conjugation operator $K$ to the transformation given above, we get
\begin{align}
    U K \left(\exp\left(-i \frac{k'}{2j}J_x^2\right)\exp\left(-i \frac{k}{2j}J_z^2\right)\exp\left(-i p J_y\right)\right) K^{-1} U^\dagger = U \left(\exp\left(i \frac{k'}{2j}J_x^2\right)\exp\left(i \frac{k}{2j}J_z^2\right)\exp\left(- i p J_y\right)\right) KK^{-1} U^\dagger.
\end{align}
Here, arguments of the first two exponential terms in the bracket are purely imaginary. Whereas the argument of the third term is real. Therefore, arguments of the first two exponential terms gets opposite sign leaving the third unchanged during conjugation operation. Let us consider $U = \exp\left(i p J_y\right) V$ then, we get
\begin{equation}
    T \; \mathbf{F}\; T^{-1} = \exp\left(i p J_y\right) V \left(\exp\left(i \frac{k'}{2j}J_x^2\right)\exp\left(i \frac{k}{2j}J_z^2\right)\exp\left(- i p J_y\right)\right) V^\dagger \exp\left(-i p J_y\right).
\end{equation}
Now, we choose a unitary operator $V = W \exp\left(i\pi J_z\right)$. Here, $\exp\left(i\pi J_z\right)$ transforms $(J_y\to -J_y, J_x\to -J_x)$. Inserting identity operator $I = \exp\left(-i\pi J_z\right)\exp\left(i\pi J_z\right)$, we get 
\begin{align}
    \begin{split}
        T \; \mathbf{F}\; T^{-1} =& \exp\left(i p J_y\right) \left(W \exp\left(i\pi J_z\right)\right) \left(\exp\left(i \frac{k'}{2j}J_x^2\right)\exp\left(i \frac{k}{2j}J_z^2\right)\cdot I \cdot \exp\left(- i p J_y\right)\right) \left(\exp\left(-i\pi J_z\right) W^\dagger\right) \exp\left(-i p J_y\right) \\
        =& \exp\left(i p J_y\right) W \exp\left(i\pi J_z\right) \left(\exp\left(i \frac{k'}{2j}J_x^2\right)\exp\left(i \frac{k}{2j}J_z^2\right)\right) \exp\left(-i\pi J_z\right) \exp\left(i p J_y\right) W^\dagger \exp\left(-i p J_y\right) \\
        =& \exp\left(i p J_y\right) W \left(\exp\left(i \frac{k'}{2j}J_x^2\right)\exp\left(i \frac{k}{2j}J_z^2\right)\right) \exp\left(i p J_y\right) W^\dagger \exp\left(-i p J_y\right).
    \end{split}
\end{align}
For case $k' = 0$, it can be shown that the choice $W=I$ satisfies Eq. (\ref{TRevOp}). Thus, the time-reversal operator for the standard QKT is given by
\begin{equation}\label{eq:stdT}
    T = \exp\left(i p J_y\right) \exp\left(i\pi J_z\right) K.
\end{equation}
However, for case $k'= k$, we need an operator that transforms $(J_x\to J_z, J_z\to -J_x)$. Hence, we choose $W = \exp\left(i\frac{\pi}{2}J_y\right)$. For this choice of $T$ operator, we get
\begin{align}
    \begin{split}
        T \; \mathbf{F}\; T^{-1} =& \exp\left(i p J_y\right) \exp\left(i\frac{\pi}{2}J_y\right) \left(\exp\left(i \frac{k}{2j}J_x^2\right)\exp\left(i \frac{k}{2j}J_z^2\right)\right) \exp\left(i p J_y\right) \exp\left(-i\frac{\pi}{2}J_y\right) \exp\left(-i p J_y\right)  \\
        =& \exp\left(i p J_y\right) \exp\left(i\frac{\pi}{2}J_y\right) \left(\exp\left(i \frac{k}{2j}J_x^2\right)\exp\left(i \frac{k}{2j}J_z^2\right)\right) \exp\left(-i\frac{\pi}{2}J_y\right)  \\
        =& \exp\left(i p J_y\right) \exp\left(i \frac{k}{2j}J_z^2\right)\exp\left(i \frac{k}{2j}J_x^2\right)  \\
        =& \; \mathbf{F}^\dagger.         
    \end{split}
\end{align}
Therefore, the time-reversal operator for case $k'=k$ is given by
\begin{equation}
    T = \exp\left(i p J_y\right) \exp\left(i\frac{\pi}{2}J_y\right) \exp\left(i\pi J_z\right) K.
\end{equation}
This operator and Eq.~(\ref{eq:stdT}) are presented in the main text.

\suppsection[]{Exact analytical solution for 2-qubits}\label{supp:sec:2qubit}
In this section, we derive time evolved linear entropy, and it's infinite-time average for any initial spin-coherent state $|\psi_0 \rangle = |\theta_0,\phi_0\rangle$ in the case of two qubits. We derive the infinite-time averaged linear entropy for the two special \citep{dogra2019quantum} states: $|\theta_0 = 0,\phi_0 = 0\rangle$ and $|\theta_0 = \pi/2,\phi_0 = -\pi/2\rangle$. The initial spin-coherent state is given by
\begin{equation}
    |\psi_0 \rangle = \otimes^2 \left[\cos\left(\frac{\theta_0}{2}\right) |0\rangle + e^{-i\phi_0} \sin\left(\frac{\theta_0}{2}\right)|1\rangle\right],
\end{equation}
where $|0\rangle = {\left[1, 0\right]}^T$ and $|1\rangle = {\left[0, 1\right]}^T$. The basis states $|\Phi_0^\pm\rangle$ and $|\Phi_1^+\rangle$ are defined in the qubit basis as follows:
\begin{align}
    |\Phi_0^\pm\rangle &= \frac{1}{\sqrt{2}} |00\rangle \mp \frac{1}{\sqrt{2}} |11\rangle = {\left[\frac{1}{\sqrt{2}}, 0, 0, \mp\frac{1}{\sqrt{2}}\right]}^T, \\
    |\Phi_1^+\rangle &= \frac{1}{\sqrt{2}} |10\rangle + \frac{1}{\sqrt{2}} |01\rangle = {\left[0, \frac{1}{\sqrt{2}}, \frac{1}{\sqrt{2}}, 0\right]}^T.
\end{align}
The initial spin-coherent state can be represented in the $\Phi$-basis as follows:
\begin{align}
    |\psi_0\rangle = \frac{1}{\sqrt{2}}\Big\lbrace &e^{-i\phi_0} \left(\cos(\theta_0)\cos(\phi_0) + i \sin(\phi_0)\right) |\Phi_0^+\rangle + e^{-i\phi_0} \sin(\phi_0) |\Phi_1^+\rangle + \left[\cos^2\left(\frac{\theta_0}{2}\right) + i \sin^2\left(\frac{\theta_0}{2}\right)\right] |\Phi_0^-\rangle \Big\rbrace.
\end{align}
The $n$-th power of the Floquet operator in the $\Phi$-basis is given by
\begin{equation}
    \mathcal{U}^n = \begin{bmatrix}
        \cos\left(\frac{n\pi}{2}\right) & -\sin\left(\frac{n\pi}{2}\right)e^{-i\frac{k_\theta}{2}} & 0 \\
        \sin\left(\frac{n\pi}{2}\right)e^{i\frac{k_\theta}{2}} & \cos\left(\frac{n\pi}{2}\right) & 0 \\
        0 & 0 & e^{-i\frac{n k_r}{2}}
    \end{bmatrix}.
\end{equation}
Thus, the time-evolved state $|\psi_n\rangle = \mathcal{U}^n |\psi_0\rangle$ can be expressed as follows:
\begin{equation}
    |\psi_n\rangle = c_0 |\Phi_0^+\rangle + c_1 |\Phi_1^+\rangle + c_2 |\Phi_0^-\rangle,
\end{equation}
where the coefficients $c_0$, $c_1$, and $c_2$ are given by
\begin{align}
    c_0 &= \frac{1}{\sqrt{2}} e^{-\frac{i}{2}(k_\theta + 2\phi_0)} \big[-\sin\left(\frac{n\pi}{2}\right)\sin(\theta_0) + e^{i\frac{k_\theta}{2}} \cos\left(\frac{n\pi}{2}\right)\big(\cos(\theta_0)\cos(\phi_0) + i \sin(\phi_0)\big)\big], \\
    c_1 &= \frac{1}{\sqrt{2}} e^{-i\phi_0} \big[\cos\left(\frac{n\pi}{2}\right)\sin(\theta_0) + e^{i\frac{k_\theta}{2}} \sin\left(\frac{n\pi}{2}\right)\big(\cos(\theta_0)\cos(\phi_0) + i \sin(\phi_0)\big)\big] \text{ and } \\
    c_2 &= \frac{1}{\sqrt{2}} e^{-\frac{i}{2}(n k_r + 2\phi_0)} \big(\cos(\phi_0) + i \cos(\theta_0)\sin(\phi_0)\big).
\end{align}
The density matrix $\rho_{12}(n) = |\psi_n\rangle \langle\psi_n|$ in the qubit basis is:
\begin{equation}
    \rho_{12}(n) = \frac{1}{2}\begin{bmatrix}
         |c_0 + c_2|^2 & (c_0 + c_2)c_1^* & (c_0 + c_2)c_1^* & -(c_0 + c_2)(c_0 - c_2)^* \\
         (c_0 + c_2)^*c_1 & |c_1|^2 & |c_1|^2 & -c_1(c_0 - c_2)^* \\
         (c_0 + c_2)^*c_1 & |c_1|^2 & |c_1|^2 & -c_1(c_0 - c_2)^* \\
         -(c_0 - c_2)(c_0 + c_2)^* & -(c_0 - c_2)c_1^* & -(c_0 - c_2)c_1^* & 1 - |c_1|^2 - c_2c_0^*
    \end{bmatrix}.
\end{equation}
The partial trace over one qubit yields the reduced density matrix (RDM) $\rho_1(n)$ given by
\begin{equation}\label{eq:2qubitrho}
    \rho_1(n) = \begin{bmatrix}
        \frac{1}{2} + \Re[c_0 c_2^*] & \Re[c_1 c_2^*] + i \Im[c_0 c_1^*] \\
        \Re[c_1 c_2^*] - i \Im[c_0 c_1^*] & \frac{1}{2} - \Re[c_0 c_2^*]
    \end{bmatrix}.
\end{equation}
Its eigenvalues are $R_\pm = \frac{1}{2} \pm \sqrt{\Re[c_0 c_2^*]^2 + \Im[c_0 c_1^*]^2 + \Re[c_1 c_2^*]^2}$ and eigenvectors are given by 
\begin{align}
    { \bigg[ \frac{ \Re[c_0 c_2^*] \pm \sqrt{{ \Re[c_0 c_2^*]}^2  +  {\Re[c_1 c_2^*]}^2 + {\Im[c_0 c_1^*]}^2 }}{\Re[c_1 c_2^*] - i \Im[c_0 c_1^*]} , 1 \bigg] }^T.
\end{align}
Then, the linear entropy is given by
\begin{equation}\label{eq:S2qubit}
    S_{(\theta_0, \phi_0)}^{(2)}(n, k_r, k_\theta) = 2R_+ R_- = \frac{1}{2} - 2\big(\Re[c_0 c_2^*]^2 + \Im[c_0 c_1^*]^2 + \Re[c_1 c_2^*]^2\big).
\end{equation}
The coefficients $\Re[c_0 c_2^*]$, $\Im[c_0 c_1^*]$, and $\Re[c_1 c_2^*]$ are given as follows:
\begin{alignat}{2}
    \Re[c_0 c_2^*] =& -\frac{1}{4}\sin\left(\frac{n\pi}{2}\right)\Big[2\cos\left(\frac{k_\theta - n k_r}{2}\right)\cos(\phi_0)\sin(\theta_0) + \sin\left(\frac{k_\theta - n k_r}{2}\right)\sin(2\theta_0)\sin(\phi_0)\Big] \notag \\
    & \quad + \frac{1}{4}\cos\left(\frac{n\pi}{2}\right)\Big[2\cos\left(\frac{n k_r}{2}\right)\cos(\theta_0) + \sin\left(\frac{n k_r}{2}\right)\sin^2(\theta_0)\sin(2\phi_0)\Big], \label{eq1} \\
    \Re[c_1 c_2^*] =& \frac{1}{4}\cos\left(\frac{n\pi}{2}\right)\Big[2\cos\left(\frac{n k_r}{2}\right)\cos(\phi_0)\sin(\theta_0) - \sin\left(\frac{n k_r}{2}\right)\sin(2\theta_0)\sin(\phi_0)\Big] \notag \\
    & \quad + \frac{1}{4}\sin\left(\frac{n\pi}{2}\right)\Big[2\cos\left(\frac{n k_r + k_\theta}{2}\right)\cos(\theta_0) + \sin\left(\frac{n k_r + k_\theta}{2}\right)\sin^2(\theta_0)\sin(2\phi_0)\Big], \label{eq2} \\
    \Im[c_0 c_1^*] =& \frac{1}{16}\sin\left(\frac{k_\theta}{2}\right)\sin(n\pi)\big(1 + 3\cos(2\theta_0) - 2\cos(2\phi_0)\sin^2(\theta_0)\big) - \frac{1}{4}\cos(\phi_0)\sin(k_\theta)\sin^2\left(\frac{n\pi}{2}\right)\sin(2\theta_0) \notag \\
    & \quad + \frac{1}{4}\big[1 + \cos(k_\theta) - (1 - \cos(k_\theta))\cos(n\pi)\big]\sin(\theta_0)\sin(\phi_0). \label{eq3}
\end{alignat}
Finally, the infinite-time average \citep{dogra2019quantum} linear entropy is given by
\begin{align}
    \langle S^{(2)}_{(\theta_0, \phi_0)}(k_r, k_\theta) \rangle &= \lim_{N\to\infty} \frac{1}{N} \sum_{n=0}^{N-1} S_{(\theta_0, \phi_0)}^{(2)}(n, k_r, k_\theta) \notag \\
    &= \frac{106 + 8\cos(2\theta_0) + 14\cos(4\theta_0) - 4\cos(2\theta_0 - 4\phi_0) + \cos(4\theta_0 - 4\phi_0) + 6\cos(4\phi_0) + \cos(4\theta_0 + 4\phi_0)}{1024} \notag \\
    &\quad + \frac{32\cos(2k_\theta)\sin^2(\theta_0)\big[\cos(2\phi_0)(3 + \cos(2\theta_0)) - 2\sin^2(\theta_0)\big] - 4\cos(2\theta_0 + 4\phi_0)}{1024} \\
    &\quad + \frac{1}{128}\big[3 + \cos(2\theta_0) + 2\cos(2\phi_0)\sin^2(\theta_0)\big]^2 + \frac{1}{16}\sin(2k_\theta)\sin(\theta_0)\sin(2\theta_0)\sin(2\phi_0). \notag
\end{align}

\suppsubsection{\texorpdfstring{Initial State: $|\theta_0 = 0, \phi_0 = 0\rangle$}{}}
This state is of a particular interest as it belongs to a period-4 cycle. Evolving this state under the Floquet operator yields the following quantum state at time $n$:
\begin{align}
    \begin{split}
        |\psi_n \rangle = \frac{\cos(\frac{n\pi}{2})}{\sqrt{2}}|\Phi_0^+\rangle + \frac{e^{\frac{i k_\theta}{2}}\sin(\frac{n\pi}{2})}{\sqrt{2}}|\Phi_1^+ \rangle + \frac{e^{-\frac{i n k_r}{2}}}{\sqrt{2}} |\Phi_0^-\rangle. 
    \end{split}
\end{align}
To analyse the entanglement between 2 qubits, we compute the single-qubit RDM $\rho_1(n) = \text{tr}_{2}\left(|\psi_n\rangle\langle\psi_n |\right)$. For odd $n$, the single-qubit RDM is given as follows:
\begin{equation}
    \rho_1(n) = \frac{1}{2}
        \begin{pmatrix}
            1  &- i^{n+1}\cos(\frac{nk_r + k_\theta}{2}) \\
            - i^{n+1} \cos(\frac{nk_r + k_\theta}{2})  &1
        \end{pmatrix}.
\end{equation}
The eigenvalues of this matrix are $\left(1 \pm \cos\left(\frac{nk_r + k_\theta}{2}\right)\right)\big/2$. For even $n$, the single-qubit RDM is given by
\begin{equation}
    \rho_1(n) = \frac{1}{2} \begin{pmatrix}
            1 + i^n\cos(\frac{n k_r}{2})   &0 \\
            0  &1 - i^n\cos(\frac{n k_r}{2})
        \end{pmatrix}.
\end{equation}
The linear entropy is then calculated using these eigenvalues as follows:
\begin{align}\label{eq:S2qubit00}
    S^{(2)}_{(0,0)}\left(n,k_r,k_\theta\right) = \begin{cases}
        \frac{1}{2}\sin^2\left(\frac{n k_r + k_\theta}{2}\right) & \text{odd } n\\
        \frac{1}{2}\sin^2\left(\frac{n k_r}{2}\right) & \text{even } n.
    \end{cases}
\end{align}
From the above expression, we can clearly observe that the linear entropy show periodicity for all $k_r = a \pi$ with $a\in \mathds{Q}$ and independent of $k_\theta$.  The infinite-time-averaged linear entropy is calculated as:
\begin{align}\label{eq:S2qubitavg}
    \begin{split}
        \langle S^{(2)}_{(0,0)}\left(k_r,k_\theta\right)\rangle &= \lim_{N\to \infty} \frac{1}{N} \sum_{n=0}^{N-1} S^{(2)}_{(0,0)}\left(n, k_r, k_\theta\right) \\ 
        &= \frac{1}{4},
    \end{split}
\end{align}

\suppsubsection{\texorpdfstring{Initial State: $|\theta_0 = \pi/2, \phi_0 = -\pi/2\rangle$}{}}
The other interesting state belongs to the fixed points $(0, \pm 1, 0)$ in the phase-space. This initial state, also referred to as the positive parity state $|++\,+\rangle$, evolves into the following state:
\begin{align}
    \begin{split}
        |\psi_n \rangle = \left[\cos(\frac{n\pi}{2}) - i e^{-\frac{i k_\theta}{2}} \sin(\frac{n\pi}{2})\right] |\Phi_0^+ \rangle + \left[e^{\frac{i k_\theta}{2}} \sin(\frac{n\pi}{2}) + i \cos(\frac{n\pi}{2})\right] |\Phi_1^+ \rangle.
    \end{split}
\end{align}
Taking the partial trace over one qubit from the state $|\psi_n \rangle$, we obtain the single-qubit RDM as follows:
\begin{align}
    \rho_1(n) = \frac{1}{2} \begin{cases}
    \begin{pmatrix}
            1  &-i \cos(k_\theta) \\
            i \cos(k_\theta)  &1
        \end{pmatrix} & \text{odd } n, \\
         \begin{pmatrix}
            1  &-i \\
            i  &1
        \end{pmatrix} & \text{even } n.
    \end{cases}
\end{align}
For odd $n$, the eigenvalues are $\cos^2\left(\frac{k_\theta}{2}\right)$ and $\sin^2\left(\frac{k_\theta}{2}\right)$. For even $n$, they are $1$ and zero. Then, the linear entropy is given by
\begin{align}\label{eq:S2qubitpi2}
    S^{(2)}_{(\frac{\pi}{2},-\frac{\pi}{2})}\left(n, k_r, k_\theta\right) = \begin{cases}
        \frac{1}{2} \sin^2\left(\frac{k_\theta}{2}\right)  & \text{odd } n,\\
        0 & \text{even } n.
    \end{cases}
\end{align}
The linear entropy for this initial state is periodic. Our computations show that if a particular initial state shows periodic nature then every initial state also shows the same periodic nature for that kick strength $k_r$. This is shown in the main text. The infinite-time-averaged linear entropy for this initial state is calculated as follows:
\begin{align}\label{eq:S2qubitavgpi2}
    \langle S^{(2)}_{(\frac{\pi}{2},-\frac{\pi}{2})}\left(k_r,k_\theta\right)\rangle = \frac{1}{4}\sin^2\left(\frac{k_\theta}{2}\right).
\end{align}
The Eqns.~(\ref{eq:2qubitrho}), (\ref{eq:S2qubit}), (\ref{eq:S2qubit00}), (\ref{eq:S2qubitavg}), (\ref{eq:S2qubitpi2}) and (\ref{eq:S2qubitavgpi2}) are presented in the main text.

\suppsection[]{Exact analytical solution for 3-qubits}\label{supp:sec:3qubit}
In this section, we derive the time evolved linear entropy for the general initial spin-coherent state $|\psi_0 \rangle = |\theta_0,\phi_0\rangle$. Further, we obtain infinite-time average linear entropy for two specific initial states, namely $|\theta_0 = 0, \phi_0 = 0\rangle$ and $|\theta_0 = \pi/2, \phi_0 = -\pi/2\rangle$. The initial spin-coherent state for the 3-qubit system is expressed as follows:
\begin{equation}
    |\psi_0\rangle = \otimes^3 \left[\cos(\frac{\theta_0}{2}) | 0\rangle + e^{-i\phi_0}\sin(\frac{\theta_0}{2})| 1 \rangle\right].
\end{equation}
The states $|\Phi_0^\pm\rangle$ and $|\Phi_1^\pm\rangle$ in the qubit basis are written as:
\begin{align}
    |\Phi_0^\pm\rangle &= \frac{1}{\sqrt{2}} |000\rangle \pm \frac{1}{\sqrt{2}} |111\rangle = {\left[\frac{1}{\sqrt{2}}, 0, 0, 0, 0, 0, 0, \mp\frac{i}{\sqrt{2}}\right]}^T, \\
    |\Phi_1^\pm\rangle &= \frac{1}{\sqrt{2}} |W\rangle \pm \frac{i}{\sqrt{2}} |\overline{W}\rangle = {\left[0, \frac{1}{\sqrt{6}}, \frac{1}{\sqrt{6}}, \pm\frac{i}{\sqrt{6}}, \frac{1}{\sqrt{6}}, \pm\frac{i}{\sqrt{6}}, \pm\frac{i}{\sqrt{6}}, 0\right]}^T,
\end{align}
where $|W\rangle = \frac{1}{\sqrt{3}} \sum_{\mathcal{P}} |001\rangle_{\mathcal{P}}$, $|\overline{W}\rangle = \frac{1}{\sqrt{3}} \sum_{\mathcal{P}} |110\rangle_{\mathcal{P}}$, and $\sum_{\mathcal{P}}$ denotes the sum over all permutations. The general initial state $|\psi_0\rangle$ in the $\Phi$-basis is given by:
\begin{align}
    |\psi_0\rangle =& \frac{1}{\sqrt{2}} \left[\cos^3\left(\frac{\theta_0}{2}\right) + i e^{-3i\phi_0}\sin^3\left(\frac{\theta_0}{2}\right)\right] |\Phi_0^+\rangle + \sqrt{\frac{3}{8}} e^{-2i\phi_0} \left[-i + e^{i\phi_0}\cot\left(\frac{\theta_0}{2}\right)\right]\sin\left(\frac{\theta_0}{2}\right)\sin(\theta_0) |\Phi_1^+\rangle \nonumber \\
    &+\frac{1}{\sqrt{2}} \left[\cos^3\left(\frac{\theta_0}{2}\right) - i e^{-3i\phi_0}\sin^3\left(\frac{\theta_0}{2}\right)\right] |\Phi_0^-\rangle + \sqrt{\frac{3}{8}} e^{-2i\phi_0} \left[i + e^{i\phi_0}\cot\left(\frac{\theta_0}{2}\right)\right]\sin\left(\frac{\theta_0}{2}\right)\sin(\theta_0) |\Phi_1^-\rangle.
\end{align}
Thus, the time-evolved state $|\psi_n\rangle = \mathcal{U}^n |\psi_0\rangle$ can be expressed as follows:
\begin{align}
    \mathcal{U}^n =& e^{-\frac{i}{3}nk_r} \begin{pmatrix}
        e^{-in\frac{\pi}{4}}\alpha_n    &-e^{-in\frac{\pi}{4}}\beta_n^*    &0   &0 \\
        e^{-in\frac{\pi}{4}}\beta_n     &e^{-in\frac{\pi}{4}}\alpha_n^* &0 &0 \\
        0         &0       &{(-1)}^n e^{n\frac{\pi}{4}} \alpha_n   &{(-1)}^n e^{n\frac{\pi}{4}}\beta^* \\
        0         &0       &-{(-1)}^n e^{n\frac{\pi}{4}} \beta_n   &{(-1)}^n e^{n\frac{\pi}{4}}\alpha_n^* 
    \end{pmatrix},
\end{align}
where,
\begin{align}
    \alpha_n =& \cos(n\gamma) + \frac{i}{4} \frac{\sin (n\gamma)}{\sin\gamma}\left[3\cos(\frac{2k_\theta}{3}) - \cos(\frac{2k_r}{3})\right]  \, \text{and} \\
        \beta_n =& \frac{\sqrt{3}}{4} \frac{\sin (n\gamma)}{\sin\gamma} \left[\cos(\frac{2k_r}{3})+\cos(\frac{2k_\theta}{3}) + 2i\sin(\frac{2k_\theta}{3})\right].
\end{align}
The time-evolved state $|\psi_n\rangle$ is expanded as follows:
\begin{align}
    |\psi_n\rangle =& c_0' |\Phi_0^+\rangle + c_1' |\Phi_1^+\rangle + c_2' |\Phi_0^-\rangle + c_3' |\Phi_1^-\rangle, 
\end{align}
where the coefficients $c_i'$ are given as follows:
\begin{alignat}{2}
    c_0' =& \frac{e^{-\frac{i}{4}(n\pi + 6\phi_0)}}{2\sqrt{2}} \left[\cos\left(\frac{\theta_0 + \phi_0}{2}\right)-i\sin\left(\frac{\theta_0 - \phi_0}{2}\right)\right] \left[2\alpha_n \cos(\theta_0)\cos(\phi_0) - \sqrt{3}\beta_n^* \sin(\theta_0) + i \alpha_n \left(\sin(\theta_0) + 2\sin(\phi_0) \right)\right], \\
    c_1' =& \frac{e^{-\frac{i}{4}(n\pi + 6\phi_0)}}{2\sqrt{2}} \left[\cos\left(\frac{\theta_0 + \phi_0}{2}\right)-i\sin\left(\frac{\theta_0 - \phi_0}{2}\right)\right] \left[2\beta_n \cos(\theta_0)\cos(\phi_0) + \sqrt{3}\alpha_n^* \sin(\theta_0) + i \alpha_n \left(\sin(\theta_0) + 2\sin(\phi_0) \right)\right], \\
    c_2' =& \frac{e^{\frac{i}{4}(5n\pi - 6\phi_0)}}{2\sqrt{2}} \left[\cos\left(\frac{\theta_0 - \phi_0}{2}\right)+i\sin\left(\frac{\theta_0 + \phi_0}{2}\right)\right] \left[2\alpha_n \cos(\theta_0)\cos(\phi_0) + \sqrt{3}\beta_n^* \sin(\theta_0) - i \alpha_n \left(\sin(\theta_0) - 2\sin(\phi_0) \right)\right], \\
    c_3' =& -\frac{e^{\frac{i}{4}(5n\pi - 6\phi_0)}}{2\sqrt{2}} \left[\cos\left(\frac{\theta_0 - \phi_0}{2}\right)+i\sin\left(\frac{\theta_0 + \phi_0}{2}\right)\right] \left[2\beta_n \cos(\theta_0)\cos(\phi_0) - \sqrt{3}\alpha_n^* \sin(\theta_0) - i \beta_n \left(\sin(\theta_0) - 2\sin(\phi_0) \right)\right].
\end{alignat} 
The two-qubit RDM $\rho_{12}(n) = \tr_{3} |\psi_n\rangle \langle \psi_n|$ is given by
\begin{align}
    \rho_{12}(n) = \begin{pmatrix}
        a_1   &a_2   &a_2   &a_3 \\
        a_2^*   &\frac{1}{3}\left({|c_1'|}^2 + {|c_3'|}^2\right)    &\frac{1}{3}\left({|c_1'|}^2 + {|c_3'|}^2\right)    &a_4 \\
        a_2^*   &\frac{1}{3}\left({|c_1'|}^2 + {|c_3'|}^2\right)    &\frac{1}{3}\left({|c_1'|}^2 + {|c_3'|}^2\right)    &a_4  \\
        a_3^*   &a_4^*   &a_4^*   &a_5
    \end{pmatrix},
\end{align}
where,
\begin{align}
    a_1 &= \frac{1}{2} {|c_0' + c_2'|}^2 + \frac{1}{6} {|c_1' + c_3'|}^2, 
    a_2 = -\frac{i}{6} (c_1' + c_3'){(c_1' - c_3')}^* + \frac{\sqrt{3}}{6} (c_0' + c_2') {(c_1' + c_3')}^*, \nonumber \\
    a_3 &= \frac{i}{2\sqrt{3}} (c_1' + c_3') {(c_0' - c_2')}^* - \frac{i}{2\sqrt{3}} (c_0' + c_2') {(c_1' - c_3')}^*, a_4 = -\frac{1}{2\sqrt{3}} (c_1' - c_3'){(c_0' - c_2')}^* -\frac{i}{6} (c_1' + c_3'){(c_1' - c_3')}^* \text{ and } \\
    a_5 &= \frac{1}{2} - \Re[c_0' {c_2'}^*]  - \frac{1}{3}\left({|c_1'|}^2 + {|c_3'|}^2\right) - \frac{1}{3}\Re[c_1' {c_3'}^*].\nonumber 
\end{align}
Then, the single qubit RDM $\rho_1(n) = \tr_{23} |\psi_n\rangle \langle \psi_n|$ is given by
\begin{align}\label{eq:3qubitrho}
    \rho_1(n) =& \begin{pmatrix}
        \frac{1}{2} + \Re[c_0' {c_2'}^*] + \frac{1}{3}\Re[c_1' {c_3'}^*]   &a_2 + a_4 \\
        {(a_2 + a_4)}^*   &\frac{1}{2} - \Re[c_0' {c_2'}^*] - \frac{1}{3}\Re[c_1' {c_3'}^*]
    \end{pmatrix}, \\
    a_2 + a_4 =& -\frac{i}{3} (c_1' + c_3'){(c_1' - c_3')}^* + \frac{\sqrt{3}}{6} (c_0' + c_2') {(c_1' + c_3')}^* -\frac{1}{2\sqrt{3}} (c_1' - c_3'){(c_0' - c_2')}^*.
\end{align}
Its eigenvalues are $\dfrac{1}{2}\pm 2{\left[{\left(\Re[c_0' {c_2'}^*] + \dfrac{1}{3}\Re[c_1' {c_3'}^*]\right)}^2 + 2{|a_2 + a_4|}^2\right]}^\frac{1}{2}$ and eigenvectors are given by
\begin{equation}
    {\left[ \frac{ \Re[c_0' {c_2'}^*] + \frac{1}{3}\Re[c_1' {c_3'}^*]  \pm \sqrt{ {\left(\Re[c_0' {c_2'}^*] + \frac{1}{3}\Re[c_1' {c_3'}^*]\right)}^2 + {| a_2 + a_4 |}^2 } }{a_2^* + a_4^*} , 1 \right]}^T.
\end{equation}
Thus, linear entropy is given by
\begin{align}\label{eq:S3qubit}
    S^{(3)}_{(\theta_0, \phi_0)}(n, k_r, k_\theta) = \frac{1}{2} - 2 {\left(\Re[c_0' {c_2'}^*] + \frac{1}{3}\Re[c_1' {c_3'}^*]\right)}^2 - 2 {|a_2 + a_4|}^2.
\end{align}
While it is, in principle, possible to calculate the infinite-time average linear entropy $\langle S^{(3)}_{(\theta_0, \phi_0)}(k_r, k_\theta) \rangle$, the resulting expression is exceedingly lengthy and thus is not explicitly presented here. 

\suppsubsection{Initial State: \texorpdfstring{$|\theta_0 = 0, \phi_0 = 0\rangle$}{}}
Here, we derive the expression for an infinite-time average linear entropy for the state $|\psi_0\rangle = |000\rangle$. The time-evolved state $|\psi_n\rangle = \mathcal{U}^n |000\rangle$ is then obtained as follows:
\begin{align}
    |\psi_n\rangle =&  \frac{1}{\sqrt{2}} \mathcal{U}^n |\Phi_0^+\rangle + \frac{1}{\sqrt{2}} \mathcal{U}^n |\Phi_0^-\rangle \nonumber \\
    =& \frac{1}{\sqrt{2}} e^{-\frac{i}{4}n\pi} \alpha_n | \Phi_0^+\rangle 
    + \frac{1}{\sqrt{2}} e^{-\frac{i}{4}n\pi} \beta_n | \Phi_1^+\rangle + \frac{1}{\sqrt{2}} (-1)^n e^{-\frac{i}{4}n\pi} \alpha_n | \Phi_0^-\rangle 
    - \frac{1}{\sqrt{2}} (-1)^n e^{-\frac{i}{4}n\pi} \beta_n | \Phi_1^-\rangle.
\end{align}
The single qubit RDM $\rho_1(n)$, obtained by performing a partial trace over any two qubits and is given by
\begin{align}
    \rho_1(n) =& \begin{pmatrix}
        |\alpha_n|^2 \cos^2\left(\frac{3n\pi}{4}\right) 
        + \frac{|\beta_n|^2}{6} \left[3 - \cos\left(\frac{3n\pi}{2}\right)\right] 
        & -\left(\frac{|\beta_n|^2}{3} + \frac{\Im[\alpha_n \beta^*_n]}{\sqrt{3}} \right) \sin\left(\frac{3n\pi}{2}\right) \\
        -\left(\frac{|\beta_n|^2}{3} + \frac{\Im[\alpha_n \beta^*_n]}{\sqrt{3}} \right) \sin\left(\frac{3n\pi}{2}\right) 
        & |\alpha_n|^2 \sin^2\left(\frac{3n\pi}{4}\right) 
        + \frac{|\beta_n|^2}{6} \left[3 + \cos\left(\frac{3n\pi}{2}\right)\right]
    \end{pmatrix}.
\end{align}
For even $n$, $\rho_1(n)$ becomes diagonal. Let $n = 2m$ with $m \in \mathds{Z}$. The odd $n$ state can be obtained via the inverse Floquet operator:
\begin{align}
    |\psi_{2m-1}\rangle = \mathcal{U}^{-1}|\psi_{2m}\rangle.
\end{align} 
For even $m$, the surviving terms are:
\begin{align}
    |\psi_{2m}\rangle = \frac{e^{-i\left(\frac{k_r}{3} + \frac{3\pi}{4}\right) 2m}}{\sqrt{2}}\left(\alpha_{2m}|000\rangle + i\beta_{2m}|\overline{W}\rangle\right).
\end{align} 
For $k_\theta \neq k_r$ (i.e., $k \neq 0$ and $k' \neq 0$), the action of $\mathcal{U}^{-1}$ does not yield a local phase factor (due to the existence of an additional $k'$-term) on $|000\rangle$ or $|\overline{W}\rangle$. As a result, $|\psi_{2m-1}\rangle$ is not locally unitarily equivalent to the state after even $n$ applications of $\mathcal{U}$. Therefore, the entanglement and concurrence do not show the step-like features seen in the standard QKT where $k_\theta \neq k_r$.

The linear entropy corresponding to this state is given by
\begin{align}\label{eq:S3qubit00}
    S^{(3)}_{(0,0)} \left(n, k_r, k_\theta\right) =& 2 \Bigg(|\alpha_n|^2 \cos^2\left(\frac{3n\pi}{4}\right) 
    + \frac{|\beta_n|^2}{6} \left[3 - \cos\left(\frac{3n\pi}{2}\right)\right]\Bigg) \Bigg(|\alpha_n|^2 \sin^2\left(\frac{3n\pi}{4}\right) 
    + \frac{|\beta_n|^2}{6} \left[3 + \cos\left(\frac{3n\pi}{2}\right)\right]\Bigg) \nonumber \\
    & - 2{\left(\frac{|\beta_n|^2}{3} + \frac{\Im[\alpha_n \beta^*_n]}{\sqrt{3}} \right)}^2 \sin^2\left(\frac{3n\pi}{2}\right).
\end{align}
Note that $n$ appears with $\gamma$ through $\alpha_n$, $\beta_n$, their conjugates and products. It also appears with multiple of $3\pi/4$ and $3\pi/2$. However, these terms are kick strengths independent and periodic. The $\alpha_n$ and $\beta_n$ both are polynomials in $\gamma$. Therefore, the linear entropy is periodic if there exists a $\gamma = a\pi$ with $a \in \mathds{Q}$ satisfying:
\begin{align}\label{Eq:periodicEq}
 \frac{1}{2}\sin\left(\frac{2k_r}{3}\right) = \cos(a\pi) \,\,\text{and} \,\, \frac{1}{3} \leq a \leq \frac{2}{3}.
\end{align}

Finally, the infinite-time average linear entropy is given by
\begin{align}\label{eq:S3qubitavg}
    \langle S^{(3)}_{(0,0)} \left(k_r, k_\theta\right) \rangle =&  \frac{1026 + 13 \cos\left(\dfrac{8k_r}{3}\right) + \left[304 - 52\cos\left(\dfrac{4k_\theta}{3}\right)\right] \cos\left(\dfrac{4k_r}{3}\right) - 112\cos\left(\dfrac{4k_\theta}{3}\right)}{64 {\left[7+\cos\left(\dfrac{4k_r}{3}\right)\right]}^2} \notag\\
    &+ \frac{8 \cos\left(\dfrac{2k_r}{3}\right) \cos\left(\dfrac{2k_\theta}{3}\right) \left[-2 + 9 \cos \left(\dfrac{4k_\theta}{3}\right) + \cos\left(\dfrac{4k_r}{3}\right) \right] - 27\cos\left(\dfrac{8k_\theta}{3}\right)}{64 {\left[7+\cos\left(\dfrac{4 k_r}{3}\right)\right]}^2}.
\end{align}
The special case $k_\theta = k_r$ of this expression is derived in the Ref. \citep{dogra2019quantum}. 

\suppsubsection{Initial State: \texorpdfstring{$|\theta_0 = \pi/2, \phi_0 = -\pi/2\rangle = |++\,+\rangle$}{}}
Here, we derive the expression for an infinite-time average linear entropy for the initial positive parity state $|\psi_0\rangle = |++\,+\rangle$. The time-evolved state $|\psi_n\rangle = \mathcal{U}^n |++\,+\rangle$ is then given by
\begin{align}
    |\psi_n\rangle =& \mathcal{U}^n \left(\frac{1}{2} |\Phi_0^+\rangle + i\frac{\sqrt{3}}{2} |\Phi_1^+\rangle\right) \nonumber \\
    =& \frac{1}{2} e^{-\frac{i}{4}n\pi} \left(\alpha_n - i\sqrt{3}\beta_n^*\right) |\Phi_0^+\rangle 
    + \frac{1}{2} e^{-\frac{i}{4}n\pi} \left(\beta_n + i\sqrt{3}\alpha_n^*\right) |\Phi_1^+\rangle.
\end{align}
Denoting $\eta_n = \frac{1}{2}(\alpha_n - i\sqrt{3}\beta_n^*)$ and $\delta_n = \frac{1}{2}(\beta_n + i\sqrt{3}\alpha_n^*)$ for brevity, the RDM $\rho_{12}(n)$ is given by
\begin{align}
    \rho_{12}(n) =& \begin{pmatrix}
        \frac{1}{2}{|\eta_n|}^2 + \frac{1}{6}{|\delta_n|}^2 & -\frac{i}{6} {|\delta_n|}^2 + \frac{\eta_n \delta_n^*}{2\sqrt{3}} & -\frac{i}{6} {|\delta_n|}^2 + \frac{\eta_n \delta_n^*}{2\sqrt{3}} & \frac{\Im[\eta_n \delta_n^*]}{2\sqrt{3}} \\
        \frac{i}{6} {|\delta_n|}^2 + \frac{\eta_n^* \delta_n}{2\sqrt{3}} & \frac{{|\delta_n|}^2}{3} & \frac{{|\delta_n|}^2}{3} & -\frac{i}{6} {|\delta_n|}^2 - \frac{\delta_n \eta_n^*}{2\sqrt{3}} \\
        \frac{i}{6} {|\delta_n|}^2 + \frac{\eta_n^* \delta_n}{2\sqrt{3}} & \frac{{|\delta_n|}^2}{3} & \frac{{|\delta_n|}^2}{3} & -\frac{i}{6} {|\delta_n|}^2 - \frac{\delta_n \eta_n^*}{2\sqrt{3}} \\
        \frac{\Im[\eta_n \delta_n^*]}{2\sqrt{3}} & \frac{i}{6} {|\delta_n|}^2 - \frac{\delta_n^* \eta_n}{2\sqrt{3}} & \frac{i}{6} {|\delta_n|}^2 - \frac{\delta_n^* \eta_n}{2\sqrt{3}} & \frac{1}{2}{|\eta_n|}^2 + \frac{1}{6}{|\delta_n|}^2
    \end{pmatrix}.
\end{align}
Then, the single qubit RDM $\rho_1(n)$ is given by
\begin{equation}
    \rho_1(n) = \begin{pmatrix}
        \frac{1}{2} & -\frac{i}{3}|\delta_n|^2 + \frac{i}{\sqrt{3}} \Im\left(\eta_n \delta_n^*\right) \\
        \frac{i}{3}|\delta_n|^2 - \frac{i}{\sqrt{3}} \Im\left(\eta_n \delta_n^*\right) & \frac{1}{2}
    \end{pmatrix}.
\end{equation}
The corresponding linear entropy is given by
\begin{align} \label{eq:S3qubitpi2}
    S^{(3)}_{\left(\frac{\pi}{2},-\frac{\pi}{2}\right)} \left(n, k_r, k_\theta\right) =& \frac{1}{2} - 2\Bigg(\frac{1}{3}|\delta_n|^2 - \frac{1}{\sqrt{3}} \Im\left(\eta_n \delta_n^*\right)\Bigg)^2 \nonumber \\
    =& \frac{1}{2} - 2\Bigg(\frac{1}{12}|\alpha_n - i\sqrt{3}\beta_n^*|^2 
    - \frac{1}{4\sqrt{3}} \Im\Big[(\alpha_n - i\sqrt{3}\beta_n^*) (\beta_n^* - i\sqrt{3}\alpha_n)\Big]\Bigg)^2.
\end{align}
For this state also $n$ appears with $\gamma$ through $\alpha_n$, $\beta_n$, their conjugates and products. Therefore, the choice of $\gamma$ alone decides the periodicity of the linear entropy. Therefore, it satisfies the Eq.~(\ref{Eq:periodicEq}). The linear entropy for this initial state is periodic. Our computations show that if a particular initial state shows periodic nature then every initial state also shows the same periodic nature for that kick strength $k_r$. This is shown in the main text.

Finally, the infinite-time average linear entropy is given by
\begin{align}\label{eq:S3qubitavgpi2}
    \begin{split}
        \langle S^{(3)}_{(\frac{\pi}{2},-\frac{\pi}{2})} \left(k_r,k_\theta\right) \rangle =&  \frac{410 + 5 \cos\left(\dfrac{8k_r}{3}\right)+ 4\left[28 - 9\cos\left(\dfrac{4k_\theta}{3}\right)\right]\cos\left(\dfrac{4k_r}{3}\right)-144\cos\left(\dfrac{4k_\theta}{3}\right)}{32 {\left[7+\cos\left(\dfrac{4k_r}{3}\right)\right]}^2} \\
        & + \frac{8 \cos\left(\dfrac{2k_r}{3}\right) \cos\left(\dfrac{2k_\theta}{3}\right) \left[10 + 9\cos\left(\dfrac{4k_\theta}{3}\right) + \cos\left(\dfrac{4k_r}{3}\right) \right] - 27\cos\left(\dfrac{8k_\theta}{3}\right)}{32 {\left[7+\cos\left(\dfrac{4k_r}{3}\right)\right]}^2}.
    \end{split}
\end{align}
The special case $k_\theta = k_r$ of this expression is derived in the Ref. \citep{dogra2019quantum}. 
The Eqns.~(\ref{eq:3qubitrho}), (\ref{eq:S3qubit}), (\ref{eq:S3qubit00}), (\ref{Eq:periodicEq}), (\ref{eq:S3qubitavg}), (\ref{eq:S3qubitpi2}) and (\ref{eq:S3qubitavgpi2}) are presented in the main text.

\suppsection[]{Exact analytical solution for 4-qubits}\label{supp:sec:4qubit}
In this section, we derive the time evolved linear entropy for the general initial spin-coherent state $|\psi_0 \rangle = |\theta_0,\phi_0\rangle$. Further, we obtain infinite-time average linear entropy for two specific initial states, namely $|\theta_0 = 0, \phi_0 = 0\rangle$ and $|\theta_0 = \pi/2, \phi_0 = -\pi/2\rangle$. The initial spin-coherent state for the 4-qubit system is expressed as follows:\begin{align}
    |\psi_0\rangle =& \otimes^4 \left[\cos(\frac{\theta_0}{2}) | 0\rangle + e^{-i\phi_0}\sin(\frac{\theta_0}{2})| 1\rangle\right].
\end{align}
The states $|\Phi_0^\pm\rangle$, $|\Phi_1^\pm\rangle$ and $|\Phi_2^+\rangle$ are written in qubit basis as follows:
\begin{align}
    |\Phi_0^\pm\rangle =& \frac{1}{\sqrt{2}} |0000\rangle \pm \frac{1}{\sqrt{2}} |1111\rangle = {\left[\frac{1}{\sqrt{2}}, 0,  0,  0,  0,  0,  0,  0,  0,  0,  0,  0,  0,  0,  0,  \pm\frac{1}{\sqrt{2}}\right]}^T, \\
    |\Phi_1^\pm\rangle =& \frac{1}{\sqrt{2}} |W\rangle \mp \frac{1}{\sqrt{2}} |\overline{W}\rangle = {\left[0,  \frac{1}{2 \sqrt{2}},  \frac{1}{2 \sqrt{2}},  0,  \frac{1}{2 \sqrt{2}},  0,  0,  \mp\frac{1}{2 \sqrt{2}},  \frac{1}{2 \sqrt{2}},  0,  0,  \mp\frac{1}{2 \sqrt{2}},  0,  \mp\frac{1}{2 \sqrt{2}},  \mp\frac{1}{2 \sqrt{2}},  0\right]}^T, \\ 
    |\Phi_2^+\rangle =& \frac{1}{\sqrt{6}} \sum_{\mathcal{P}}  |0011\rangle = {\left[0, 0, 0, \frac{1}{\sqrt{6}}, 0, \frac{1}{\sqrt{6}}, \frac{1}{\sqrt{6}}, 0, 0, \frac{1}{\sqrt{6}}, \frac{1}{\sqrt{6}}, 0, \frac{1}{\sqrt{6}}, 0, 0, 0\right]}^T,
\end{align}
where $|W\rangle = \frac{1}{2} \sum_{\mathcal{P}} |0001\rangle_{\mathcal{P}}$, $|\overline{W}\rangle = \frac{1}{2} \sum_{\mathcal{P}} |1110\rangle_{\mathcal{P}}$ and $\sum_{\mathcal{P}}$ sums over all possible permutations. Now, the general initial state $|\psi_0\rangle$ in the above basis is given by
\begin{align}
    |\psi_0\rangle =& \frac{1}{\sqrt{2}} \left[\cos^4\left(\frac{\theta_0}{2}\right) + e^{-4i\phi_0}\sin^4\left(\frac{\theta_0}{2}\right)\right] |\Phi_0^+\rangle + \frac{1}{\sqrt{2}} e^{-2i\phi_0} \sin(\theta_0) \left[\cos(\theta_0) \cos(\phi_0)+ i \sin(\phi_0)\right] |\Phi_1^+\rangle + \sqrt{\frac{3}{8}} e^{-2i\phi_0} \sin^2(\theta_0) | \Phi_2^+ \rangle  \nonumber \\
    &+ \frac{1}{\sqrt{2}} \left[\cos^4\left(\frac{\theta_0}{2}\right) - e^{-4i\phi_0}\sin^4\left(\frac{\theta_0}{2}\right)\right] |\Phi_0^-\rangle + \frac{1}{\sqrt{2}} e^{-2i\phi_0} \sin(\theta_0) \left[ \cos(\phi_0) + i \cos(\theta_0) \sin(\phi_0)\right] |\Phi_1^-\rangle.
\end{align}
The $n$-th power of the Floquet operator in the $\Phi$-basis is given by
\begin{align}
    \mathcal{U}^n = \begin{pmatrix}
        e^{-\frac{i}{2}n (k_r + \pi)} \alpha'_n   &0    &i e^{-\frac{i}{2}n (k_r + \pi)} {\beta'_n}^* &0   &0\\
        0    &{(-1)}^n e^{-\frac{i}{2}n (k_r + \pi)}   &0     &0      &0 \\
        i e^{-\frac{i}{2}n (k_r + \pi)} \beta'_n  &0    &e^{-\frac{i}{2}n (k_r + \pi)} {\alpha'_n}^*  &0   &0\\
        0   &0   &0   &\cos(\frac{n\pi}{2}) e^{-\frac{3i}{4} n k_r}   &-\sin(\frac{n\pi}{2})e^{-\frac{3i}{4} k_\theta}e^{-\frac{3i}{4} n k_r}\\
        0   &0   &0   &-\sin(\frac{n\pi}{2})e^{-\frac{3i}{4} k_\theta} e^{-\frac{3i}{4} n k_r}    &\cos(\frac{n\pi}{2}) e^{-\frac{3i}{4} n k_r}
    \end{pmatrix},
\end{align}
where,
\begin{align}
    \alpha'_n =& \cos(n \gamma) + \frac{i}{4} \frac{\sin(n\gamma)}{\sin(\gamma)} \left[3\cos(k_\theta) - \cos(k_r)\right] \, \text{and}  \\
    \beta'_n =&  \frac{\sqrt{3}}{4} \frac{\sin(n\gamma)}{\sin(\gamma)} \left[ \cos(k_r) + \cos(k_\theta) + 2i \sin(k_\theta)\right].
\end{align}
Thus, the time-evolved state $|\psi_n \rangle = \mathcal{U}^n |\psi_0\rangle$ can be expressed as follows:
\begin{align}
    |\psi_n\rangle =& c_0'' |\Phi_0^+\rangle + c_1'' |\Phi_1^+\rangle + c_2'' |\Phi_2^+\rangle + c_3'' |\Phi_0^-\rangle + c_4'' |\Phi_1^-\rangle,
\end{align}
where the coefficients $c_i^{\prime\prime}$ are given by
\begin{align}
    c_0'' =& \frac{e^{-\frac{i}{2}(nk_r+n\pi+8\phi_0)}}{2\sqrt{2}}  \left\{2\alpha_n \left[\cos^4\left(\frac{\theta_0}{2}\right) + e^{-4i\phi_0}\sin^4\left(\frac{\theta_0}{2}\right)\right] + i \sqrt{3} e^{2i\phi_0} \beta_n^* \sin^2(\theta_0) \right\}, \\
    c_1'' =& {(-1)}^n \frac{e^{-2i\phi_0}}{\sqrt{2}}  \sin(\theta_0) \left[\cos(\theta_0) \cos(\phi_0)+ i \sin(\phi_0)\right], \\
    c_2'' =& \frac{e^{-\frac{i}{2}(nk_r+n\pi+8\phi_0)}}{2\sqrt{2}}  \left\{2i\beta_n \left[e^{4i\phi_0}\cos^4\left(\frac{\theta_0}{2}\right) + \sin^4\left(\frac{\theta_0}{2}\right)\right] + \sqrt{3} e^{2i\phi_0} \beta_n^* \sin^2(\theta_0) \right\}, \\
    c_3'' =& \frac{e^{-\frac{i}{4}(3k_\theta +3nk_r + 16\phi_0)}}{\sqrt{2}} \left\{ e^{\frac{3ik_\theta}{4}} \cos\left(\frac{n\pi}{2}\right) \left[e^{4i\phi_0}\cos^4\left(\frac{\theta_0}{2}\right) - \sin^4\left(\frac{\theta_0}{2}\right)\right]  - e^{2i\phi_0} \sin\left(\frac{n\pi}{2}\right) \sin(\theta_0) \left[\cos(\phi_0) + i \cos(\theta_0)\sin(\phi_0)\right] \right\},\\
    c_4'' =& \frac{e^{-\frac{3ink_r}{4}}}{\sqrt{2}}  \left\{ e^{\frac{3ik_\theta}{4}} \cos\left(\frac{n\pi}{2}\right) \left[\cos^4\left(\frac{\theta_0}{2}\right) - e^{-4i\phi_0}\sin^4\left(\frac{\theta_0}{2}\right)\right]  + e^{-2i\phi_0} \cos\left(\frac{n\pi}{2}\right) \sin(\theta_0) \left[\cos(\phi_0) + i \cos(\theta_0)\sin(\phi_0)\right] \right\}.
\end{align}
The two-qubit RDM $\rho_{12}(n) = \tr_{34} |\psi_n\rangle \langle \psi_n|$ is given by
\begin{align}
    \rho_{12}(n) = \begin{pmatrix}
        b_1   &b_2   &b_2   &b_3 \\
        b_2^*   &\frac{{|c_1''|}^2 + {|c_4''|}^2}{4} + \frac{{|c_2''|}^2}{3}    &\frac{{|c_1''|}^2 + {|c_4''|}^2}{4} + \frac{{|c_2''|}^2}{3}    &b_4 \\
        b_2^*   &\frac{{|c_1''|}^2 + {|c_4''|}^2}{4} + \frac{{|c_2''|}^2}{3}    &\frac{{|c_1''|}^2 + {|c_4''|}^2}{4} + \frac{{|c_2''|}^2}{3}    &b_4  \\
        b_3^*   &b_4^*   &b_4^*   &b_5
    \end{pmatrix},
\end{align}
where,
\begin{alignat}{2}
    b_1 =& \frac{1}{6} {|c_1''|}^2 + \frac{1}{2} {|c_0'' + c_3''|}^2 + \frac{1}{4} {|c_1'' + c_4''|}^2, \notag \\
    b_2 =&  \frac{\sqrt{3}}{6} (c_1'' + c_4'') {c_2''}^* - \frac{\sqrt{3}}{12} {(c_1'' - c_4'')}^* c_2'' + \frac{1}{4} (c_0'' + c_3'') {(c_1'' + c_4'')}^*, \notag  \\
    b_3 =&  \frac{\sqrt{3}}{6} (c_0'' + c_3'') {c_2''}^* + \frac{\sqrt{3}}{6} {(c_0'' - c_3'')}^* c_2'' - \frac{1}{4} (c_1'' + c_4'') {(c_1'' - c_4'')}^*,  \\
    b_4 =& -\frac{1}{4} (c_1'' - c_4'') {(c_0'' - c_3'')}^* + \frac{\sqrt{3}}{12} (c_1'' + c_4'') {c_2''}^* - \frac{\sqrt{3}}{6} {(c_1'' + c_4'')}^* c_2'', \notag \\
    b_5 =&  \frac{1}{2} - \Re[c_0''{c_3''}^*] - \frac{1}{3} {|c_2''|}^2 - \frac{1}{4} {|c_1'' + c_4''|}^2, \notag
\end{alignat}
and the single qubit RDM $\rho_1(n)$ is given by
\begin{align}\label{eq:4qubitrho}
    \rho_1(n) = \begin{pmatrix}
        \frac{1}{2}\left(1+2\,\Re[c_0'' {c_3''}^*]+2\,\Re[c_1'' {c_4''}^*]\right) &p_{12}'' \\
        {p_{12}''}^*   &\frac{1}{2}\left(1-2\,\Re[c_0'' {c_3''}^*]-2\,\Re[c_1'' {c_4''}^*]\right)
    \end{pmatrix},
\end{align}
where
\begin{equation}
    p_{12}'' = \frac{1}{4}(c_1''-c_4'') {(c_3''-c_0'')}^* + \frac{1}{4}(c_0''+c_3'') {(c_1''+c_4'')}^* + \frac{\sqrt{3}}{4}c_2'' {(c_4''-c_1'')}^* + \frac{\sqrt{3}}{4}{c_2''}^* (c_4''+c_1'').
\end{equation}
Its eigenvalues are $\frac{1}{2} \pm \sqrt{ {\Re[c_0'' {c_3''}^*]+\,\Re[c_1'' {c_4''}^*]}^2 + {| p_{12}'' |}^2}$ and eigenvectors are given by
\begin{equation}
    { \left[ \frac{\Re[c_0'' {c_3''}^*]+\,\Re[c_1'' {c_4''}^*] \pm \sqrt{ {\Re[c_0'' {c_3''}^*]+\,\Re[c_1'' {c_4''}^*]}^2 + {| p_{12}'' |}^2} }{{p_{12}''}^*} , 1 \right]}^T.
\end{equation}  
Then the corresponding linear entropy is given by
\begin{align}\label{eq:S4qubit}
    S^{(4)}_{(\theta_0,\phi_0)}(n, k_r, k_\theta) =& \frac{1}{2} - 2 {\left(\Re[c_0'' {c_2''}^*] + \Re[c_1'' {c_4''}^*]\right)}^2  - 2 {|p_{12}''|}^2.
\end{align}
Here also, in principle, it is possible to calculate the infinite-time average linear entropy $\langle S^{(4)}_{(\theta_0, \phi_0)}(k_r, k_\theta) \rangle$. However, the resulting expression is exceedingly lengthy and thus is not explicitly presented here. 

\suppsubsection{The initial state: \texorpdfstring{$|\theta_0 = 0, \phi_0 = 0\rangle$}{}}
Here, we derive the expression for an infinite-time average linear entropy for the state $|\psi_0\rangle = |0000\rangle$. The time-evolved state $|\psi_n\rangle = \mathcal{U}^n |0000\rangle$ is then obtained as follows:
\begin{align}
    \begin{split}
        |\psi_n\rangle =&  \frac{1}{\sqrt{2}} \mathcal{U}^n |\Phi_0^+\rangle + \frac{1}{\sqrt{2}} \mathcal{U}^n|\Phi_0^-\rangle \\
        =&  e^{-\frac{i}{2}n (k_r + \pi)} \left[\frac{1}{\sqrt{2}}\alpha'_n |\Phi_0^+\rangle + \frac{i}{\sqrt{2}} \beta'_n |\Phi_2^+\rangle + \frac{1}{\sqrt{2}} e^{-\frac{i}{4}n k_r}\cos(\frac{n\pi}{2})|\Phi_0^-\rangle + \frac{1}{\sqrt{2}} e^{-\frac{i}{4} (n k_r - 3 k_\theta)}\sin(\frac{n\pi}{2})|\Phi_1^-\rangle \right].
    \end{split}
\end{align}
The two qubit RDM $\rho_{12}(n)$ is obtained by carrying out partial trace over any two qubits and is given by 
\begin{align}
    \rho_{12}(n) = \begin{pmatrix}
        b_1   &b_2   &b_2   &b_3 \\
        b_2^*   &\frac{1 -{(-1)}^n}{16} + \frac{{|\beta_n|}^2}{6}    &\frac{1 -{(-1)}^n}{16} + \frac{{|\beta_n|}^2}{6}    &b_4 \\
        b_2^*   &\frac{1 -{(-1)}^n}{16} + \frac{{|\beta_n|}^2}{6}    &\frac{1 -{(-1)}^n}{16} + \frac{{|\beta_n|}^2}{6}    &b_4  \\
        b_3^*   &b_4^*   &b_4^*   &b_5
    \end{pmatrix},
\end{align}
where,
\begin{align}
    b_1 =& -\frac{{|\beta_n|}^2}{6} + \frac{1}{16} \left[7 + {(-1)}^n + 8\Re[\alpha_n' \delta_n]\cos\left(\frac{n\pi}{2}\right)\right], \,\,
    b_2 =  \frac{1}{24}  \left(-2i\sqrt{3} \varepsilon_n \beta_n^* + \left(3\alpha_n + i\sqrt{3}\beta_n\right)\varepsilon_n \right)\sin\left(\frac{n\pi}{2}\right), \notag \\
    b_3 =&  \frac{1 -{(-1)}^n}{16} + i \frac{\sqrt{3}}{12} \beta_n \alpha_n^* - i \frac{\sqrt{3}}{12} \left[\alpha_n \beta_n^* + \Re[\beta_n \delta_n^*] \cos\left(\frac{n\pi}{2}\right) \right], \,\,
    b_4 =  \frac{1}{24} \left(3\varepsilon_n\alpha_n^* - i\sqrt{3} \varepsilon_n \beta_n^* + 2 i \sqrt{3} \beta_n\varepsilon_n^* \right) \sin\left(\frac{n\pi}{2}\right), \text{ and }  \\
    b_5 =& -\frac{{|\beta_n|}^2}{6} + \frac{1}{16} \left[7 + {(-1)}^n - 8\Re[\alpha_n' \delta_n]\cos\left(\frac{n\pi}{2}\right)\right]. \notag
\end{align}
The single qubit RDM $\rho_{1}(n)$ is given by 
\begin{equation}\label{Eq:4qbitRho}
    \rho_1(n) = \frac{1}{4} \begin{pmatrix}
        2(1+\Re[\alpha'_n \delta_n^*])  \cos\left(\frac{n\pi}{2}\right) &\Re[\varepsilon_n^* (\alpha'_n + i\sqrt{3}\beta'_n )]  \sin\left(\frac{n\pi}{2}\right) \\
        \Re[\varepsilon_n^* (\alpha'_n + i\sqrt{3}\beta'_n )]  \sin\left(\frac{n\pi}{2}\right)    &2(1-\Re[\alpha'_n \delta_n^*])\cos\left(\frac{n\pi}{2}\right)
    \end{pmatrix},
\end{equation}
where, $\delta_n = e^{-\frac{i}{4}nk_r}$ and $\varepsilon_n = e^{-\frac{i}{4}(n k_r - 3k_\theta)}$. The eigenvalues of $\rho_1(n)$ for odd $n$ are $1/2\pm \Re[\varepsilon_n^* (\alpha'_n + i\sqrt{3}\beta'_n )]$. For even $n$, the eigenvalues are $1/2\pm \Re[\alpha'_n \delta_n^*]$. The corresponding linear entropy is given by
\begin{align}\label{eq:S4qubit00}
    S^{(4)}_{(0,0)} \left(n, k_r, k_\theta\right) = \frac{1}{2} - \frac{1+\cos(n\pi)}{4} {\Re[\alpha'_n \delta_n^*]}^2 + \frac{-1+\cos(n\pi)}{16} {\Re[\varepsilon_n^* (\alpha'_n + i\sqrt{3}\beta'_n )]}^2,
\end{align}
and the infinite-time average linear entropy is obtained as follows:
\begin{align}\label{eq:S4qubitavg}
    \langle S^{(4)}_{(0,0)} \left(k_r,k_\theta\right) \rangle =\frac{160+25\cos(2k_r)-9\cos(2k_\theta)}{64\left[7+\cos(2k_r)\right]}.
\end{align}
The special case $k_\theta = k_r$ of this expression is derived in the Ref. \citep{dogra2019quantum}. 
To derive the condition for periodicity, we need to consider the following two non-trivial components:
\begin{align}
    \begin{split}
        {(\rho_1(n))}_{11} =& \frac{1}{2} (1+\Re[\alpha'_n \delta_n^*])  \cos\left(\frac{n\pi}{2}\right)\; \text{ and } \;
        {(\rho_1(n))}_{12} = \frac{1}{4} \Re[\varepsilon_n^* (\alpha'_n + i\sqrt{3}\beta'_n )],
    \end{split}
\end{align}
where,
\begin{align}
    \begin{split}
        \Re[\alpha'_n \delta_n^*] =& \cos\left(\frac{nk_r}{4}\right) \cos(n\gamma) + \frac{\cos(k_r) - 3\cos(k_\theta)}{4\sin(\gamma)} \sin\left(\frac{nk_r}{4}\right) \sin(n\gamma) \\
        \Re[\varepsilon_n^* (\alpha'_n + i\sqrt{3}\beta'_n )] =& - \frac{1}{2\sin(\gamma)} \left[\cos(k_r)+3\cos(k_\theta)\right] \sin(n\gamma) \sin\left(\frac{nk_r - 3k_\theta}{4}\right) \\
        &- \frac{1}{2\sin(\gamma)} \cos\left(\frac{nk_r - 3k_\theta}{4}\right) \left[3\sin(k_\theta) \sin(n\gamma) - 2\cos(n\gamma) \sin(\gamma)\right].
    \end{split}
\end{align}
Note that $n$ appears only with $\gamma$ and $k_r/4$. Then, the above expressions are periodic if $\gamma$ and $k_r/4$ both are rational multiple of $\pi$. Therefore, the linear entropy is periodic if there exist $a, b \in \mathds{Q}$, such that
\begin{align}\label{Eq:4qubitPeriocity}
    \begin{split}
 \frac{1}{2}\sin(k_r) = \cos(a\pi), \; \frac{k_r}{4} = b\pi \;\text{ and }\; \frac{1}{3} \leq a \leq \frac{2}{3}.
    \end{split}
\end{align}

\suppsubsection{The initial state: \texorpdfstring{$|\theta_0 = \pi/2, \phi_0 = -\pi/2\rangle$}{}}
Here, we derive the expression for an infinite-time average linear entropy for the initial positive parity state $|\psi_0\rangle = |++++\rangle$. The time-evolved state $|\psi_n\rangle = \mathcal{U}^n |++++\rangle$ is then given by
\begin{equation}
    |++++\rangle = \frac{1}{\sqrt{8}}|\Phi_0^+\rangle + \frac{i}{\sqrt{2}}|\Phi_1^+\rangle -\sqrt{\frac{3}{8}}|\Phi_2^+\rangle.  
\end{equation}
The time evolved state $|\psi_n\rangle = \mathcal{U}^n|++++\rangle$ is given by 
\begin{align}
    |\psi_n\rangle =& {(-1)}^n \left[ \frac{1}{\sqrt{2}} e^{i\delta'_n} \left(\frac{\alpha'_n - i\sqrt{3}{\beta'_n}^*}{2}\right) |\Phi_0^+\rangle + i\frac{1}{\sqrt{2}}|\Phi_1^+\rangle  - \frac{1}{\sqrt{2}} e^{i\delta'_n} \left(\frac{\sqrt{3}{\alpha'_n}^* - i\beta'_n}{2}\right) |\Phi_2^+\rangle \right],
\end{align}
where, $\delta'_n = n(\pi - k_r)/2$. Then, the two qubit RDM $\rho_{12}(n)$ obtained by carrying out partial trace over any two qubits, is given by 
\begin{align}
    \rho_{12}(n) = \begin{pmatrix}
        b_1   &b_2   &b_2   &b_3 \\
        b_2^*   &\frac{1}{24} \left(7 - 4{|\xi_n|}^2\right)    &\frac{1}{24} \left(7 - 4{|\xi_n|}^2\right)    &b_4 \\
        b_2^*   &\frac{1}{24} \left(7 - 4{|\xi_n|}^2\right)    &\frac{1}{24} \left(7 - 4{|\xi_n|}^2\right)    &b_4  \\
        b_3^*   &b_4^*   &b_4^*   &b_5
    \end{pmatrix},
\end{align}
where, 
\begin{align}
    b_1 =& \frac{3}{8} - \frac{{|\xi_n|}^2}{6},
    b_2 =  -\frac{1}{24} e^{-i\delta_n'} \left(-i e^{2i\delta_n'}(3\xi_n - i \sqrt{3}\chi_n)  +2\sqrt{3} \chi_n^*\right), 
    b_3 =  -\frac{1}{8} + \frac{\sqrt{3}}{6}\Im[\xi_n \chi_n^*], \\
    b_4 =&  \frac{1}{24} e^{-i\delta_n'} \left[-3i \xi_n^* + \sqrt{3}\left(-2e^{2i\delta_n'}\chi_n + \chi_n^*\right)\right], 
    b_5 =   \frac{3}{8} - \frac{{|\xi_n|}^2}{6}, 
    \xi_n =  \frac{\alpha'_n - i\sqrt{3}{\beta'_n}^*}{2} \,\, \text{and}\,\, \chi_n = \frac{\sqrt{3}{\alpha'_n}^* - i\beta'_n}{2}. \notag
\end{align}
The single qubit RDM $\rho_1(n)$ is given by
\begin{align}
    \rho_1(n) =& \begin{pmatrix}
        1/2 &-\frac{i}{8} \left[e^{i\delta_n'} \left(\xi_n -i\sqrt{3}\chi_n\right) + e^{-i\delta_n'} {\left(\xi_n -i\sqrt{3}\chi_n\right)}^* \right] \\
        \frac{i}{8} \left[e^{i\delta_n'} \left(\xi_n -i\sqrt{3}\chi_n\right) + e^{-i\delta_n'} {\left(\xi_n -i\sqrt{3}\chi_n\right)}^* \right] & 1/2
    \end{pmatrix}. \nonumber
\end{align}
The corresponding linear entropy is given by
\begin{align}\label{eq:S4qubitpi2}
    S^{(4)}_{(\frac{\pi}{2},-\frac{\pi}{2})} \left(n, k_r, k_\theta\right) = \frac{1}{2} - \frac{1}{32} {\left[e^{i\delta_n'} \left(\xi_n -i\sqrt{3}\chi_n\right) + e^{-i\delta_n'} {\left(\xi_n -i\sqrt{3}\chi_n\right)}^* \right]}^2.
\end{align}
Since $\chi_n$ and $\xi_n$ are linear functions of $\alpha_n'$ and $\beta_n'$, the periodicity of the corresponding terms is given by $\gamma$. In addition, $\delta_n'$ is a linear function of $k_r$. Hence, for this state also the linear entropy is periodic provided it satisfies the Eq.~(\ref{Eq:4qubitPeriocity}).

Finally, the infinite-time average linear entropy then is obtained as follows:
\begin{align}\label{eq:S4qubitavgpi2}
    \langle S^{(4)}_{(\frac{\pi}{2},-\frac{\pi}{2})} \left(k_r, k_\theta\right) \rangle = \frac{3}{8} -  \frac{{\left[\cos(k_r) + 3\cos(k_\theta)\right]}^2}{16 {\left[7+\cos(2k_r)\right]}}.
\end{align}
The special case $k_\theta = k_r$ of this expression is derived in the Ref. \citep{dogra2019quantum}. 
The Eqns.~(\ref{eq:4qubitrho}), (\ref{eq:S4qubit}), (\ref{eq:S4qubit00}), (\ref{eq:S4qubitavg}), (\ref{Eq:4qubitPeriocity}), (\ref{eq:S4qubitpi2}) and (\ref{eq:S4qubitavgpi2}) are presented in the main text.

\end{document}